\input amstex
\documentstyle{amsppt}
\nopagenumbers
\nologo
\catcode`@=11
\redefine\output@{%
  \def\break{\penalty-\@M}\let\par\endgraf
  \shipout\vbox{%
    \ifplain@
      \let\makeheadline\relax \let\makefootline\relax
    \else
      \iffirstpage@ \global\firstpage@false
        \let\rightheadline\frheadline
        \let\leftheadline\flheadline
      \else
        \ifrunheads@ 
        \else \let\makeheadline\relax
        \fi
      \fi
    \fi
    \makeheadline \pagebody \makefootline
  }%
  \advancepageno \ifnum\outputpenalty>-\@MM\else\dosupereject\fi
}
\font\cpr=cmr7
\newcount\xnumber
\footline={\xnumber=\pageno
\divide\xnumber by 7
\multiply\xnumber by -7
\advance\xnumber by\pageno
\ifnum\xnumber>0\hfil\else\vtop{\vskip 0.5cm
\noindent\cpr CopyRight \copyright\ Sharipov R.A.,
1997, 2003.}\hfil\fi}
\def\setfirstpage{\global\firstpage@true}
\catcode`\@=\active
\def\startpage#1{\pageno=#1}
\fontdimen3\tenrm=3pt
\fontdimen4\tenrm=0.7pt

\def\leaderfill{\leaders\hbox to 0.3em{\hss.\hss}\hfill}
\font\tvbf=cmbx12
\font\tvrm=cmr12
\font\etbf=cmbx8
\def\vtrule{\vrule height 12pt depth 6pt}
\def\divr{\operatorname{div}}
\def\grad{\operatorname{grad}}
\def\const{\operatorname{const}}
\def\rot{\operatorname{rot}}
\def\MatGrSO{\operatorname{SO}}
\def\MatGrO{\operatorname{O}}
\def\Span{\operatorname{Span}}
\def\tr{\operatorname{tr}}
\let\ch=\cosh
\let\sh=\sinh
\let\th=\tanh
\Monograph
\pagewidth{10.1cm}
\pageheight{14.5cm}
\loadbold
\document
\global\hoffset=0cm\global\voffset=0cm
\centerline{\etbf RUSSIAN FEDERAL COMMITTEE}
\centerline{\etbf FOR HIGHER EDUCATION}
\bigskip
\centerline{\etbf BASHKIR STATE UNIVERSITY}
\vskip 3cm
\centerline{SHARIPOV\ R.\,A.}
\vskip 1.5cm
\centerline{\tvbf CLASSICAL \ ELECTRODYNAMICS}
\vskip 0.2cm
\centerline{\tvbf AND \ THEORY \ OF \ RELATIVITY}
\vskip 1.3cm
\centerline{\tvrm the manual}
\vskip 4.0cm
\centerline{Ufa 1997}
\newpage
\vbox to 13.5cm{
UDC 517.9\par
Sharipov R. A. {\bf Classical Electrodynamics and Theory of
Relativity}: the manual / Publ\. of Bashkir State University
--- Ufa, 1997. --- pp\.~163. --- ISBN 5-7477-0180-0.
\bigskip
\bigskip
     This book is a manual for the course of electrodynamics
and theory of relativity. It is recommended primarily for students 
of mathematical departments. This defines its style: I use
elements of vectorial and tensorial analysis, differential
geometry, and theory of distributions in it.\par
     In preparing Russian edition of this book I used
computer typesetting on the base of \AmSTeX\ package and
I used cyrillic fonts of Lh-family distributed by CyrTUG
association of Cyrillic \TeX\ users. English edition is also
typeset by \AmSTeX.\par
     This book is published under the approval by Methodic
Commission of Mathematical Department of Bashkir State University.
\bigskip
\noindent
Referees:\ \ \ \
\vtop{\hsize 7.5cm\noindent Chair of Algebra and Geometry
of Bashkir State Pedagogical University (BGPI),\vskip 0.1cm
\noindent Prof. V. A. Baikov, Ufa State University for
\hphantom{\qquad}\linebreak
Aviation and Technology (UGATU).}
\vfil
\line{ISBN 5-7477-0180-0\hss\copyright\ Sharipov R.A., 1997}
\line{English Translation\hss\copyright\ Sharipov R.A., 2003}}
\newpage
\ \bigskip\medskip
\centerline{\bf CONTENTS.}
\bigskip
\line{CONTENTS.\ \leaderfill\ 3.}
\medskip
\line{PREFACE.\ \leaderfill\ 5.}
\medskip
\line{CHAPTER~\uppercase\expandafter{\romannumeral 1}.
ELECTROSTATICS AND MAGNETO-\hss}
\line{\qquad STATICS.\ \leaderfill\ 7.}
\medskip
\line{\S\,1. Basic experimental facts and unit systems.\ \leaderfill\ 7.}
\line{\S\,2. Concept of near action.\ \leaderfill\ 13.}
\line{\S\,3. Superposition principle.\ \leaderfill\ 15.}
\line{\S\,4. Lorentz force and Biot-Savart-Laplace law.
\ \leaderfill\ 18.}
\line{\S\,5. Current density and the law of charge conservation.
\ \leaderfill\ 21.}
\line{\S\,6. Electric dipole moment.\ \leaderfill\ 24.}
\line{\S\,7. Magnetic moment.\ \leaderfill\ 26.}
\line{\S\,8. Integral equations of static electromagnetic field.\
\leaderfill\ 31.}
\line{\S\,9. Differential equations of static electromagnetic field.\
\leaderfill\ 41.}
\bigskip
\line{CHAPTER~\uppercase\expandafter{\romannumeral 2}.
CLASSICAL ELECTRODYNAMICS.\ \ \leaderfill\ 43.}
\medskip
\line{\S\,1. Maxwell equations.\ \leaderfill\ 43.}
\line{\S\,2. Density of energy and energy flow for electromagnetic\hss}
\line{\qquad field.\ \leaderfill\ 46.}
\line{\S\,3. Vectorial and scalar potentials of electromagnetic\hss}
\line{\qquad field.\ \leaderfill\ 54.}
\line{\S\,4. Gauge transformations and Lorentzian gauge.\ \leaderfill\ 56.}
\line{\S\,5. Electromagnetic waves.\ \leaderfill\ 59.}
\line{\S\,6. Emission of electromagnetic waves.\ \leaderfill\ 60.}
\bigskip
\line{CHAPTER~\uppercase\expandafter{\romannumeral 3}.
SPECIAL THEORY OF RELATIVITY.\ \leaderfill\ 68.}
\medskip
\line{\S\,1. Galileo transformations.\ \leaderfill\ 68.}
\line{\S\,2. Lorentz transformations.\ \leaderfill\ 73.}
\line{\S\,3. Minkowsky space.\ \leaderfill\ 77.}
\line{\S\,4. Kinematics of relative motion.\ \leaderfill\ 82.}
\line{\S\,5. Relativistic law of velocity addition.\ \leaderfill\ 90.}
\line{\S\,6. World lines and private time.\ \leaderfill\ 91.}
\line{\S\,7. Dynamics of material point.\ \leaderfill\ 95.}
\line{\S\,8. Four-dimensional form of Maxwell equations.
\ \leaderfill\ 100.}
\line{\S\,9. Four-dimensional vector-potential.\ \leaderfill\ 107.}
\line{\S\,10. The law of charge conservation.\ \leaderfill\ 112.}
\line{\S\,11. Note on skew-angular and curvilinear\hss}
\line{\qquad \ coordinates.\ \leaderfill\ 115.}
\bigskip
\line{CHAPTER~\uppercase\expandafter{\romannumeral 4}.
LAGRANGIAN FORMALISM\hss}
\line{\qquad \ IN THEORY OF RELATIVITY.\ \leaderfill\ 119.}
\medskip
\line{\S\,1. Principle of minimal action for particles
and fields.\ \leaderfill\ 119.}
\line{\S\,2. Motion of particle in electromagnetic field.\
\leaderfill\ 124.}
\line{\S\,3. Dynamics of dust matter.\ \leaderfill\ 128.}
\line{\S\,4. Action functional for dust matter.\ \leaderfill\ 133.}
\line{\S\,5. Equations for electromagnetic field.\ \leaderfill\ 141.}
\bigskip
\line{CHAPTER~\uppercase\expandafter{\romannumeral 5}.
GENERAL THEORY OF RELATIVITY.\ \leaderfill\ 145.}
\medskip
\line{\S\,1. Transition to non-flat metrics and curved Minkowsky\hss}
\line{\qquad space.\ \leaderfill\ 145.}
\line{\S\,2. Action for gravitational field. Einstein equation.\
\leaderfill\ 147.}
\line{\S\,3. Four-dimensional momentum conservation law\hss}
\line{\qquad for fields.\ \leaderfill\ 153.}
\line{\S\,4. Energy-momentum tensor for electromagnetic field.\
\leaderfill\ 155.}
\line{\S\,5. Energy-momentum tensor for dust matter.\
\leaderfill\ 157.}
\line{\S\,6. Concluding remarks.\ \leaderfill\ 160.}
\bigskip
\line{REFERENCES. \leaderfill\ 162.}
\bigskip
\line{CONTACTS. \leaderfill\ 163.}

\newpage
\ \bigskip
\newtoks\truehead
\truehead=\headline
\headline{\hfill}
\centerline{\bf PREFACE.}
\bigskip
     Theory of relativity is a physical discipline
which arose in the beginning of
\uppercase\expandafter{\romannumeral 20}-th century.
It has dramatically changed traditional notion
about the structure of the Universe. Effects predicted
by this theory becomes essential only when we describe
processes at high velocities close to light velocity
$$
c=2.998\cdot 10^5\text{\it km}/\text{\it sec}.
$$
In \uppercase\expandafter{\romannumeral 19}-th century
there was the only theory dealing with such processes,
this was theory of electromagnetism. Development
of theory of electromagnetism in \uppercase
\expandafter{\romannumeral 19}-th century became a
premise for arising theory of relativity.\par
     In this book I follow historical sequence of
events. In Chapter~\uppercase\expandafter{\romannumeral 1}
electrostatics and magnetostatics are explained starting
with first experiments on interaction of charges and currents.
Chapter~\uppercase\expandafter{\romannumeral 2} is devoted to
classical electrodynamics based on Maxwell equations.\par
     In the beginning of
Chapter~\uppercase\expandafter{\romannumeral 3} Lorentz
transformations are derived as transformations keeping form
of Maxwell equations. Physical interpretation of such transformation
requires uniting space and time into one four-dimensional
continuum (Minkowsky space) where there is no fixed 
direction for time axis. Upon introducing four-dimensional
space-time in Chapter~\uppercase\expandafter{\romannumeral 3}
classical electrodynamics is rederived in the form invariant
with respect to Lorentz transformations.\par
     In Chapter~\uppercase\expandafter{\romannumeral 4}
variational approach to describing electromagnetic field and
other material fields in special relativity is considered. Use
of curvilinear coordinates in Minkowsky space
and appropriate differential-geometric methods prepares
background for passing to general relativity.\par
     In Chapter~\uppercase\expandafter{\romannumeral 5}
Einstein's theory of gravitation (general relativity)
is considered, this theory interprets gravitational field as
curvature of space-time itself.\par
     This book is addressed to Math\. students. Therefore
I paid much attention to logical consistence of given material.
References to physical intuition are minimized: in those places,
where I need additional assumptions which do not follow from
previous material, detailed comment is given.\par
\headline=\truehead
     I hope that assiduous and interested reader with
sufficient preliminary background could follow all mathematical
calculations and, upon reading this book, would get pleasure of
understanding how harmonic is the nature of things.\par
     I am grateful to N.~T.~Ahtyamov, D.~I.~Borisov,
Yu.~P.~Ma\-shentseva, and A.~I.~Utarbaev for reading and correcting
Russian version of book.
\bigskip\bigskip
\line{\vbox{\hsize 7.5cm\noindent November, 1997;\newline November,
2003.}\hss R.~A.~Sharipov.}
\newpage
\setfirstpage
\topmatter
\title\chapter{1}
ELECTROSTATICS AND MAGNETOSTATICS
\endtitle
\endtopmatter
\leftheadtext{CHAPTER \uppercase\expandafter{\romannumeral 1}.
ELECTROSTATICS AND MAGNETOSTATICS.}
\document

\head
\S\,1. Basic experimental facts and unit systems.\endhead
\rightheadtext{\S\,1. Basic facts and unit systems.}
     Quantitative description of any physical phenomenon requires
measurements. In mechanics we have three basic quantities and three
basic units of measure: for mass, for length, and for time.
\medskip
\hbox to \hsize{\hss
\vbox{\hsize 10cm
\offinterlineskip\settabs\+\indent
\vtrule
\hskip 1.8cm &\vtrule 
\hskip 2.2cm &\vtrule 
\hskip 2.2cm &\vtrule 
\hskip 2.8cm &\vtrule 
\cr\hrule 
\+\vtrule
\hss Quantity\hss &\vtrule
\hss Unit\hss &\vtrule
\hss Unit\hss &\vtrule
\hss Relation\hss &\vtrule\cr
\vskip -0.2cm
\+\vtrule
\hss &\vtrule
\hss in SI\hss&\vtrule
\hss in SGS\hss&\vtrule
\hss of units\hss&\vtrule\cr\hrule
\+\vtrule
\hss mass\hss&\vtrule
\hss {\it kg}\hss&\vtrule
\hss {\it g}\hss&\vtrule
\hss {\it $1$\,kg\,=\,$10^3$\,g}\hss&\vtrule\cr\hrule
\+\vtrule
\hss length\hss&\vtrule
\hss {\it m}\hss&\vtrule
\hss {\it cm}\hss&\vtrule
\hss {\it $1$\,m\,=\,$10^2$\,cm}\hss&\vtrule\cr\hrule
\+\vtrule
\hss time\hss&\vtrule
\hss {\it sec}\hss&\vtrule
\hss {\it sec}\hss&\vtrule
\hss {\it $1$\,sec\,=\,$1$\,sec}\hss&\vtrule\cr\hrule
}\hss}
\medskip
\par
     Units of measure for other quantities are derived from
the above basic units. Thus, for instance, for measure unit
of force due to Newton's second law we get:
\roster
\item {\it N = kg $\cdot$ m $\cdot$ sec$^{-2}$} in SI,
\item {\it dyn = g $\cdot$ cm $\cdot$ sec$^{-2}$} in SGS.
\endroster
Unit systems SI and SGS are two most popular unit systems in
physics. Units for measuring mechanical quantities (velocity,
acceleration, force, energy, power) in both systems are defined
in quite similar way. Proportions relating units for these
quantities can be derived from proportions for basic quantities
(see table above). However, in choosing units for electric and
magnetic quantities these systems differ essentially.
\par
     Choice of measure unit for electric charge in SGS is based
on Coulomb law describing interaction of two charged point.
\definition{\bf Coulomb law}
\parshape 6 0cm 10.1cm 0cm 10.1cm
4.5cm 5.6cm 4.5cm 5.6cm 4.5cm 5.6cm 4.5cm 5.6cm
Two charged points with charges of the same sign are repulsing,
while points with charges of opposite signs are attracting with
force proportional to quantities of their charges and inverse
proportional to square of distance between them:
$$
F \sim \frac{Q_1\,Q_2}{r^2}.\tag1.1
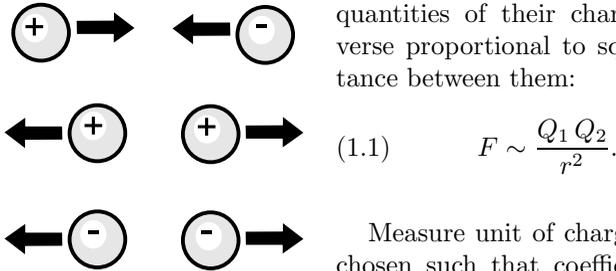
$$
\enddefinition
\parshape 5
4.5cm 5.6cm 4.5cm 5.6cm 4.5cm 5.6cm 4.5cm 5.6cm
0cm 10.1cm
Measure unit of charge in SGS is chosen such that 
\vadjust{\vskip 15pt\hbox to 0pt{\kern -25pt\includegraphics{fig1.eps}
\hss}\vskip -4pt\hbox{\kern 40pt{\it Fig\.~1.1 }\hss}
\vskip -23pt}coefficient in formula \thetag{1.1} is equal to unity.
Hence we have the following relation:
$$
\text{\it unit of charge in SGS = $dyn^{\,1\!/2}$ $\cdot$ cm=
g$^{\,1\!/2}$ $\cdot$ cm$^{3/2}$ $\cdot$ sec$^{-1}$}.
$$
Coulomb law itself then is written in form of the equality
$$
F = \frac{Q_1\,Q_2}{r^2}.\tag1.2
$$\par
Force $F$ defined by the relationship \thetag{1.2} is very strong.
However, in everyday life it does not reveal itself. This is due
to the {\it screening}. The numbers of positive and negative charges
in nature are exactly balanced. Atoms and molecules, which
constitute all observable matter around us, have the same amount of
positive and negative charges. Therefore they are electrically neutral
in whole. Force \thetag{1.2} reveals itself in form of chemical links
only when atoms are pulled together.\par
     Electric current arises as a result of motion of charged points.
This occurs in metallic conductor, which usually have lengthy form
(form of wire). Current in such conductor is determined by the
{\it amount of charge passing through it within the unit of
time}. Therefore for unit of current we have:
$$
\gather
\text{\it unit of current in SGS}
=\text{\it unit of charge in SGS $\cdot$ sec$^{-1}$}=\\
=\text{\it g$^{\,1\!/2}$ $\cdot$ cm$^{\,3/2}$ $\cdot$ sec$^{-2}$}.
\endgather
$$\par
Let's consider straight conducting rod of the length $l$. Current
in it leads to misbalance of charges in its ends. Charges of
definite sign move to one end of the rod, while lack of these
charges in the other end of the rod is detected as the charge
of opposite sign. Then Coulomb force \thetag{1.2} arises that
tends to recover balance of charges in electrically neutral
rod. This means that in such rod current could not flow in
constant direction during long time. Another situation we have
with conductor of the form of ring or circuit. Here current
does not break the balance of charges. Direct current can flow
in it during unlimitedly long time. Circular conductor itself
thereby remains electrically neutral and no Coulomb forces
arise.\par
\parshape 11 0cm 10.1cm 0cm 10.1cm 0cm 10.1cm 0cm 10.1cm
0cm 10.1cm 0cm 10.1cm 0cm 10.1cm 0cm 10.1cm 0cm 10.1cm
5cm 5.1cm 5cm 5.1cm
     In spite of absence of Coulomb forces, in experiments
the interaction of two circular conductors with currents was
detected. This interaction has other nature, it is not due 
to electrical, but due to magnetic forces. The magnitude
of magnetic forces depends essentially on the shape and
mutual arrangement of circular conductors. In order to reveal
quantitative characteristics for magnetic forces one should
maximally simplify the geometry of conductors. For this purpose
they are deformed so that each possesses straight rod-shaped
part of sufficiently big length $l$. These rod-shaped parts are
arranged parallel to each other with the distance $r$ between
them. In the limit, when $l$ is much larger than $r$, this
configuration of conductors can be treated as a pair of infinitely
long parallel conductors.
\vadjust{\vskip 15pt\hbox to 0pt{\kern -25pt\includegraphics{fig2.eps}
\hss}\vskip 30pt\hbox{\kern 45pt{\it Fig\.~1.2 }\hss}
\vskip -57pt}In experiments it was found that such conductors do
interact according to the following law.
\definition{\bf Ampere law}
\parshape 4 5cm 5.1cm 5cm 5.1cm 5cm 5.1cm 0cm 10.1cm
Force of interaction of \ two \ infinite \
parallel conductors with currents per unit length of them is proportional
ti the values of currents in them and inverse proportional to the distance
between them:
$$
\frac{F}{l}\sim\frac{I_1\,I_2}{r}.\tag1.3
$$
Two co-directed currents attract each other, while opposite directed
currents repulse each other.
\enddefinition
     The unit of current in SGS was already introduced above.
Therefore coefficient of proportionality in formula \thetag{1.3}
is unique quantity that should be determined in experiment. Here
is the measure unit for this coefficient: {\it sec$^{\,2}$\,
$\cdot$\,cm$^{-2}$}. It coincides with inverse square of velocity.
Therefore formula \thetag{1.3} in SGS is written as
$$
\frac{F}{l}=\frac{2}{c^2}\frac{I_1\,I_2}{r}.\tag1.4
$$
Constant $c$ in \thetag{1.4} is a velocity constant. The value
of this constant is determined experimentally:
$$
c\approx 2.998\cdot 10^{10}\text{\it\ cm}/\text{\it sec}.\tag1.5
$$
As we shall see below, constant $c$ in \thetag{1.5} coincides with
velocity of light in vacuum. Numeric coefficient $2$ in \thetag{1.4}
is introduced intentionally for to provide such coincidence.\par
     In SI measure unit of current $1$\,{\it A} (one {\it ampere}) is a basic
unit. It is determined such that formula \thetag{1.3} is written as
$$
\frac{F}{l}=\frac{2\,\mu_0}{4\pi}\frac{I_1\,I_2}{r}.
\tag1.6
$$
Here $\pi=3.14\dots$ is exact (though it is irrational) mathematical
constant with no measure unit. Constant $\mu_0$ is called {\it magnetic
susceptibility} of vacuum. It has the measure unit:
$$
\mu_0=4\pi\cdot 10^{-7}\text{\it N $\cdot$ A$^{-2}$}.\tag1.7
$$
But, in contrast to constant $c$ in \thetag{1.5}, it is exact constant.
Its value should not be determined experimentally. One could choose it
to be equal to unity, but the above value \thetag{1.7} for this constant
was chosen by convention when SI system was established. Due to this
value of constant \thetag{1.7} current of $1$ {\it ampere} appears to be
in that range of currents, that really appear in industrial and household
devices.
Coefficient $4\pi$ in denominator \thetag{1.6} is used in order to simplify
some other formulas, which are more often used for engineering calculations
in electric technology.\par
    Being basic unit in SI, unit of current {\it ampere} is used for
defining unit of charge of $1$ {\it coulomb}: {\it $1$C = $1$A $\cdot$
$1$sec}. Then coefficient of proportionality in Coulomb law \thetag{1.1}
appears to be not equal to unity. In SI Coulomb law is written as
$$
F=\frac{1}{4\pi\epsilon_0}\frac{Q_1 Q_2}{r^2}.
\tag1.8
$$
Constant $\epsilon_0$ is called dielectric permittivity of vacuum.
In contrast to constant $\mu_0$ in \thetag{1.7} this is physical
constant determined experimentally:
$$
\epsilon_0\approx8.85 \cdot 10^{-12}\text{\it\ C$^{\,2}$ $\cdot$
N$^{-1}$ $\cdot$ m$^{-2}$}.
\tag1.9
$$
Constants \thetag{1.5}, \thetag{1.7}, and \thetag{1.9} are related
to each other by the following equality:
$$
c=\frac{1}{\sqrt{\mathstrut\,\epsilon_0\,\mu_0\,}}
\approx 2.998\cdot 10^{8}\text{\it\ m}/\text{\it sec}.
\tag1.10
$$\par
    From the above consideration we see that SGS and SI systems differ
from each other not only in the scale of units, but in formulas for
two fundamental laws: Coulomb law and Ampere law. SI better suits for
engineering calculations. However, derivation of many formulas in this
system appears more huge than in SGS. Therefore below in this book we
use SGS system.\par
     Comparing Coulomb law and Ampere law we see that electrical and
magnetic forces reveal themselves in quite different way. However,
they have common origin: they both are due to electric charges.
Below we shall see that their relation is much more close. Therefore
theories of electricity and magnetism are usually united into one
theory of electromagnetic phenomena. Theory of electromagnetism is
a theory with one measurable constant: this is light velocity $c$.
Classical mechanics (without Newton's theory of gravitation) has no
measurable constants. Newton's theory of gravitation has one constant:
$$
\gamma\approx 6.67 \cdot 10^{-8}\text{\it\ cm$^3$ $\cdot$
g$^{-1}$ $\cdot$ sec$^{-2}$}.\tag1.11
$$
This theory is based on Newton's fourth law formulated as follows.
\definition{\bf Universal law of gravitation} Two point masses
attract each other with the force proportional to their masses and
inverse proportional to the square of distance between them.
\enddefinition
     Universal law of gravitation is given by the same formula
$$
F=\gamma\frac{M_1\,M_2}{r^2}\tag1.12
$$
in both systems: in SGS and in SI.\par
    According to modern notion of nature classical mechanics and
Newton's theory of gravitation are approximate theories. Currently
they are replaced by special theory of relativity and general
theory of relativity. Historically they appeared as a result of
development of the theory of electromagnetism. Below we keep this
historical sequence in explaining all three theories.
\proclaim{\bf Exercise 1.1} On the base of above facts find
quantitative relation of measure units for charge and current in
SGS and SI.
\endproclaim\par
\head
\S\,2. Concept of near action.
\endhead
     Let's consider pair of charged bodies, which are initially
fixed, and let's do the following mental experiment with them.
When we start moving second body apart from first one, the distance
$r$ begins increasing and consequently force of Coulomb interaction
\thetag{1.2} will decrease. In this situation we have natural
question: how soon after second body starts moving second body
will feel change of Coulomb force of interaction\,? There are two
possible answers to this question:
\roster
\rosteritemwd=4pt
\item immediately;
\item with some delay depending on the distance between bodies.
\endroster\par
    First answer is known as concept of {\it distant action}.
Taking this concept we should take formula \thetag{1.2} as
absolutely exact formula applicable for charges at rest and for
moving charges as well.\par
    Second answer is based on the concept of {\it near action}.
According to this concept, each interaction (and electric interaction
among others) can be transmitted immediately only to the point of
space infinitesimally close to initial one. Transmission of any action
to finite distance should be considered as a process of successive
transmission from point to point. This process always leads to
some finite velocity of transmission for any action. In the framework
of the concept of near action Coulomb law \thetag{1.2} is treated as
approximate law, which is exact only for the charges at rest that
stayed at rest during sufficiently long time so that process of
transmission of electric interaction has been terminated.\par
    Theory of electromagnetism has measurable constant $c$ (light
velocity \thetag{1.5}), which is first pretender for the role of
transmission velocity of electric and magnetic interactions. For this
reason electromagnetic theory is much more favorable as compared to
Newton's theory of gravitation.\par
\vskip 0pt plus 0.5pt minus 0.5pt
    The value of light velocity is a very large quantity. If we settle
an experiment of measuring Coulomb force at the distances of the order of
$r\approx 10\text{\it\ cm}$, for the time of transmission of interaction
we would get times of the order of $t\approx 3\cdot 10^{-10}\text{\it\
sec}$. Experimental technique of \uppercase\expandafter{\romannumeral
19}-th century was unable to detect such a short interval of time.
Therefore the problem of choosing concept could not be solved
experimentally. In \uppercase\expandafter{\romannumeral 19}-th century
it was subject for contests. The only argument against the concept of
distant action that time, quite likely, was its straightness, its
self-completeness, and hence its scarcity.\par
\vskip 0pt plus 0.5pt minus 0.5pt
     In present time concept of near action is commonly accepted. Now
we have the opportunity for testing it experimentally in the scope of
electromagnetic phenomena. Let's study this concept more attentively.
According to the concept of near action, process of transmitting
interaction to far distance exhibits an inertia. Starting at one point,
where moving charge is placed, for some time this process exist in
hidden form with no influence to both charges. In order to describe
this stage of process we need to introduce new concept. This concept
is a {\it field}.\par
\vskip 0pt plus 0.5pt minus 0.5pt
     {\it Field} is a material entity able to fill the whole space
and able to act upon other material bodies transmitting mutual
interaction of them.\par
\vskip 0pt plus 0.5pt minus 0.5pt
     The number of fields definitely known to scientists is not big.
There are only four fundamental fields: {\it strong field, weak
field, electromagnetic field}, and {\it gravitational field}. Strong
and weak fields are very short distance fields, they reveal themselves
only in atomic nuclei, in collisions and decay of elementary particles,
and in stellar objects of extremely high density, which are called
neutron stars. Strong and weak interactions and fields are not
considered in this book.\par
    There are various terms using the word field: {\it vector field,
tensor field, spinor field, gauge field,} and others. These are
mathematical terms reflecting some definite properties of real
physical fields.\par
\head
\S\,3. Superposition principle.
\endhead
    Let's apply concept of near action to Coulomb law for two charged
points. Coulomb force in the framework of this concept can be
interpreted as follows: first charge produces electric field around
itself, and this field acts upon other charge. Result of such action
is detected as a force $F$ applied to second charge. Force is vectorial
quantity. Let's denote by $\bold F$ vector of force and take into
account the direction of this vector determined by verbal statement of
Coulomb law above. This yields
$$
\bold F = Q_1\,Q_2\,\frac{\bold r_2-\bold r_1}
{|\bold r_2-\bold r_1|^3}.
\tag3.1
$$
Here $\bold r_1$ and $\bold r_2$ are radius-vectors of points, where
charges $Q_1$ and $Q_2$ are placed. Let's consider vector $\bold E$
determined as the ratio $\bold E=\bold F/Q_2$. For this vector from
formula \thetag{3.1} we derive
$$
\bold E = Q_1\,\frac{\bold r_2-\bold r_1}
{|\bold r_2-\bold r_1|^3}.\tag3.2
$$\par
     Vector $\bold E$ depends upon the position of first charge and
upon its value. It depends also on the position of second charge,
but it doesn't depend on the value of second charge. One can take
vector $\bold E$ for quantitative measure of electric field produced
by first charge $Q_1$ at the point $\bold r_2$, where second charge
is placed. Vector $\bold E$ can be determined by formula \thetag{3.2}
or it can be measured experimentally. For this purpose one should
place test charge $q$ to the point $\bold r_2$ and one should measure
Coulomb force $\bold F$ acting upon this test charge. Then vector
$\bold E$ is determined by division of $\bold F$ by the value of
test charge $q$:
$$
\bold E = \bold F/q.\tag3.3
$$\par
    Now consider more complicated situation. Suppose that charges
$Q_1,\dots,Q_n$ are placed at the points $\bold r_1,\dots,\bold r_n$.
They produce electric field around them, and this field acts upon
test charge $q$ placed at the point $\bold r$. This action reveals
as a force $\bold F$ applied to the charge $q$. Again we can define
vector $\bold E$ of the form \thetag{3.3} and take it for the
quantitative measure of electric field at the point $\bold r$.
This vector is called {\it vector of intensity of electric field}
or simply {\it vector of electric field} at that point.\par
    Generally speaking, in this case one cannot be a priori sure
that vector $\bold E$ does not depend on the quantity of test charge
$q$. However, there is the following experimental fact.
\definition{\bf Superposition principle} Electric field $\bold E$
at the point $\bold r$ produced by a system of point charges
$Q_1,\dots,Q_n$ is a vectorial sum of electric fields that would
be produced at this point by each charge $Q_1,\dots,Q_n$ separately.
\enddefinition\par
     Superposition principle combined with Coulomb law leads to the
following formula for the intensity of electric field produced by
a system of point charges at the point $\bold r$:
$$
\bold E(\bold r) = \sum^n_{i=1}Q_i\,\frac{\bold r-\bold r_i}
{\,|\bold r-\bold r_i|^3}.\tag3.4
$$
Using superposition principle, one can pass from point charges to
continuously distributed charges. Suppose that the number of point
charges tends to infinity: $n\to\infty$. In such limit sum in formula
\thetag{3.4} is replaced by integral over $3$-dimensional space: 
$$
\bold E(\bold r)=\int\rho(\tilde{\bold r})\,\frac{\bold r-
\tilde{\bold r}}{\,|\bold r-\tilde{\bold r}|^3}\,
d^{3}\tilde{\bold r}.
\tag3.5
$$
Here $\rho(\tilde{\bold r})$ is spatial density of charge at the point
$\tilde{\bold r}$. This value designates the amount of charge per unit
volume.\par
     In order to find force acting on test charge $q$ we should invert
formula \thetag{3.3}. As a result we obtain
$$
\bold F = q\,\bold E(\bold r).\tag3.6
$$
Force acting on a charge $q$ in electric field is equal to the product
of the quantity of this charge by the vector of intensity of field at the
point, where charge is placed. However, charge $q$ also produces electric
field. Does it experience the action of its own field\,? For point charges
the answer to this question is negative. This fact should be treated as
a supplement to principle of superposition. Total force acting on a system
of distributed charges in electric field is determined by the following
integral:
$$
\bold F = \int\rho(\bold r)\,\bold E(\bold r)\,d^{3}\bold r.
\tag3.7
$$
Field $\bold E(\bold r)$ in \thetag{3.7} is external field produced
by external charges. Field of charges with density $\rho(\bold r)$
is not included into $\bold E(\bold r)$.\par
     Concluding this section, note that formulas \thetag{3.4} and
\thetag{3.5} hold only for charges at rest, which stayed at rest for
sufficiently long time so that process of interaction transmitting
reached the point of observation $\bold r$. Fields produced by such
systems of charges are called {\it static fields}, while branch of
theory of electromagnetism studying such fields is called
{\it electrostatics}.\par
\head
\S\,4. Lorentz force and Biot-Savart-Laplace law.
\endhead
\rightheadtext{\S\,4. Lorentz force and \dots}
\parshape 15 0cm 10.1cm 0cm 10.1cm
4.45cm 5.65cm 4.45cm 5.65cm 4.45cm 5.65cm 4.45cm 5.65cm
4.45cm 5.65cm 4.45cm 5.65cm 4.45cm 5.65cm 4.45cm 5.65cm
4.45cm 5.65cm 4.45cm 5.65cm 4.45cm 5.65cm 4.45cm 5.65cm
0cm 10.1cm
     Ampere law of interaction of parallel conductors with currents
is an analog of Coulomb law for magnetic interactions. According to
near action principle, force $F$ arises as a result of action of
magnetic field produced by a current in first conductor upon second
conductor. However, parallel conductors cannot be treated as point
objects: formula \thetag{1.4} holds only for $l\gg r$.
\vadjust{\vskip 35pt\hbox to 0pt{\kern -45pt\includegraphics{fig3.eps}\hss}\vskip -10pt\hbox{\kern 25pt{\it Fig\.~4.1 }
\hss}\vskip -87pt\hbox{\kern 103pt$vt$\hss}\vskip 39pt}In order
to get quantitative measure of magnetic field at some point $\bold r$
let's consider current $I_2$ in \thetag{1.4} as a flow of charged
particles of charge $q$ each, and each moving along conductor with
constant velocity $v$. If we denote by $\nu$ the number of such
particles per unit length of conductor, then in the whole length $l$
we would have $N=\nu\,l$ particles. Then during time interval $t$
we would have $n=\nu\,v\,t$ particles passing through a fixed
cross-section of the conductor. They carry charge amounting to $Q=q\,
\nu\,v\,t$. Therefore for current $I_2$ in second conductor we get
$$
I_2=Q/t=q\,\nu\,v.
$$
Upon calculating force acting on a segment of conductor of the length  
$l$ by formula \thetag{1.4} we should divide it by the number of particles
$N$ contained in this segment. Then for the force per each particle we
derive
$$
F=\frac{2}{c^2}\frac{I_1\,I_2\,l}{r\,N}=
\frac{2}{c^2}\frac{I_1\,q\,v}{r}.
\tag4.1
$$
Formula determines \thetag{4.1} qualitative dependence of $F$ on $q$ and
on $v$: each charged particle moving in magnetic field experiences a force
proportional to its charge $q$ and to the magnitude of its velocity vector
$v=|\bold v|$, i\.\,e\. we have proportionality
$$
F\sim q\,v.
\tag4.2
$$
Force and velocity both are vectorial quantities. Simplest way to relate
two vectors $\bold F$ and $\bold v$ is to use vector product of $\bold v$
with some third vectorial quantity $\bold H$:
$$
\bold F=\frac{q}{c}\,[\bold v,\,\bold H(\bold r)].
\tag4.3
$$
Here $c$ is scalar constant equal to light velocity. Vectorial quantity
$\bold H(\bold r)$ is a quantitative measure of magnetic field at the
point $\bold r$. It is called {\it intensity of magnetic field} at that
point. Scalar factor $1/c$ in \thetag{4.3} is used for to make $\bold H$
to be measured by the same units as intensity of electric field $\bold E$
in \thetag{3.6}. Force $\bold F$ acting on a point charge in magnetic
field is called {\it Lorentz force}. Total Lorentz force acting on a charge
in electromagnetic field is a sum of two components: electric component and
magnetic component:
$$
\bold F=q\,\bold E+\frac{q}{c}\,[\bold v,\,\bold H].
\tag4.4
$$
Formula \thetag{4.4} extends formula \thetag{3.6} for the case of general
electromagnetic fields. It holds not only for static but for time-dependent
(non-static) fields. Surely the above derivation of formula \thetag{4.4} is
empiric. Actually, one should treat formula \thetag{4.4} as experimental
fact that do not contradict to another experimental fact \thetag{1.4} within
theory being developed.\par
     Let's turn back to our conductors. Formula \thetag{4.3} can be
interpreted in terms of currents. Each segment of unit length of a
conductor with current $I$ in magnetic field $\bold H$ experiences
the force
$$
\frac{\bold F}{l}=\frac{I}{c}\,[\boldsymbol\tau,\,\bold H]
\tag4.5
$$
acting on it. Here $\boldsymbol\tau$ is unit vector tangent to
conductor and directed along current in it. Total force acting on
circular conductor with current $I$ is determined by contour integral
$$
\bold F=\oint\frac{I}{c}\,[\boldsymbol\tau(s),\,\bold H(\bold r(s))]
\,ds,
\tag4.6
$$
where $s$ is natural parameter on contour (length) and $\bold r(s)$
is vector-function determining shape of contour in parametric form.
\par
     Let's consider the case of two parallel conductors. Force $\bold F$
now can be calculated by formula \thetag{4.5} assuming that first conductor
produces magnetic field $\bold H(\bold r)$ that acts upon second conductor.
Auxiliary experiment shows that vector $\bold H$ is perpendicular to the
plane of these two parallel conductors. The magnitude of magnetic field
$H=|\bold H|$ can be determined by formula \thetag{4.1}:
$$
H=\frac{2}{c}\frac{I_1}{r}.
\tag4.7
$$
Here $r$ is the distance from observation point to the conductor producing
field at that point.\par
    Magnetic field produced by conductor with current satisfies
superposition principle. In particular, field of infinite straight line
conductor \thetag{4.7} is composed by fields produced by separate segments
of this conductor. One cannot measure magnetic field of separate
segment experimentally since one cannot keep constant current is such
separate segment for sufficiently long time. But theoretically one can
consider infinitesimally small segment of conductor with current of the
length $ds$. And one can write formula for magnetic field produced by
such segment of conductor:
$$
d\bold H(\bold r)=\frac{1}{c}\frac{[I\,\boldsymbol\tau,\,
\bold r-\tilde{\bold r}]}{|\bold r-\tilde{\bold r}|^3}\,ds.
\tag4.8
$$
Here $\boldsymbol\tau$ is unit vector determining spatial orientation
of infinitesimal conductor. It is always taken to be directed along
current $I$. In practice, when calculating magnetic fields produced by
circular conductors, formula \thetag{4.8} is taken in integral form:
$$
\bold H(\bold r)=\oint\frac{1}{c}\frac{[I\,\boldsymbol\tau(s),\,
\bold r-\tilde{\bold r}(s)]}{|\bold r-\tilde{\bold r}(s)|^3}\,ds.
\tag4.9
$$
Like in \thetag{4.6}, here $s$ is natural parameter on the contour
and $\tilde{\bold r}(s)$ is vectorial function determining shape of
this contour. Therefore $\boldsymbol\tau(s)=d\tilde{\bold r}(s)/ds$.
The relationship \thetag{4.8} and its integral form \thetag{4.9}
constitute Biot-Savart-Laplace law for circular conductors with
current.\par
    Biot-Savart-Laplace law in form \thetag{4.8} cannot be tested
experimentally. However, in integral form \thetag{4.9} for each
particular conductor it yields some particular expression for
$\bold H(\bold r)$. This expression then can be verified in
experiment.\par
\proclaim{\bf Exercise 4.1} Using relationships \thetag{4.6} and
\thetag{4.9}, derive the law of interaction of parallel conductors
with current in form \thetag{1.4}.
\endproclaim
\proclaim{\bf Exercise 4.2} Find magnetic field of the conductor
with current having the shape of circle of the radius $a$.
\endproclaim\par
\head
\S\,5. Current density\\and the law of charge conservation.
\endhead
\rightheadtext{\S\,5. Current density and \dots}
    Conductors that we have considered above are kind of idealization.
They are linear, we assume them having no thickness. Real conductor
always has some thickness. This fact is ignored when we consider long
conductors like wire. However, in some cases thickness of a conductor
cannot be ignored. For example, if we consider current in electrolytic
bath or if we study current in plasma in upper layers of atmosphere.
Current in bulk conductors can be distributed non-uniformly within
volume of conductor. The concept of {\it current density} $\bold j$ is
best one for describing such situation.\par
     Current density is vectorial quantity depending on a point of
conducting medium: $\bold j=\bold j(\bold r)$. Vector of current density
$\bold j(\bold r)$ indicate the direction of charge transport at the
point $\bold r$. Its magnitude $j=|\,\bold j\,|$ is determined by the
amount of charge passing through unit area perpendicular to vector
$\bold j$ per unit time.
     Let's mark mentally some restricted domain $\Omega$ within bulk
conducting medium. Its boundary is smooth closed surface. Due to the
above definition of current density the amount of charge flowing out
from marked domain per unit time is determined by surface integral
over the boundary of this domain, while charge enclosed within this
domain is given by spatial integral:
$$
\xalignat 2
&\quad Q=\int\limits_\Omega\rho\,d^{3}\bold r,
&
&J=\int\limits_{\partial\Omega}\bigl<\,\bold j,\,
\bold n\bigr>\,dS.\tag5.1
\endxalignat
$$
Here $\bold n$ is unit vector of external normal to the surface
$\partial\Omega$ restricting domain $\Omega$.\par
     Charge conservation law is one more fundamental experimental
fact reflecting the nature of electromagnetism. In its classical
form it states that charges cannot appear from nowhere and cannot
disappear as well, they can only move from one point to another.
Modern physics insert some correction to this statement: charges
appear and can disappear in processes of creation and annihilation
of pairs of elementary particles consisting of particle and
corresponding antiparticle. However, even in such creation-annihilation
processes total balance of charge is preserved since total charge
of a pair consisting of particle and antiparticle is always equal to
zero. When applied to integrals \thetag{5.1} charge conservation law
yields: $\dot Q=-J$. This relationship means that decrease
of charge enclosed within domain $\Omega$ is always due to
charge leakage through the boundary and conversely increase of charge
is due to incoming flow through the boundary of this domain. Let's
write charge conservation law in the following form:
$$
\quad\frac{d}{dt}\left(\vphantom{\int}\right.\shave{\int\limits_\Omega}
\rho\,d^{3}\bold r\left.\vphantom{\int}\right)+
\int\limits_{\partial\Omega}\bigl<\,\bold j,\,
\bold n\bigr>\,dS=0.
\tag5.2
$$\par
     Current density $\bold j$ is a vector depending on a point of
conducting medium. Such objects in differential geometry are called
{\it vector fields}. Electric field $\bold E$ and magnetic field
$\bold H$ are other examples of vector fields. Surface integral $J$
in \thetag{5.1} is called {\it flow of vector field} $\bold j$ through
the surface $\partial\Omega$. For smooth vector field any surface
integral like $J$ can be transformed to spatial integral by means of
Ostrogradsky-Gauss formula. When applied to \thetag{5.2}, this yields
$$
\int\limits_\Omega
\left(\frac{\partial\rho}{\partial t}
+\divr\bold j\right)\,d^{3}\bold r=0.
\tag5.3
$$
Note that $\Omega$ in \thetag{5.3} is an arbitrary domain that we marked
mentally within conducting medium. This means that the expression being
integrated in \thetag{5.3} should be identically zero:
$$
\frac{\partial\rho}{\partial t}+\divr\bold j=0.
\tag5.4
$$
The relationships \thetag{5.2} and \thetag{5.4} are integral and
differential forms of charge conservation law respectively. The
relationship \thetag{5.4} also is known as {\it continuity equation}
for electric charge.\par
     When applied to bulk conductors with distributed current $\bold j$
within them, formula \thetag{4.6} is rewritten as follows:
$$
\bold F=\int\frac{1}{c}\,[\,\bold j(\bold r),\,\bold H(\bold r)]\,
\,d^{3}\bold r.
\tag5.5
$$
Biot-Savart-Laplace law for such conductors also is written in terms
of spatial integral in the following form:
$$
\bold H(\bold r)=\int\frac{1}{c}\frac{[\,\bold j(\tilde{\bold r}),
\,\bold r-\tilde{\bold r}]}{\,|\bold r-\tilde{\bold r}|^3}\,
d^{3}\tilde{\bold r}.
\tag5.6
$$
In order to derive formulas \thetag{5.5} and \thetag{5.6} from formulas
\thetag{4.6} and \thetag{4.8} one should represent bulk conductor as a
union of linear conductors, then use superposition principle and pass to
the limit by the number of linear conductors $n\to\infty$.\par
\head
\S\,6. Electric dipole moment.
\endhead
     Let's consider some configuration of distributed charge with
density $\rho(\bold r)$ which is concentrated within some restricted
domain $\Omega$. Let $R$ be maximal linear size of the domain $\Omega$.
Let's choose coordinates with origin within this domain $\Omega$ and
let's choose observation point $\bold r$ which is far enough from
the domain of charge concentration: $|\bold r|\gg R$. In order to find
electric field $\bold E(\bold r)$ produced by charges in $\Omega$ we
use formula \thetag{3.5}:
$$
\bold E(\bold r)=\int\limits_\Omega\rho(\tilde{\bold r})\,
\frac{\bold r-\tilde{\bold r}}{\,|\bold r-\tilde{\bold r}|^3}\,
d^{3}\tilde{\bold r}.
\tag6.1
$$
Since domain $\Omega$ in \thetag{6.1} is restricted, we have inequality
$|\tilde{\bold r}|\leq R$. Using this inequality along with $|\bold r|\gg
R$, we can write Taylor expansion for the fraction in the expression under
integration in \thetag{6.1}. As a result we get power series in powers
of ratio $\tilde{\bold r}/|\bold r|$:
$$
\frac{\bold r-\tilde{\bold r}}{\,|\bold r-\tilde{\bold r}|^3}=
\frac{\bold r}{\,|\bold r|^3}+\frac{1}{\,|\bold r|^2}\cdot\left(
3\frac{\bold r}{|\bold r|}\cdot\left<\frac{\bold r}{|\bold r|},\,
\frac{\tilde{\bold r}}{|\bold r|}\right>-
\frac{\tilde{\bold r}}{|\bold r|}\right)+\ldots\,.
\hskip-1em
\tag6.2
$$
Substituting \thetag{6.2} into \thetag{6.1}, we get the \pagebreak
following expression for the vector of electric field $\bold E(\bold r)$
produced by charges in $\Omega$:
$$
\bold E(\bold r)=Q\frac{\bold r}{\,|\bold r|^3}+
\frac{3\,\bigl<\bold r,\,\bold D\bigr>\,\bold r -
|\bold r|^2\,\bold D}{|\bold r|^5}+\ldots\,.
\hskip-2em
\tag6.3
$$
First summand in \thetag{6.3} is Coulomb field of point charge $Q$ placed
at the origin, where $Q$ is total charge enclosed in the domain $\Omega$.
It is given by integral \thetag{5.1}.\par
     Second summand in \thetag{6.3} is known as field of point dipole
placed at the origin. Vector $\bold D$ there is called {\it dipole moment}.
For charges enclosed within domain $\Omega$ it is given by integral
$$
\bold D=\int\limits_\Omega\rho(\tilde{\bold r})\,
\tilde{\bold r}\,d^{3}\tilde{\bold r}.\tag6.4
$$
For point charges dipole moment is determined by sum
$$
\bold D=\sum^n_{i=1}Q_i\,\tilde{\bold r}_i.\tag6.5
$$\par
For the system of charges concentrated near origin, which is
electrically neutral in whole, the field of point dipole
$$
\bold E(\bold r)=
\frac{3\,\bigl<\bold r,\,\bold D\bigr>\,\bold r
-|\bold r|^2\,\bold D}
{|\bold r|^5}
\tag6.6
$$
is leading term in asymptotics for electrostatic field 
\thetag{3.4} or \thetag{3.5} as $\bold r\to\infty$. Note
that for the system of charges with $Q=0$ dipole moment
$\bold D$ calculated by formulas \thetag{6.4} and \thetag{6.5}
is invariant quantity. This quantity remains unchanged when we
move all charges to the same distance at the same direction
without changing their mutual orientation: $\tilde{\bold r}\to
\tilde{\bold r}+\bold r_0$.\par
\proclaim{\bf Exercise 6.1} Concept of charge density is applicable
to point charges as well. However, in this case $\rho(\bold r)$ is
not ordinary function. It is distribution. For example point charge
$Q$ placed at the point $\bold r=0$ is represented by density
$\rho(\bold r)=Q\,\delta(\bold r)$, where $\delta(\bold r)$ is Dirac's
delta-function. Consider the density
$$
\rho(\bold r)=\bigl<\bold D,\,\grad\delta(\bold r)\bigr>=
\sum^3_{i=1}D^i\,\frac{\partial\delta(\bold r)}{\partial r^i}.
\tag6.7
$$
Applying formula \thetag{5.1}, calculate total charge $Q$ corresponding
to this density \thetag{6.7}. Using formula \thetag{6.4} calculate dipole
moment for distributed charge \thetag{6.7} and find electrostatic field
produced by this charge. Compare the expression obtained with \thetag{6.6}
and explain why system of charges described by the above density
\thetag{6.7} is called {\it point dipole}.
\endproclaim
\proclaim{\bf Exercise 6.2} Using formula \thetag{3.7} find the force
acting on point dipole in external electric field $\bold E(\bold r)$.
\endproclaim\par
\head
\S\,7. Magnetic moment.
\endhead
     Let's consider situation similar to that of previous section.
Suppose some distributed system of currents is concentrated in some
restricted domain near origin. Let $R$ be maximal linear size of this
domain $\Omega$. Current density $\bold j(\bold r)$ is smooth
vector-function, which is nonzero only within $\Omega$ and which
vanishes at the boundary $\partial\Omega$ and in outer space.
Current density $\bold j(\bold r)$ is assumed to be stationary, i\.\,
e\. it doesn't depend on time, and it doesn't break charge balance,
i\.\,e\. $\rho(\bold r)=0$. Charge conservation law applied to this
situation yields
$$
\divr\bold j=0.
\tag7.1
$$
In order to calculate magnetic field $\bold H(\bold r)$ we use
Biot-Savart-Laplace law written in integral form \thetag{5.6}:
$$
\bold H(\bold r)=\int\limits_\Omega\frac{1}{c}\frac{[\,
\bold j(\tilde{\bold r}),\,\bold r-\tilde{\bold r}]}
{\,|\bold r-\tilde{\bold r}|^3}\,d^{3}\tilde{\bold r}.
\tag7.2
$$
Assuming that $|\bold r|\gg R$, we take Taylor expansion \thetag{6.2}
and substitute it into \thetag{7.2}. As a result we get
$$
\aligned
&\bold H(\bold r)=\int\limits_\Omega\frac{[\,\bold j(
\tilde{\bold r}),\,\bold r]}{c\,|\bold r|^3}\,d^{3}
\tilde{\bold r}\,\,+\\
&+\int\limits_\Omega\frac{3\,\bigl<\bold r,\,
\tilde{\bold r}\bigr>\,[\,\bold j(\tilde{\bold r}),\,\bold r]-
|\bold r|^2\,[\,\bold j(\tilde{\bold r}),\,\tilde{\bold r}]}
{c\,|\bold r|^5}\,d^{3}\tilde{\bold r}\,+\,\ldots\,.
\hskip-2em
\endaligned
\tag7.3
$$
\proclaim{\bf Lemma 7.1} First integral in \thetag{7.3} is
identically equal to zero.
\endproclaim
\demo{Proof} Denote this integral by $\bold H_1(\bold r)$.
Let's choose some arbitrary constant vector $\bold e$ and consider
scalar product 
$$
\bigl<\bold H_1,\,\bold e\bigr>=\int\limits_\Omega\frac{
\bigl<\bold e,\,[\,\bold j(\tilde{\bold r}),\,\bold r]\bigr>}
{c\,|\bold r|^3}\,d^{3}\tilde{\bold r}=
\int\limits_\Omega\frac{
\bigl<\,\bold j(\tilde{\bold r}),\,[\,\bold r,\,\bold e]\bigr>}
{c\,|\bold r|^3}\,d^{3}\tilde{\bold r}.
\tag7.4
$$
Then define vector $\bold a$ and function $f(\tilde{\bold r})$
as follows:
$$
\xalignat 2
&\bold a=\frac{[\bold r,\,\bold e]}{c\,|\bold r|^3},
&&f(\tilde{\bold r})=\bigl<\bold a,\,\tilde{\bold r}\bigr>.
\endxalignat
$$
Vector $\bold a$ does not depend on $\tilde{\bold r}$, therefore
in calculating integral \thetag{7.4} we can take it for constant
vector. For this vector \pagebreak we derive $\bold a=\grad f$.
Substituting this formula into the \thetag{7.4}, we get
$$
\aligned
&\bigl<\bold H_1,\,\bold e\bigr>=\int\limits_\Omega\bigl<\,
\bold j,\,\grad f\bigr>\,d^{3}\tilde{\bold r}=\\
&=\int\limits_\Omega\divr(f\,\bold j)\,d^{3}\tilde{\bold r}-
\int\limits_\Omega f\,\divr\bold j\,d^{3}\tilde{\bold r}.
\endaligned
\tag7.5
$$
Last integral in \thetag{7.5} is equal to zero due to \thetag{7.1}.
Previous  integral is transformed to surface integral by means of
Ostrogradsky-Gauss formula. It is also equal to zero since
$\bold j(\tilde{\bold r})$ vanishes at the boundary of domain $\Omega$.
Therefore 
$$
\bigl<\bold H_1,\,\bold e\bigr>=\int\limits_{\partial\Omega}
f\,\bigl<\,\bold j,\,\bold n\bigr>\,dS=0.
\tag7.6
$$
Now vanishing of vector $\bold H_1(\bold r)$ follows from formula
\thetag{7.6} since $\bold e$ is arbitrary constant vector. Lemma~7.1
is proved.\qed\enddemo\par
     Let's transform second integral in \thetag{7.3}. First of all
we denote it by $\bold H_2(\bold r)$. Then, taking an arbitrary
constant vector $\bold e$, we form scalar product $\bigl<\bold H_2,
\,\bold e\bigr>$. This scalar product can be brought to
$$
\bigl<\bold H_2,\,\bold e\bigr>=
\frac{1}{\,c\,|\bold r|^5}\int\limits_\Omega
\bigl<\,\bold j(\tilde{\bold r}),\,\bold b(\tilde{\bold r})
\bigr>\,d^{3}\tilde{\bold r},
\tag7.7
$$
where $\bold b(\tilde{\bold r})=3\bigl<\bold r,\,\tilde{\bold r}
\bigr>\,[\bold r,\,\bold e]-|\bold r|^2\,[\tilde{\bold r},\,
\bold e]$. If one adds gradient of arbitrary function $f(\tilde{\bold
r})$ to $\bold b(\tilde{\bold r})$, this wouldn't change integral in
\thetag{7.7}. Formulas \thetag{7.5} and \thetag{7.6} form an example
of such invariance. Let's specify function $f(\tilde{\bold r})$,
choosing it as follows:
$$
f(\tilde{\bold r})=-\frac{3}{2}\bigl<\bold r,\,\tilde{\bold r}\bigr>\,
\bigl<\tilde{\bold r},\,[\bold r,\bold e]\bigr>.
\tag7.8
$$
For gradient of function \thetag{7.8} by direct calculations we find
$$
\aligned
&\grad f(\tilde{\bold r})=
-\frac{3}{2}\bigl<\tilde{\bold r},\,[\bold r,\bold e]\bigr>\,\bold r-
\frac{3}{2}\bigl<\bold r,\,\tilde{\bold r}\bigr>\,[\bold r,\bold e]=\\
&=-3\,\bigl<\bold r,\,\tilde{\bold r}\bigr>\,[\bold r,\bold e]-
\frac{3}{2}\left(\bold r\,\bigl<\tilde{\bold r},\,[\bold r,\bold e]\bigr>
-[\bold r,\bold e]\,\bigl<\bold r,\,\tilde{\bold r}\bigr>\right).
\endaligned
$$
Now let's use well-known identity $[\bold a,\,[\bold b,\,\bold c]]=
\bold b\,\bigl<\bold a,\,\bold c\bigr>-\bold c\,\bigl<\bold a,\,
\bold b\bigr>$. Assuming that $\bold a=\tilde{\bold r}$,
$\bold b=\bold r$, and $\bold c=[\bold r,\bold e]$, we transform
the above expression for $\grad f$ to the following form:
$$
\grad f(\tilde{\bold r})=
-3\,\bigl<\bold r,\,\tilde{\bold r}\bigr>\,[\bold r,\bold e]-
\frac{3}{2}\,[\tilde{\bold r},\,[\bold r,\,[\bold r,\,\bold e]]].
\tag7.9
$$
Right hand side of \thetag{7.9} contains triple vectorial product.
In order to transform it we use the identity $[\bold a,\,[\bold b,
\,\bold c]]=\bold b\,\bigl<\bold a,\,\bold c\bigr>-\bold c\,\bigl<\bold a,\,
\bold b\bigr>$ again, now assuming that $\bold a=\bold r$, $\bold b=\bold r$,
and $\bold c=\bold e$:
$$
\grad f(\tilde{\bold r})=
-3\,\bigl<\bold r,\,\tilde{\bold r}\bigr>\,[\bold r,\bold e]-
\frac{3}{2}\,\bigl<\bold r,\bold e\bigr>\,[\tilde{\bold r},\,\bold r]
+\frac{3}{2}\,|\bold r|^2\,[\tilde{\bold r},\,\bold e].
$$
Let's add this expression for $\grad f$ to vector $\bold b(\tilde{\bold
r})$. Here is resulting new expression for this vector:
$$
\bold b(\tilde{\bold r})=-\frac{3}{2}\,\bigl<\bold r,\bold e\bigr>\,
[\tilde{\bold r},\,\bold r]+\frac{1}{2}\,|\bold r|^2\,[\tilde{\bold r},\,
\bold e].
\tag7.10
$$
Let's substitute \thetag{7.10} into formula \thetag{7.7}. This yields
$$
\bigl<\bold H_2,\,\bold e\bigr>=
\int\limits_\Omega
\frac{-3\bigl<\bold r,\bold e\bigr>\,\bigl<\bold r,\,
[\,\bold j(\tilde{\bold r}),\,\tilde{\bold r}]\bigr>+
|\bold r|^2\,\bigl<\bold e,\,[\,\bold j(\tilde{\bold r}),\,
\tilde{\bold r}]\bigr>}{\,2\,c\,|\bold r|^5}
\,d^{3}\tilde{\bold r}.
$$
Note that quantities $\bold j(\tilde{\bold r})$ and $\tilde{\bold r}$
\pagebreak enter into this formula in form of vector product $[\,\bold
j(\tilde{\bold r}),\,\tilde{\bold r}]$ only. Denote by $\bold M$ the
following integral: 
$$
\bold M=\int\limits_\Omega
\frac{\,[\tilde{\bold r},\,\bold j(\tilde{\bold r})]\,}
{2\,c}\,d^{3}\tilde{\bold r}.
\tag7.11
$$
Vector $\bold M$ given by integral \thetag{7.11} is called {\it magnetic
moment} for currents with density $\bold j(\tilde{\bold r})$. In terms of
$\bold M$ the above relationship for scalar product $\bigl<\bold H_2,\,
\bold e\bigr>$ is written as follows:
$$
\bigl<\bold H_2,\,\bold e\bigr>=
\frac{3\,\bigl<\bold r,\bold e\bigr>\,\bigl<\bold r,\,
\bold M\bigr>-|\bold r|^2\,\bigl<\bold e,\,\bold M\bigr>}
{|\bold r|^5}.
\tag7.12
$$
If we remember that $\bold e$ in formula \thetag{7.12} is an arbitrary
constant vector, then from \thetag{7.3} and lemma~7.1 we can conclude
that the field of point magnetic dipole
$$
\bold H(\bold r)=\frac{3\,\bigl<\bold r,\,\bold M\bigr>\,\bold r
-|\bold r|^2\,\bold M}{|\bold r|^5}
\tag7.13
$$
is leading term in asymptotical expansion of static magnetic field
\thetag{4.9} and \thetag{5.6} as $\bold r\to\infty$.\par
    Like electric dipole moment $\bold D$ of the system with zero
total charge $Q=0$, magnetic moment $\bold M$ is invariant with
respect to displacements $\bold r\to\bold r+\bold r_0$ that don't
change configuration of currents within system. Indeed, under such
displacement integral \thetag{7.11} is incremented by
$$
\triangle\bold M=\int\limits_\Omega
\frac{\,[\bold r_0,\,\bold j(\tilde{\bold r})]\,}
{2\,c}\,d^{3}\tilde{\bold r}=0.
\tag7.14
$$
Integral in formula \thetag{7.14} is equal to zero by the same
reasons as in proof of lemma~7.1.\par
\proclaim{\bf Exercise 7.1} Consider localized system of currents
$\bold j(\bold r)$ with current density given by the following
distribution:
$$
\bold j(\bold r)=-c\,[\bold M,\,\grad\delta(\bold r)].
\tag7.15
$$
Verify the relationship \thetag{7.1} for the system of currents
\thetag{7.15} and find its magnetic moment $\bold M$. Applying
formula \thetag{5.6}, calculate magnetic field of this system
of currents and explain why this system of currents is called
{\it point magnetic dipole}.
\endproclaim
\proclaim{\bf Exercise 7.2} Using formula \thetag{5.5}, find
the force acting upon point magnetic dipole in external magnetic
field $\bold H(\bold r)$.
\endproclaim
\proclaim{\bf Exercise 7.3} By means of the following formula for the
torque
$$
\Cal M=\int\frac{1}{c}\,[\bold r,\,[\,\bold j(\bold r),\,\bold H]]\,
d^{3}\bold r
$$
find torque $\Cal M$ acting upon point magnetic dipole \thetag{7.15}
in homogeneous magnetic field $\bold H=\const$.
\endproclaim\par
\head
\S\,8. Integral equations\\for static electromagnetic field.
\endhead
\rightheadtext{\S\,8. Integral equations \dots}
     Remember that we introduced the concept of flow of vector field
through a surface in considering charge conservation law (see integral
$J$ in \thetag{5.1}). Now we consider flows of vector fields $\bold
E(\bold r)$ and $\bold H(\bold r)$, i\.\,e\. for electric field and
magnetic field:
$$
\xalignat 2
&\Cal E=\int\limits_S\bigl<\bold E,\,
\bold n\bigr>\,dS,
&&\Cal H=\int\limits_S\bigl<\bold H,\,
\bold n\bigr>\,dS.
\tag8.1
\endxalignat
$$
Let $S$ be closed surface enveloping some domain $\Omega$, i\.\,e\.
$S=\partial\Omega$. Electrostatic field $\bold E$ is determined by
formula \thetag{3.5}. Let's substitute \thetag{3.5} into first
integral \thetag{8.1} and then let's change order of integration
in resulting double integral:
$$
\Cal E=\int\rho(\tilde{\bold r})
\int\limits_{\partial\Omega}\frac{\bigl<\bold r-
\tilde{\bold r},\,\bold n(\bold r)\bigr>}
{\,|\bold r-\tilde{\bold r}|^3}\,dS
\,d^{3}\tilde{\bold r}.
\tag8.2
$$
Inner surface integral in \thetag{8.2} is an integral of explicit
function. This integral can be calculated explicitly:
$$
\int\limits_{\partial\Omega}\frac{\bigl<\bold r-
\tilde{\bold r},\,\bold n(\bold r)\bigr>}
{\,|\bold r-\tilde{\bold r}|^3}\,dS=
\left\{
\aligned
&0\text{, if $\tilde{\bold r}\not\in\overline{\Omega}$,}\\
&4\pi\text{, if $\tilde{\bold r}\in\Omega$.}
\endaligned\right.
\tag8.3
$$
Here by $\overline{\Omega}=\Omega\cup\partial\Omega$ we denote
closure of the domain $\Omega$.\par
    In order to prove the relationship \thetag{8.3} let's consider
vector field $\bold m(\bold r)$ given by the following formula:
$$
\bold m(\bold r)=\frac{\bold r-\tilde{\bold r}}
{\,|\bold r-\tilde{\bold r}|^3}.
\tag8.4
$$
Vector field $\bold m(\bold r)$ is smooth everywhere except for
one special point $\bold r=\tilde{\bold r}$. In all regular points
of this vector field by direct calculations we find $\divr\bold m=0$.
If $\tilde{\bold r}\not\in\overline{\Omega}$ special point of the
field $\bold m$ is out of the domain $\Omega$. Therefore in this case
we can apply Ostrogradsky-Gauss formula to \thetag{8.3}:
$$
\int\limits_{\partial\Omega}\bigl<\bold m,\,\bold n\bigr>\,
dS=\int\limits_\Omega\divr\bold m\,d^{3}\bold r=0.
$$
This proves first case in formula \thetag{8.3}. In order to prove second
case, when $\tilde{\bold r}\in\Omega$, we use tactical maneuver. Let's
consider spherical $\epsilon$-neighborhood $O=O_\epsilon$ of special
point $\bold r=\tilde{\bold r}$. For sufficiently small $\epsilon$ this
neighborhood $O=O_\epsilon$ is completely enclosed into the domain
$\Omega$. Then from zero divergency condition $\divr\bold m=0$ for the
field given by formula \thetag{8.4} we derive
$$
\int\limits_{\partial\Omega}\bigl<\bold m,\,\bold n\bigr>\,dS=
\int\limits_{\partial O}\bigl<\bold m,\,\bold n\bigr>\,dS=4\pi.
\tag8.5
$$
The value of last integral over sphere $\partial O$ in \thetag{8.5}
is found by direct calculation, which is not difficult. Thus, formula
\thetag{8.3} is proved. Substituting \thetag{8.3} into \thetag{8.2}
we get the following relationship:
$$
\int\limits_{\partial\Omega}\bigl<\bold E,\,\bold n\bigr>\,dS=
4\pi\int\limits_\Omega\rho(\bold r)\,d^{\,3}\bold r.
\tag8.6
$$
This relationship \thetag{8.6} can be formulated as a theorem.

\proclaim{\bf Theorem \rm(on the flow of electric field)} Flow of
electric field through the boundary of restricted domain is equal
to total charge enclosed within this domain multiplied by $4\pi$.
\endproclaim\par
     Now let's consider flow of magnetic field $\Cal H$ in \thetag{8.1}.
Static magnetic field is determined by formula \thetag{5.6}. Let's
substitute $\bold H(\bold r)$ given by \thetag{5.6} into second integral
\thetag{8.1}, then change the order of integration in resulting double
integral:
$$
\Cal H=\int\int\limits_{\partial\Omega}
\frac{1}{c}\frac{\bigl<[\,\bold j(\tilde{\bold r}),\,
\bold r-\tilde{\bold r}],\,\bold n(\bold r)\bigr>}
{\,|\bold r-\tilde{\bold r}|^3}\,dS\,d^{\,3}\tilde{\bold r}.
\tag8.7
$$
It's clear that in calculating inner integral over the surface $\partial
\Omega$ vector $\bold j$ can be taken for constant. Now consider the field 
$$
\bold m(\bold r)=\frac{\,[\,\bold j,\,\bold r-\tilde{\bold r}]\,}
{c\,|\bold r-\tilde{\bold r}|^3}.
\tag8.8
$$
Like \thetag{8.4}, this vector field \thetag{8.8} has only one singular
point $\bold r=\tilde{\bold r}$. Divergency of this field is equal to zero,
this fact can be verified by direct calculations. As appears in this case,
singular point makes no effect to the value of surface integral in
\thetag{8.7}. Instead of \thetag{8.3} in this case we have the following
formula:
$$
\int\limits_{\partial\Omega}
\frac{1}{c}\frac{\bigl<[\,\bold j,\,\bold r-\tilde{\bold r}],\,
\bold n(\bold r)\bigr>}{\,|\bold r-\tilde{\bold r}|^3}\,dS=0.
\tag8.9
$$
For $\tilde{\bold r}\not\in\overline{\Omega}$ the relationship 
\thetag{8.9} follows from $\divr\bold m=0$ by applying Ostrogradsky-Gauss
formula. For $\tilde{\bold r}\in\Omega$ we have the relationship similar
to the above relationship \thetag{8.5}:
$$
\int\limits_{\partial\Omega}\bigl<\bold m,\,\bold n\bigr>\,dS=
\int\limits_{\partial O}\bigl<\bold m,\,\bold n\bigr>\,dS=0.
\tag8.10
$$
However, the value of surface integral over sphere $\partial O$ in this
case is equal to zero since vector $\bold m(\bold r)$ is perpendicular
to normal vector $\bold n$ at all points of sphere $\partial O$. As a
result of substituting \thetag{8.9} into \thetag{8.7} we get the
relationship
$$
\int\limits_{\partial\Omega}\bigl<\bold H,\,\bold n\bigr>\,dS=0,
\tag8.11
$$
which is formulated as the following theorem.
\proclaim{\bf Theorem (\rm on the flow of magnetic field)} Total flow
of magnetic field through the boundary of any restricted domain is equal
to zero.
\endproclaim\par
\par
\parshape 15 0cm 10.1cm 0cm 10.1cm 0cm 10.1cm 0cm 10.1cm 0cm 10.1cm
4.5cm 5.6cm 4.5cm 5.6cm 4.5cm 5.6cm 4.5cm 5.6cm
4.5cm 5.6cm 4.5cm 5.6cm 4.5cm 5.6cm 4.5cm 5.6cm 
0cm 10.1cm 0cm 10.1cm
     Let $\bold r(s)$ be vectorial parametric equation of some closed
spatial curve $\Gamma$ being the rim for some open surface $S$, i\.\,e\.
$\Gamma=\partial S$. Open surface $S$ means that $S$ and $\Gamma$ have
empty intersection. By $\overline{S}$ we denote the closure of the
surface $S$. Then $\overline{S}=S\cup\Gamma$.
Taking $s$ for natural parameter on $\Gamma$, we define {\it circulation}
for electric and magnetic fields in form of the following contour integrals:
$$
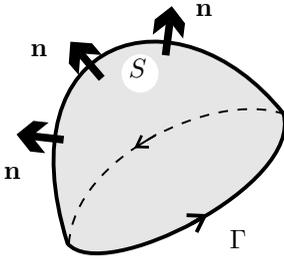
\aligned
&\goth e=\oint\limits_\Gamma\bigl<\bold E,\,
\boldsymbol\tau\bigr>\,ds,\\
&\goth h=\oint\limits_\Gamma\bigl<\bold H,\,
\boldsymbol\tau\bigr>\,ds.
\endaligned
\tag8.12
$$
Substituting \thetag{3.5} into \thetag{8.12} and
\vadjust{\vskip 15pt\hbox to 0pt{\kern -25pt\includegraphics{fig4.eps}
\hss}\vskip -28pt\hbox{\kern 45pt{\it Fig\.~8.1 }\hss}
\vskip -41pt\hbox{\kern 85pt$\Gamma$\hss}\vskip -38pt
\hbox{\kern 0pt$\bold n$\hss}\vskip -50pt\hbox{\kern 47pt$S$\hss}
\vskip -20pt\hbox{\kern 10pt$\bold n$\hss}\vskip -27pt
\hbox{\kern 72pt$\bold n$\hss}\vskip 117pt}%
changing the order of integration in resulting double integral, 
we get the following equality for circulation of electric field:
$$
\goth e=\int\rho(\tilde{\bold r})
\oint\limits_\Gamma\frac{\bigl<\bold r(s)-
\tilde{\bold r},\,\boldsymbol\tau(s)\bigr>}
{\,|\bold r(s)-\tilde{\bold r}|^3}\,ds
\,d^{\,3}\tilde{\bold r}.
\tag8.13
$$\par
Due to \thetag{8.13} we need to consider vector field
\thetag{8.4} again. For $\tilde{\bold r}\not\in\Gamma$, taking
into account $\Gamma=\partial S$ and applying Stokes formula, we
can transform contour integral in \thetag{8.13} to surface
integral:
$$
\oint\limits_\Gamma\frac{\bigl<\bold r(s)-
\tilde{\bold r},\,\boldsymbol\tau(s)\bigr>}
{\,|\bold r(s)-\tilde{\bold r}|^3}\,ds=
\int\limits_S
\bigl<\rot\bold m,\,\bold n\bigr>\,dS=0.
\tag8.14
$$
Values of integral \thetag{8.14} at those points $\tilde{\bold r}\in
\Gamma$ are of no matter since when substituting \thetag{8.14} into
integral \thetag{8.13} such points constitute a set of zero measure.
\par
     Vanishing of integral \thetag{8.14} for $\tilde{\bold r}\not
\in\Gamma$ follows from $\rot\bold m=0$, this equality can be verified
by direct calculations. Singular point $\bold r=\tilde{\bold r}$ of
vector field \thetag{8.4} is unessential since surface $S$, for which
$\Gamma$ is a boundary, can be deformed so that $\tilde{\bold r}\not
\in S$. The result of substituting \thetag{8.14} into \thetag{8.13}
can be written as an equation:
$$
\oint\limits_{\partial S}\bigl<\bold E,\,
\boldsymbol\tau\bigr>\,ds=0.
\tag8.15
$$
\proclaim{\bf Theorem \rm(on the circulation of electric field)}
Total circulation of static electric field along the boundary of any
restricted open surface is equal to zero.
\endproclaim
     Formula like \thetag{8.15} is available for magnetic field as well.
Here is this formula that determines circulation of magnetic field:
$$
\oint\limits_{\partial S}\bigl<\bold H,\,
\boldsymbol\tau\bigr>\,ds=\frac{4\,\pi}{c}
\int\limits_S\bigl<\,\bold j,\,
\bold n\bigr>\,dS.
\tag8.16
$$
Corresponding theorem is stated as follows.
\proclaim{\bf Theorem \rm(on the circulation of magnetic field)}
Circulation of static magnetic field along boundary of restricted
open surface is equal to total electric current penetrating this
surface multiplied by fraction $4\,\pi/c$.
\endproclaim
     Integral over the surface $S$ now is in right hand side of
formula \thetag{8.16} explicitly. Therefore surface spanned over
the contour $\Gamma$ now is fixed. We cannot deform this surface
as we did above in proving theorem on circulation of electric field.
This leads to some technical complication of the proof. Let's
consider $\varepsilon$-blow-up of surface $S$. This is domain
$\Omega(\varepsilon)$ being union of all $\varepsilon$-balls
surrounding all point $\bold r\in S$. This domain encloses surface
$S$ and contour $\Gamma=\partial S$. If $\varepsilon\to 0$,
domain $\Omega(\varepsilon)$ contracts to $S$.\par
     Denote by $D(\varepsilon)=\Bbb R^3\setminus\Omega(\varepsilon)$
exterior of the domain $\Omega(\varepsilon)$ and then consider the
following modification of formula \thetag{5.6} that expresses
Biot-Savart-Laplace law for magnetic field:
$$
\bold H(\bold r)=\lim_{\varepsilon\to 0}
\int\limits_{D(\varepsilon)}
\frac{1}{c}
\frac{[\,\bold j(\tilde{\bold r}),\,\bold r-\tilde{\bold r}]}
{\,|\bold r-\tilde{\bold r}|^3}\,d^{\,3}\tilde{\bold r}.
\tag8.17
$$
Let's substitute \thetag{8.17} into integral \thetag{8.12} and
change the order of integration in resulting double integral.
As a result we get
$$
\goth h=\lim_{\varepsilon\to 0}\int\limits_{D(\varepsilon)}
\oint\limits_\Gamma
\frac{1}{c}\frac{\bigl<[\,\bold j(\tilde{\bold r}),\,
\bold r(s)-\tilde{\bold r}],\,\boldsymbol\tau(s)\bigr>}
{\,|\bold r(s)-\tilde{\bold r}|^3}\,ds\,d^{\,3}\tilde{\bold r}.
\tag8.18
$$
In inner integral in \thetag{8.18} we see vector field \thetag{8.8}.
Unlike vector filed \thetag{8.4}, rotor of this field is nonzero:
$$
\rot\bold m=
\frac{3\,\bigl<\bold r-\tilde{\bold r},\,\bold j\bigr>\,
(\bold r-\tilde{\bold r})-|\bold r-\tilde{\bold r}|^2\,\bold j}
{c\,|\bold r-\tilde{\bold r}|^5}.
\tag8.19
$$
Using Stokes formula and taking into account \thetag{8.19}, we can
transform contour integral \thetag{8.18} to surface integral:
$$
\align
&\oint\limits_\Gamma
\frac{1}{c}\frac{\bigl<[\,\bold j(\tilde{\bold r}),\,
\bold r(s)-\tilde{\bold r}],\,\boldsymbol\tau(s)\bigr>}
{\,|\bold r(s)-\tilde{\bold r}|^3}\,ds=\\
&=\int\limits_S
\frac{3\,\bigl<\bold r-\tilde{\bold r},\,\bold j(\tilde{\bold r})
\bigr>\,\bigl<\bold r-\tilde{\bold r},\,\bold n(\bold r)\bigr>-
|\bold r-\tilde{\bold r}|^2\,\bigl<\,\bold j(\tilde{\bold r}),\,
\bold n(\bold r)\bigr>}{c\,|\bold r-\tilde{\bold r}|^5}\,dS.
\endalign
$$
Denote by $\widetilde{\bold m}(\tilde{\bold r})$ vector field
of the following form:
$$
\widetilde{\bold m}(\tilde{\bold r})=
\frac{3\,\bigl<\tilde{\bold r}-\bold r,\,\bold n(\bold r)\bigr>\,
(\tilde{\bold r}-\bold r)-|\tilde{\bold r}-\bold r|^2\,
\bold n(\bold r)}{c\,|\tilde{\bold r}-\bold r|^5}.
$$
In terms of the field $\widetilde{\bold m}(\tilde{\bold r})$
formula for $\goth h$ is written as
$$
\goth h=\lim_{\varepsilon\to 0}\int\limits_{D(\varepsilon)}
\int\limits_S
\bigl<\widetilde{\bold m}(\tilde{\bold r}),\,\bold j(\tilde{\bold r})
\bigr>\,dS\,d^{\,3}\tilde{\bold r}.
$$
Vector field $\widetilde{\bold m}(\tilde{\bold r})$ in this formula
has cubic singularity $|\tilde{\bold r}-\bold r|^{-3}$ at the
point $\tilde{\bold r}=\bold r$. Such singularity is not integrable
in $\Bbb R^3$ (if we integrate with respect to $d^{\,3}
\tilde{\bold r}$). This is why we use auxiliary domain $D(\varepsilon)$
and limit as $\varepsilon\to 0$.\par
    Let's change the order of integration in resulting double integral
for circulation $\goth h$. This leads to formula
$$
\int\limits_S
\int\limits_{D(\varepsilon)}
\bigl<\widetilde{\bold m}(\tilde{\bold r}),\,
\bold j(\tilde{\bold r})\bigr>
\,d^{\,3}\tilde{\bold r}\,dS
=\int\limits_S
\int\limits_{D(\varepsilon)}
\bigl<\grad f(\tilde{\bold r}),\,
\bold j(\tilde{\bold r})\bigr>
\,d^{\,3}\tilde{\bold r}\,dS,
$$
since vector field $\widetilde{\bold m}(\tilde{\bold r})$
apparently is gradient of the function $f(\tilde{\bold r})$:
$$
f(\tilde{\bold r})=
-\frac{\bigl<\tilde{\bold r}-\bold r,\,\bold n(\bold r)\bigr>}
{c\,|\tilde{\bold r}-\bold r|^3}.
\tag8.20
$$
Function $f(\tilde{\bold r})$ vanishes as $\tilde{\bold r}\to\infty$.
Assume that current density also vanishes as $\tilde{\bold r}\to\infty$.
Then due to the same considerations as in proof of lemma~7.1 and due
to formula \thetag{7.1} spatial integral in the above formula can be
transformed to surface integral:
$$
\goth h=\lim_{\varepsilon\to 0}\,
\int\limits_S
\int\limits_{\partial D(\varepsilon)}
f(\tilde{\bold r})\,
\bigl<\,\bold j(\tilde{\bold r}),\,
\tilde{\bold n}(\tilde{\bold r})\bigr>
\,d\widetilde S\,dS.
\tag8.21
$$
Let's change the order of integration in \thetag{8.21} then
take into account common boundary $\partial D(\varepsilon)
=\partial\Omega(\varepsilon)$. Outer normal to the surface
$\partial D(\varepsilon)$ coincides with inner normal to
$\partial\Omega(\varepsilon)$. This coincidence and explicit
form of function \thetag{8.20} lead to the following expression
for circulation of magnetic field $\goth h$:
$$
\goth h=\lim_{\varepsilon\to 0}\,
\int\limits_{\partial\Omega(\varepsilon)}
\frac{\bigl<\,\bold j(\tilde{\bold r}),\,
\tilde{\bold n}(\tilde{\bold r})\bigr>}{c}
\int\limits_S
\frac{\bigl<\tilde{\bold r}-\bold r,\,\bold n(\bold r)\bigr>}
{|\tilde{\bold r}-\bold r|^3}\,
\,dS\,d\widetilde S.
\tag8.22
$$
Let's denote by $V(\tilde{\bold r})$ inner integral in formula
\thetag{8.22}:
$$
V(\tilde{\bold r})=
\int\limits_S
\frac{\bigl<\tilde{\bold r}-\bold r,\,\bold n(\bold r)\bigr>}
{|\tilde{\bold r}-\bold r|^3}\,dS.
\tag8.23
$$
Integral \thetag{8.23} is well-known in mathematical physics.
It is called {\it potential of double layer}. There is the
following lemma, proof of which can be found in \cite{1}.
\proclaim{\bf Lemma 8.1} Double layer potential \thetag{8.23} is
restricted function in $\Bbb R^3\setminus\overline{S}$. At each
inner point $\tilde{\bold r}\in S$ there are side limits 
$$
V_{\pm}(\tilde{\bold r})=\lim_{\bold r\to\pm S}V(\bold r),
$$
inner limit $V_{-}(\tilde{\bold r})$ as $\bold r$ tends to
$\tilde{\bold r}\in S$ from inside along normal vector $\bold n$,
and outer limit $V_{+}(\tilde{\bold r})$ as $\bold r$ tends to
$\tilde{\bold r}\in S$ from outside against the direction of normal
vector $\bold n$. Thereby $V_{+}-V_{-}=4\pi$ for all points
$\tilde{\bold r}\in S$.
\endproclaim
\parshape 14 0cm 10.1cm 0cm 10.1cm 0cm 10.1cm 0cm 10.1cm
0cm 10.1cm 0cm 10.1cm 0cm 10.1cm 0cm 10.1cm 0cm 10.1cm
0cm 10.1cm 0cm 10.1cm 0cm 10.1cm 0cm 10.1cm 
5.1cm 5cm
    In order to calculate limit in formula \thetag{8.22} we need to
study the geometry of $\varepsilon$-blow-up of the surface $S$. On
Fig\.~8.2 below we see cross-section of the domain $\Omega(\varepsilon)$
obtained from the surface $S$ shown on Fig\.~8.1. For sufficiently small
$\varepsilon$ boundary of the domain $\Omega(\varepsilon)$ is composed
of three parts:
$$
\partial\Omega(\varepsilon)=S_0\cup S_{+}\cup S_{-}.
\hskip -1.5em
\tag8.24
$$
Surface $S_0$ is a part of $\varepsilon$-blow-up of the contour $\Gamma$
Area of this surface $S_0$ satisfies the relationship
$$
S_0\sim\varepsilon\pi L\text{\ \ as\ \ }\varepsilon\to 0,
\tag8.25
$$
where $L$ is length of contour $\Gamma$. Surfaces $S_{+}$ and $S_{-}$
are obtained as a result of normal shift of surface $S$ to the distance
$\varepsilon$ along normal vector $\bold n$, and to the same distance
against normal vector $\bold n$. \vadjust{\vskip 67pt\hbox
to 0pt{\kern -25pt\includegraphics{fig5.eps}\hss}\vskip -7pt
\hbox{\kern 40pt{\it Fig\.~8.2 }\hss}\vskip -45pt
\hbox{\kern 58pt$S_0$\hss}\vskip -33pt\hbox{\kern 65pt$S_{-}$\hss}
\vskip -31pt\hbox{\kern 55pt$S$\hss}\vskip -31pt
\hbox{\kern 62pt$\bold n$\hss}
\vskip -62pt
\hbox{\kern 25pt$\bold n$\kern 81pt$S_{+}$\hss}
\vskip 69pt}\par
\parshape 7 5.1cm 5cm 5.1cm 5cm 5.1cm 5cm 5.1cm 5cm 5.1cm 5cm
5.1cm 5cm 0cm 10.1cm
    Substituting \thetag{8.24} into \thetag{8.22} we break surface
integral over $\partial\Omega(\varepsilon)$ into three parts. Since
double layer potential and function $|\,\bold j(\tilde{\bold r})|$
are restricted, we get the relationship
$$
\lim_{\varepsilon\to 0}\,
\int\limits_{S_0}
V(\tilde{\bold r})\,
\frac{\bigl<\,\bold j(\tilde{\bold r}),\,
\tilde{\bold n}(\tilde{\bold r})\bigr>}{c}
\,d\widetilde S=0.
\tag8.26
$$
For other two summand we also can calculate limits as
$\varepsilon\to 0$:
$$
\int\limits_{S_{\pm}}
V(\tilde{\bold r})\,
\frac{\bigl<\,\bold j(\tilde{\bold r}),\,
\tilde{\bold n}(\tilde{\bold r})\bigr>}{c}
\,d\widetilde S\longrightarrow
\pm\int\limits_S V_{\pm}(\bold r)\,
\frac{\bigl<\,\bold j(\bold r),\,
\bold n(\bold r)\bigr>}{c}\,dS.
\hskip -1.5em
\tag8.27
$$
We shall not load reader with the proof of formulas \thetag{8.24},
\thetag{8.25} and \thetag{8.27}, which are sufficiently obvious.
Summarizing \thetag{8.26} and \thetag{8.27} and taking into account
lemma~8.1, we obtain
$$
\pagebreak
\goth h=\frac{4\pi}{c}\int\limits_S\bigl<\,\bold j(\bold r),\,
\bold n(\bold r)\bigr>\,dS.
\tag8.28
$$
This relationship \thetag{8.28} completes derivation of formula
\thetag{8.16} and proof of theorem on circulation of magnetic
field in whole.
\proclaim{\bf Exercise 8.1} Verify the relationship $\divr\bold m=0$
for vector fields \thetag{8.4} and \thetag{8.8}.
\endproclaim
\proclaim{\bf Exercise 8.2} Verify the relationship \thetag{8.19} for
vector field given by formula \thetag{8.8}.
\endproclaim
\proclaim{\bf Exercise 8.3} Calculate $\grad f$ for the function
\thetag{8.20}.
\endproclaim
\head
\S\,9. Differential equations\\for static electromagnetic field.
\endhead
\rightheadtext{\S\,9. Differential equations \dots}
    In \S\,8 we have derived four integral equations for electric
and magnetic fields. They are used to be grouped into two pairs.
Equations in first pair have zero right hand sides:
$$
\xalignat 2
&\int\limits_{\partial\Omega}\bigl<\bold H,\,\bold n\bigr>\,dS=0,
&&\oint\limits_{\partial S}\bigl<\bold E,\,
\boldsymbol\tau\bigr>\,ds=0.
\tag9.1
\endxalignat
$$
Right hand sides of equations in second pair are non-zero.
They are determined by charges and currents:
$$
\aligned
&\int\limits_{\partial\Omega}\bigl<\bold E,\,\bold n\bigr>\,dS=
4\pi\int\limits_\Omega\rho\,d^{\,3}\bold r,\\
\vspace{2ex}
&\oint\limits_{\partial S}\bigl<\bold H,\,
\boldsymbol\tau\bigr>\,ds=
\frac{4\,\pi}{c}\int\limits_S\bigl<\,\bold j,\,
\bold n\bigr>\,dS.
\endaligned
\tag9.2
$$
Applying Ostrogradsky-Gauss formula and Stokes formula, one can
transform surface integrals to spatial ones, and contour integrals
to surface integrals. Then, since $\Omega$ is arbitrary domain and
$S$ is arbitrary open surface, integral equations \thetag{9.1} and
\thetag{9.2} can be transformed to differential equations:
$$
\xalignat 2
&\divr\bold H=0,
&&\rot\bold E=0,\tag9.3\\
&\divr\bold E=4\pi\rho,
&&\rot\bold H=\frac{4\pi}{c}\,\bold j.
\tag9.4
\endxalignat
$$
When considering differential equations \thetag{9.3} and \thetag{9.4},
we should add conditions for charges and currents being stationary:
$$
\xalignat 2
&\frac{\partial\rho}{\partial t}=0,\qquad
&&\frac{\partial\bold j}{\partial t}=0.
\tag9.5
\endxalignat
$$
The relationship \thetag{7.1} then is a consequence of \thetag{9.5} and
charge conservation law.\par
     Differential equations \thetag{9.3} and \thetag{9.4} form
complete system of differential equations for describing stationary
electromagnetic fields. When solving them functions $\rho(\bold r)$
and $\bold j(\bold r)$ are assumed to be known. If they are not known,
one should have some additional equations relating $\rho$ and $\bold j$
with $\bold E$ and $\bold H$. These additional equations describe
properties of medium (for instance, continuous conducting medium is
described by the equation $\bold j=\sigma\,\bold E$, where $\sigma$
is conductivity of medium).\par
\newpage
\setfirstpage
\topmatter
\title\chapter{2}
Classical electrodynamics
\endtitle
\endtopmatter
\leftheadtext{CHAPTER \uppercase\expandafter{\romannumeral 2}.
CLASSICAL ELECTRODYNAMICS.}
\document
\head
\S\,1. Maxwell equations.
\endhead
     Differential equations \thetag{9.3} and \thetag{9.4}, which
we have derived in the end of Chapter~\uppercase\expandafter{\romannumeral
1}, describe fields generated by stationary charges and currents.
They are absolutely unsuitable if we are going to describe the
process of haw electromagnetic interaction is transmitted in space.
Note that the notion of field was introduced within framework of
the concept of near action for describing the object that transmit
interaction of charges and currents. For static fields this property
is revealed in a very restrictive form, i\.\,e\. we use fields only
to divide interaction of charges and currents into two processes:
creation of a field by charges and currents is first process, action
of this field upon other currents and charges is second process.
Dynamic properties of the field itself appears beyond our consideration.
\par
     More exact equations describing process of transmitting
electromagnetic interaction in its time evolution were suggested by
Maxwell. They are the following ones:
$$
\xalignat 2
&\divr\bold H=0,
&&\rot\bold E=-\frac{1}{c}\,\frac{\partial\bold H}{\partial t},
\tag1.1\\
\vspace{1ex}
&\divr\bold E=4\pi\rho,
&&\rot\bold H=\frac{4\pi}{c}\,\bold j+
\frac{1}{c}\,\frac{\partial\bold E}{\partial t}.
\tag1.2
\endxalignat
$$
It is easy to see that equations \thetag{1.1} and \thetag{1.2}
are generalizations for the \thetag{9.3} and \thetag{9.4}
from Chapter~\uppercase\expandafter{\romannumeral 1}. They are
obtained from latter ones by modifying right hand sides. Like
equations \thetag{9.3} and \thetag{9.4} in
Chapter~\uppercase\expandafter{\romannumeral 1}, Maxwell equations
\thetag{1.1} and \thetag{1.2} can be written in form of integral
equations:
$$
\gather
\aligned
&\int\limits_{\partial\Omega}\bigl<\bold H,\,\bold n\bigr>\,dS=0,\\
\vspace{1ex}
&\oint\limits_{\partial S}\bigl<\bold E,\,\boldsymbol\tau\bigr>
\,ds=-\frac{1}{c}\frac{d}{dt}\int\limits_S\bigl<
\,\bold H,\,\bold n\bigr>\,dS,
\hskip 2em
\endaligned
\tag1.3\\
\vspace{2ex}
\aligned
&\int\limits_{\partial\Omega}\bigl<\bold E,\,\bold n\bigr>\,dS=
4\pi\int\limits_\Omega\rho\,d^3\bold r,\\
\vspace{1ex}
&\oint\limits_{\partial S}\bigl<\bold H,\,
\boldsymbol\tau\bigr>\,ds=
\frac{4\,\pi}{c}\int\limits_S\bigl<\,\bold j,\,
\bold n\bigr>\,dS
+\frac{1}{c}\frac{d}{dt}\int\limits_S\bigl<
\,\bold E,\,\bold n\bigr>\,dS.
\hskip -2em
\endaligned
\tag1.4
\endgather
$$\par
     Consider contour integral in second equation \thetag{1.3}.
Similar contour integral is present in second equation \thetag{1.4}.
However, unlike circulation of magnetic field, circulation of
electric field
$$
\goth e=\oint\limits_{\partial S}\bigl<\bold E,\,\boldsymbol\tau
\bigr>\,ds\tag1.5
$$
possess its own physical interpretation. If imaginary contour
$\Gamma=\partial S$ in space is replaced by real circular conductor,
then electric field with nonzero circulation induces electric current
in conductor. The quantity $\goth e$ from \thetag{1.5} in this case
is called {\it electromotive force} of the field $\bold E$ in contour.
Electromotive force $\goth e\neq 0$ in contour produce the same
effect as linking electric cell with voltage $\goth e$ into this
contour. Experimentally it reveals as follow: alternating magnetic
field produces electric field with nonzero circulation, this induces
electric current in circular conductor. This phenomenon is known
as {\it electromagnetic induction}. It was first discovered by
Faraday. Faraday gave qualitative description of this phenomenon
in form of the following induction law.
\proclaim{\bf Faraday's law of electromagnetic induction} Electromo\-tive
force of induction in circular conductor is proportional to the rate of
changing of magnetic flow embraced by this conductor.
\endproclaim\par
     Faraday's induction law was a hint for Maxwell when choosing
right hand side in second equation \thetag{1.1}. As for similar term
in right hand side of second equation \thetag{1.2}, Maxwell had
written it by analogy. Experiments and further development of
technology proved correctness of Maxwell equations.\par
    Note that charge conservation law in form of relationship
\thetag{5.4} from Chapter~\uppercase\expandafter{\romannumeral 1}
is a consequence of Maxwell equations. One should calculate
divergency of both sides of second equation \thetag{1.2}:
$$
\divr\rot\bold H=\frac{4\pi}{c}\,\divr\bold j+
\frac{1}{c}\,\frac{\partial\divr\bold E}{\partial t},
$$
then one should apply the identity $\divr\rot\bold H=0$. When combined
with the first equation \thetag{1.2} this yields exactly the relationship
\thetag{5.4} from Chapter~\uppercase\expandafter{\romannumeral 1}.
\par
     Equations \thetag{1.1} and \thetag{1.2} form complete system for
describing arbitrary electromagnetic fields. In solving them functions
$\rho(\bold r,t)$ and $\bold j(\bold r,t)$ should be given, or they
should be determined from medium equations. Then each problem of
electrodynamics mathematically reduces to some boundary-value problem
or mixed initial-value/boundary-value problem for Maxwell equations
optionally completed by medium equations. \pagebreak In this section we
consider
only some very special ones among such problems. Our main goal is
to derive important mathematical consequences from Maxwell equations
and to interpret their physical nature.
\head
\S\,2. Density of energy and energy flow\\ for electromagnetic field.
\endhead
\rightheadtext{\S\,2. Density of energy and energy flow\dots}
     Suppose that in bulk conductor we have a current with density
$\bold j$, and suppose that this current is produced by the flow
of charged particles with charge $q$. If $\nu$ is the number of
such particles per unit volume and if $\bold v$ is their velocity,
then $\bold j=q\,\nu\,\bold v$. Recall that current density is a
charge passing through unit area per unit time (see \S\,5 in
Chapter~\uppercase\expandafter{\romannumeral 1}).\par
     In electromagnetic field each particle experiences Lorentz
force determined by formula \thetag{4.4} from
Chapter~\uppercase\expandafter{\romannumeral 1}. Work of this force
per unit time is equal to $\bigl<\bold F,\,\bold v\bigr>=q\,
\bigl<\bold E,\,\bold v\bigr>$. Total work produced by electromagnetic
field per unit volume is obtained if one multiplies this quantity
by $\nu$, then $w=q\,\nu\,\bigl<\bold E,\,\bold v\bigr>=\bigl<\bold E,
\,\bold j\bigr>$. This work increases kinetic energy of particles
(particles are accelerated by field). Otherwise this work is used
for to compensate forces of viscous friction that resist motion
of particles. In either case total power spent by electromagnetic
field within domain $\Omega$ is determined by the following integral:
$$
W=\int\limits_\Omega\bigl<\bold E,\,\bold j\bigr>\,d^3\bold r.
\tag2.1
$$
Let's transform integral \thetag{2.1}. Let's express current density
$\bold j$ through $\bold E$ and $\bold H$ using second equation
\thetag{1.2} for this purpose:
$$
\bold j=\frac{c}{4\,\pi}\rot\bold H-\frac{1}{4\,\pi}
\frac{\partial\bold E}{\partial t}.
\tag2.2
$$
Substituting this expression \thetag{2.2} into formula \thetag{2.1},
we get
$$
W=\frac{c}{4\,\pi}\int\limits_\Omega
\bigl<\bold E,\,\rot\bold H\bigr>\,d^3\bold r-
\frac{1}{8\,\pi}\int\limits_\Omega
\frac{\partial}{\partial t}\bigl<\bold E,\,\bold E\bigr>
\,d^3\bold r.
\tag2.3
$$
In order to implement further transformations in formula \thetag{2.3}
we use well-known identity $\divr\,[\bold a,\,\bold b]=\bigl<\bold b,
\,\rot\bold a\bigr>-\bigl<\bold a,\,\rot\bold b\bigr>$. Assuming
$\bold a=\bold H$ and $\bold b=\bold E$, for $W$ we get
$$
W=\frac{c}{4\,\pi}\int\limits_\Omega\divr [\bold H,\,\bold E]\,
d^3\bold r+\frac{c}{4\,\pi}\int\limits_\Omega
\bigr<\bold H,\,\rot\bold E\bigr>\,d^3\bold r-
\frac{d}{dt}\int\limits_\Omega
\frac{|\bold E|^2}{8\,\pi}\,d^3\bold r.
$$
First integral in this expression can be transformed by means of
Ostrogradsky-Gauss formula, while for transforming $\rot\bold E$
one can use Maxwell equations \thetag{1.1}:
$$
W+\int\limits_{\partial\Omega}\frac{c}{4\,\pi}
\bigl<[\bold E,\,\bold H],\,\bold n\bigr>\,dS+
\frac{d}{dt}\int\limits_\Omega
\frac{|\bold E|^2+|\bold H|^2}{8\,\pi}\,d^3\bold r=0.
\tag2.4
$$
Let's denote by $\bold S$ and $\varepsilon$ vectorial field and scalar
field of the form
$$
\xalignat 2
&\bold S=\frac{c}{4\,\pi}\,[\bold E,\bold H],
&&\varepsilon=\frac{|\bold E|^2+|\bold H|^2}{8\,\pi}.
\tag2.5
\endxalignat
$$
The quantity $\varepsilon$ in \thetag{2.5} is called {\it density of
energy} of electromagnetic field. Vector $\bold S$ is known as {\it
density of energy flow}. It also called {\it Umov-Pointing vector}.
Under such interpretation of quantities \thetag{2.5} the relationship
\thetag{2.4} can be treated as the equation of energy balance. First
summand in \thetag{2.4} is called dissipation power, this is the
amount of energy dissipated per unit time at the expense of
transmitting it to moving charges. Second summand is the amount of
energy that flows from within domain $\Omega$ to outer space per
unit time. These two forms of energy losses lead to diminishing
the energy stored by electromagnetic field itself within domain
$\Omega$ (see third summand in \thetag{2.4}).\par
     Energy balance equation \thetag{2.4} can be rewritten in
differential form, analogous to formula \thetag{5.4} from
Chapter~\uppercase\expandafter{\romannumeral 1}:
$$
\frac{\partial\varepsilon}{\partial t}+\divr\bold S+w=0.
\tag2.6
$$
Here $w=\bigr<\bold E,\,\bold j\bigr>$ is a density of energy
dissipation. Note that in some cases $w$ and integral \thetag{2.1}
in whole can be negative. In such a case we have energy pumping
into electromagnetic field. This energy then flows to outer space
through boundary of the domain $\Omega$. This is the process of
radiation of electromagnetic waves from the domain $\Omega$. It
is realized in antennas (aerials) of radio and TV transmitters. 
If we eliminate or restrict substantially the energy leakage
from the domain $\Omega$ to outer space, then we would have the
device like microwave oven, where electromagnetic field is used
for transmitting energy from radiator to beefsteak.\par
     Electromagnetic field can store and transmit not only the
energy, but the momentum as well. In order to derive momentum
balance equations let's consider again the current with density
$\bold j$ due to the particles with charge $q$ which move with
velocity $\bold v$. Let $\nu$ be concentration of these particles,
i\.\,e\. number of particles per unit volume. Then $\bold j=q\,\nu\,
\bold v$ and $\rho=q\,\nu$. Total force acting on all particles within
domain $\Omega$ is given by integral
$$
\bold F=\int\limits_\Omega\rho\,\bold E\,d^3\bold r+
\int\limits_\Omega\frac{1}{c}\,[\,\bold j,\,\bold H]\,d^3\bold r.
\tag2.7
$$
In order to derive formula \thetag{2.7} one should multiply Lorentz
force acting on each separate particle by the number of particles
per unit volume $\nu$ and then integrate over the domain $\Omega$.
\par
     Force $\bold F$ determines the amount of momentum transmitted
from electromagnetic field to particles enclosed within domain
$\Omega$. Integral \thetag{2.7} is vectorial quantity. For further
transformations of this integral let's choose some constant unit vector
$\bold e$ and consider scalar product of this vector $\bold e$ and
vector $\bold F$:
$$
\bigl<\bold F,\,\bold e\bigr>=\int\limits_\Omega\rho\,
\bigl<\,\bold E,\,\bold e\bigr>\,d^3\bold r+
\int\frac{1}{c}\,\bigl<\bold e,\,[\,\bold j,\,\bold H]\bigr>
\,d^3\bold r.
\tag2.8
$$
Substituting $(2.2)$ into $(2.8)$, we get
$$
\aligned
\bigl<\bold F,\,\bold e\bigr>&=\int\limits_\Omega
\rho\,\bigl<\bold E,\,\bold e\bigr>\,d^3\bold r+
\frac{1}{4\,\pi}\int\limits_\Omega
\bigl<\bold e,\,[\rot\bold H,\,\bold H]\bigr>\,d^3\bold r\,-\\
\vspace{1ex}
&-\frac{1}{4\,\pi\,c}\int\limits_\Omega
\bigl<\bold e,\,\left[{\partial\bold E}/{\partial t},
\,\bold H\right]\bigr>\,d^3\bold r.
\endaligned
\tag2.9
$$
Recalling well-known property of mixed product, we do cyclic
transposition of multiplicands in second integral \thetag{2.9}.
Moreover, we use obvious identity $\left[{\partial\bold E}
/{\partial t},\,\bold H\right]=\partial\left[\bold E,\,
\bold H\right]/\partial t-\left[\bold E,\,{\partial\bold H}
/{\partial t}\right]$. This yields the following expression
for $\bigl<\bold F,\,\bold e\bigr>$:
$$
\aligned
&\bigl<\bold F,\,\bold e\bigr>=\int\limits_\Omega
\rho\,\bigl<\,\bold E,\,\bold e\bigr>\,d^3\bold r+
\frac{1}{4\,\pi}\int\limits_\Omega
\bigl<\rot\bold H,\,[\bold H,\,\bold e]\bigr>
\,d^3\bold r\,-\\
\vspace{1ex}
&-\frac{1}{4\,\pi\,c}\,\frac{d}{dt}
\int\limits_\Omega
\bigl<\bold e,\,[\bold E,\,\bold H]\bigr>\,d^3\bold r+
\frac{1}{4\,\pi\,c}
\int\limits_\Omega\bigl<\bold e,\,\left[\bold E,\,
{\partial\bold H}/{\partial t}\right]\bigr>\,d^3\bold r.
\endaligned
$$
Now we apply second equation of the system \thetag{1.1} written
as ${\partial\bold H}/{\partial t}=-c\,\rot\bold E$. Then we get
formula
$$
\aligned
\bigl<\bold F,\,\bold e\bigr>&+\frac{d}{dt}
\int\limits_\Omega\frac{\bigl<\bold e,\,[\bold E,\,\bold H]\bigr>}
{4\,\pi\,c}\,d^3\bold r=\int\limits_\Omega
\rho\,\bigl<\bold E,\,\bold e\bigr>\,d^3\bold r\,+\\
\vspace{1ex}
&+\int\limits_\Omega
\frac{\bigl<\rot\,\bold H,\,[\bold H,\,\bold e]\bigr>
+\bigl<\rot\,\bold E,\,[\bold E,\,\bold e]\bigr>}{4\,\pi}
\,d^3\bold r.
\endaligned
\tag2.10
$$
In order to transform last two integrals in \thetag{2.10} we
use the following three identities, two of which we already
used earlier:
$$
\aligned
[\bold a,\,[\bold b,\,\bold c]]&=\bold b\,\bigl<\bold a,\,\bold c\bigr>
-\bold c\,\bigl<\bold a,\,\bold b\bigr>,\\
\vspace{0.5ex}
\divr\,[\bold a,\,\bold b]&=\bigl<\bold b,\,\rot\,\bold a\bigr>
-\bigl<\bold a,\,\rot\,\bold b\bigr>,\\
\vspace{0.5ex}
\rot\,[\bold a,\,\bold b]&=\bold a\,\divr\,\bold b-\bold b\,
\divr\bold a-\{\bold a,\,\bold b\}.
\endaligned
\tag2.11
$$
Here by curly brackets we denote commutator of two vector fields
$\bold a$ and $\bold b$ (see \cite{2}). Traditionally square brackets
are used for commutator, but here by square brackets we denote
vector product of two vectors. From second identity \thetag{2.11} we
derive
$$
\bigl<\rot\bold H,\,[\bold H,\,\bold e]\bigr>=
\divr\,[\bold H,\,[\bold H,\,\bold e]]+\bigl<\bold H,
\,\rot\,[\bold H,\,\bold e]\bigr>.
$$
In order to transform $\rot[\bold H,\,\bold e]$ we use third
identity \thetag{2.11}: $\rot[\bold H,\,\bold e]=
-\bold e\,\divr\bold H-\{\bold H,\,\bold e\}$. Then 
$$
\align
\bigl<\bold H,\,\rot\,[&\bold H,\,\bold e]\bigr>=
-\bigl<\bold H,\,\bold e\bigr>\,\divr\bold H
+\sum^3_{i=1} H_i\sum^3_{j=1}
e^j\,\frac{\partial H^i}{\partial r^j}=\\
&=-\bigl<\bold H,\,\bold e\bigr>\,\divr\bold H+\frac{1}{2}\,
\bigl<\bold e,\,\grad|\bold H|^2\bigr>.
\endalign
$$
Let's combine two above relationships and apply first identity
\thetag{2.11} for to transform double vectorial product
$[\bold H,\,[\bold H,\,\bold e]]$ in first of them. As a result
we obtain
$$
\aligned
\bigl<\rot\bold H,\,[\bold H,\,\bold e]\bigr>&=\divr\bigl(\bold H\,
\bigl<\bold H,\,\bold e\bigr>\bigl)-\divr\bigl(\bold e\,|\bold H|^2\bigl)
\,-\\
&-\bigl<\bold H,\,\bold e\bigr>\,\divr\bold H+
\frac{1}{2}\bigl<\bold e,\,\grad|\bold H|^2\bigr>.
\endaligned
$$
But $\divr\bigl(\bold e\,|\bold H|^2\bigl)=\bigl<\bold e,\,\grad|\bold H|^2
\bigr>$. Hence as a final result we get
$$
\aligned
\bigl<\rot\bold H,\,[\bold H,\,\bold e]\bigr>&=
-\bigl<\bold H,\,\bold e\bigr>\,\divr\bold H\,+\\
&+\divr\left(\bold H\,\bigl<\bold H,\,\bold e\bigr>-\frac{1}{2}\,
\bold e\,|\bold H|^2\right).
\endaligned
\tag2.12
$$
Quite similar identity can be derived for electric field $\bold E$:
$$
\aligned
\bigl<\rot\bold E,\,[\bold E,\,\bold e]\bigr>&=
-\bigl<\bold E,\,\bold e\bigr>\,\divr\bold E\,+\\
&+\divr\left(\bold E\,\bigl<\bold E,\,\bold e\bigr>-\frac{1}{2}\,
\bold e\,|\bold E|^2\right).
\endaligned
\tag2.13
$$
The only difference is that due to Maxwell equations $\divr\bold H=0$,
while divergency of electric field $\bold E$ is nonzero: $\divr\bold E
=4\pi\rho$.\par
     Now, if we take into account \thetag{2.12} and \thetag{2.13},
formula \thetag{2.10} can be transformed to the following one:
$$
\aligned
&\bigl<\bold F,\,\bold e\bigr>-
\int\limits_{\partial\Omega}
\frac{\bigl<\bold E,\,\bold e\bigr>\bigl<\bold n,\,\bold E\bigr>+
\bigl<\bold H,\,\bold e\bigr>\bigl<\bold n,\,\bold H\bigr>}
{4\,\pi}\,dS\,+\\
\vspace{1ex}
&+\int\limits_{\partial\Omega}\frac{(|\bold E|^2+|\bold H|^2)\,
\bigl<\bold e,\,\bold n\bigr>}{8\,\pi}\,dS
+\frac{d}{dt}\int\limits_\Omega\frac{\bigl<\bold e,\,
[\bold E,\,\bold H]\bigr>}{4\,\pi\,c}\,d^3\bold r=0.
\endaligned
$$
Denote by $\sigma$ linear operator such that the result of applying
\pagebreak this operator to some arbitrary vector $\bold e$ is given
by formula
$$
\sigma\,\bold e=-\frac{\bold E\,\bigl<\bold E,\,\bold e
\bigr>+\bold H\,\bigl<\bold H,\,\bold e\bigr>}{4\,\pi}+
\frac{|\bold E|^2+|\bold H|^2}{8\,\pi}\,\bold e.
\tag2.14
$$
Formula \thetag{2.14} defines tensorial field $\boldsymbol\sigma$
of type $(1,1)$ with the following components:
$$
\sigma^i_j=\frac{|\bold E|^2+|\bold H|^2}{8\,\pi}\,\delta^i_j-
\frac{E^i\,E_j+H^i\,H_j}{4\,\pi}.
\tag2.15
$$
Tensor $\boldsymbol\sigma$ with components \thetag{2.15} is called
tensor of the {\it density of momentum flow}. It is also known as
{\it Maxwell tensor}. Now let's define vector of {\it momentum density}
$\bold p$ by formula
$$
\bold p=\frac{[\bold E,\,\bold H]}{4\,\pi\,c}.
\tag2.16
$$
In terms of the notations \thetag{2.15} and \thetag{2.16} the above
relationship for $\bigl<\bold F,\,\bold e\bigr>$ is rewritten as
follows:
$$
\bigl<\bold F,\,\bold e\bigr>+\int\limits_{\partial\Omega}
\bigl<\sigma\,\bold e,\,\bold n\bigr>\,dS+
\frac{d}{dt}\int\limits_\Omega
\bigl<\bold p,\,\bold e\bigr>\,d^3\bold r=0.
\tag2.17
$$
Operator of the density of momentum flow $\boldsymbol\sigma$ is
symmetric, i\.\,e\. $\bigl<\sigma\,\bold e,\,\bold n\bigr>=
\bigl<\bold e,\,\sigma\,\bold n\bigr>$. Due to this property
and because $\bold e$ is arbitrary vector we can rewrite
\thetag{2.17} in vectorial form:
$$
\bold F+\int\limits_{\partial\Omega}\sigma
\,\bold n\, dS+\frac{d}{dt}\int\limits_\Omega
\bold p\,d^3\bold r=0.
\tag2.18
$$
This equation \thetag{2.18} is the equation of momentum balance
for electromagnetic field. Force $\bold F$, given by formula
\thetag{2.7} determines loss of momentum stored in electromagnetic
field due to transmitting it to moving particles. Second term in
\thetag{2.18} determines loss of momentum due to its flow through
the boundary of the domain $\Omega$. These two losses lead to
diminishing the momentum stored by electromagnetic field within
domain $\Omega$ (see third summand in \thetag{2.18}).\par
    The relationship \thetag{2.18} can be rewritten in differential
form. For this purpose we should define vectorial divergency for
tensorial field $\boldsymbol\sigma$ of the type $(1,1)$. Let
$$
\boldsymbol\mu=\divr\sigma\text{,\ \ where }
\mu_j=\sum^3_{i=1}\frac{\partial\sigma^i_j}{\partial r^i}.
\tag2.19
$$
Then differential form of \thetag{2.18} is written as 
$$
\frac{\partial\bold p}{\partial t}+\divr\sigma+
\bold f=0,
\tag2.20
$$
where $\bold f=\rho\,\bold E+[\,\bold j,\,\bold H]/c$ is a density of
Lorentz force, while vectorial divergency is determined according to
\thetag{2.19}.\par
     Thus, electromagnetic field is capable to accumulate within itself
the energy and momentum:
$$
\xalignat 2
&\ \Cal E=\int\limits_\Omega\frac{|\bold E|^2+|\bold H|^2}{8\,\pi}
\,d^3\bold r,
&&\bold P=\int\limits_\Omega\frac{[\bold E,\,\bold H]}{4\,\pi\,c}
\,d^3\bold r.
\hskip -2em
\tag2.21
\endxalignat
$$
It is also capable to transmit energy and momentum to material bodies.
This confirms once more our assertion that electromagnetic field itself
is a material entity. It is not pure mathematical abstraction convenient
for describing interaction of charges and currents, but real physical
object.\par
\proclaim{\bf Exercise 2.1} Verify that relationships \thetag{2.11}
hold. Check on the derivation of \thetag{2.12} and \thetag{2.13}.
\endproclaim\par
\head
\S\,3. Vectorial and scalar potentials of electromagnetic field.
\endhead
\rightheadtext{\S\,3. Vectorial and scalar potentials \dots}
     In section~2 we have found that electromagnetic field possess
energy and momentum \thetag{2.21}. This is very important consequence
of Maxwell equations \thetag{1.1} and \thetag{1.2}. However we have not
studied Maxwell equations themselves. This is system of four equations,
two of them are scalar equations, other two are vectorial equations.
So they are equivalent to eight scalar equations. However we have only
six undetermined functions in them: three components of vector $\bold E$
and three components of vector $\bold H$. So observe somewhat like
excessiveness in Maxwell equations.\par
     One of the most popular ways for solving systems of algebraic
equations is to express some variable through other ones by solving
one of the equations in a system (usually most simple equation) and
then substituting the expression obtained into other equations. Thus
we exclude one variable and diminish the number of equations in a
system also by one. Sometimes this trick is applicable to differential
equations as well. Let's consider Maxwell equation $\divr\bold H=0$.
Vector field with zero divergency is called {\it vortex field}. For
vortex fields the following theorem holds (see proof in book \cite{3}).
\proclaim{\bf Theorem on vortex field} Each vortex field is a rotor
of some other vector field.
\endproclaim
     Let's write the statement of this theorem as applied to magnetic
field. It is given by the following relationship:
$$
\bold H=\rot\bold A.
\tag3.1
$$
Vector field $\bold A$, whose existence is granted by the above theorem,
is called {\it vector-potential} of electromagnetic field.\par
     Let's substitute vector $\bold H$ as given by \thetag{3.1} into
second Maxwell equation \thetag{1.1}. This yields the equality
$$
\rot\bold E+\frac{1}{c}\frac{\partial}{\partial t}\rot\bold A=
\rot\left(\bold E+\frac{1}{c}\frac{\partial\bold A}{\partial t}
\right)=0.
\tag3.2
$$
Vector field with zero rotor is called {\it potential field}.
It is vector field $\bold E+(\partial\bold A/\partial t)/c$ in
formula \thetag{3.2} which is obviously potential field. Potential
fields are described by the following theorem (see proof in book
\cite{3}).
\proclaim{\bf Theorem on potential field} Each potential field
is a gradient of some scalar field.
\endproclaim
     Applying this theorem to vector field \thetag{3.2},
we get the relationship determining {\it scalar potential}
of electromagnetic field $\varphi$:
$$
\bold E+\frac{1}{c}\frac{\partial\bold A}{\partial t}=
-\grad\varphi.
\tag3.3
$$
Combining \thetag{3.1} and \thetag{3.3}, we can express electric
and magnetic fields $\bold E$ and $\bold H$ through newly introduced
fields $\bold A$ and $\varphi$:
$$
\aligned
&\bold E=-\grad\varphi-\frac{1}{c}\frac{\partial\bold A}
{\partial t},\\
&\bold H=\rot\bold A.
\endaligned
\tag3.4
$$\par
     Upon substituting \thetag{3.4} into first pair of Maxwell equations
\thetag{1.1} we find them to be identically fulfilled. As for second pair
of Maxwell equations, substituting \thetag{3.4} into these equations, we
get
$$
\aligned
&-\triangle\varphi-\frac{1}{c}\frac{\partial}{\partial t}
\divr\bold A=4\,\pi\,\rho,\\
&\grad\divr\bold A-\triangle\bold A+\frac{1}{c}
\frac{\partial}{\partial t}\grad\varphi+
\frac{1}{c^2}\frac{\partial^2\bold A}{\partial t^2}=
\frac{4\,\pi\,\bold j}{c}.
\endaligned
\tag3.5
$$
In deriving \thetag{3.5} we used relationships
$$
\aligned
&\divr\grad\varphi=\triangle\varphi,\\
&\rot\rot\bold A=\grad\divr\bold A-\triangle\bold A.
\endaligned
\tag3.6
$$
Second order differential operator $\triangle$ is called 
{\it Laplace operator}. In rectangular Cartesian coordinates
it is defined by formula
$$
\triangle=\sum^3_{i=1}\left(\frac{\partial}{\partial r^i}
\right)^2=\frac{\partial^2}{\partial x^2}+\frac{\partial^2}
{\partial y^2}+\frac{\partial^2}{\partial z^2}.
\tag3.7
$$\par
     In order to simplify the equations \thetag{3.5} we rearrange
terms in them. As a result we get 
$$
\aligned
&\frac{1}{c^2}\frac{\partial^2\varphi}{\partial t^2}-
\triangle\varphi=4\,\pi\,\rho+
\frac{1}{c}\frac{\partial}{\partial t}
\left(\frac{1}{c}\frac{\partial\varphi}{\partial t}+
\divr\bold A\right),\\
&\frac{1}{c^2}\frac{\partial^2\bold A}{\partial t^2}-
\triangle\bold A=\frac{4\,\pi\,\bold j}{c}-
\grad\left(\frac{1}{c}\frac{\partial\varphi}{\partial t}+
\divr\bold A\right).
\endaligned
\tag3.8
$$
Differential equations \thetag{3.8} are Maxwell equations written
in terms of $\bold A$ and $\varphi$. This is system of two equations
one of which is scalar equation, while another is vectorial equation.
As we can see, number of equations now is equal to the number of
undetermined functions in them.\par
\head
\S\,4. Gauge transformations and Lorentzian gauge.
\endhead
\rightheadtext{\S\,4. Gauge transformations \dots}
     Vectorial and scalar potentials $\bold A$ and $\varphi$ were
introduced in \S\,3 as a replacement for electric and magnetic
fields $\bold E$ and $\bold H$. However, fields $\bold A$ and
$\varphi$ are not physical fields. Physical fields $\bold E$
and $\bold H$ are expressed through $\bold A$ and $\varphi$ according
to formulas \thetag{3.4}, but backward correspondence is not unique,
i\.\,e\. fields $\bold A$ and $\varphi$ are not uniquely determined
by physical fields $\bold E$ and $\bold H$. Indeed, let's consider
transformation 
$$
\aligned
&\tilde{\bold A}=\bold A+\grad\psi,\\
&\tilde\varphi=\varphi-\frac{1}{c}\frac{\partial\psi}
{\partial t},
\endaligned
\tag4.1
$$
where $\psi(\bold r,t)$ is an arbitrary function. Substituting
\thetag{4.1} into formula \thetag{3.4}, we immediately get
$$
\xalignat 2
&\tilde{\bold E}=\bold E,
&&\tilde{\bold H}=\bold H.
\endxalignat
$$
This means that physical fields $\bold E$, $\bold H$ determined
by fields $\tilde{\bold A}$, $\tilde\varphi$ and by fields
$\bold A$, $\varphi$ do coincide. Transformation \thetag{4.1}
that do not change physical fields $\bold E$ and $\bold H$
is called {\it gauge transformation}.\par
     We use gauge transformations \thetag{4.1} for further
simplification of Maxwell equations \thetag{3.8}. Let's
consider the quantity enclosed in brackets in right hand sides
of the equations \thetag{3.8}:
$$
\frac{1}{c}\frac{\partial\varphi}{\partial t}+
\divr\bold A=
\frac{1}{c}\frac{\partial\tilde\varphi}{\partial t}+
\divr\tilde{\bold A}
+\left(\frac{1}{c^2}\frac{\partial^2\psi}{\partial t^2}-
\triangle\psi\right).
\tag4.2
$$
Denote by $\square$ the following differential operator:
$$
\square=\frac{1}{c^2}\frac{\partial^2}{\partial t^2}-
\triangle.
\tag4.3
$$
Operator \thetag{4.3} is called {\it d'Alambert operator} or
{\it wave operator}. Differential equation $\square u=v$
is called {\it d'Alambert equation}.\par
     Using gauge freedom provided by gauge transformation \thetag{4.1},
we can fulfill the following condition:
$$
\frac{1}{c}\frac{\partial\varphi}{\partial t}+
\divr\bold A=0.
\tag4.4
$$
For this purpose we should choose $\psi$ solving d'Alambert equation
$$
\square\psi=-\left(
\frac{1}{c}\frac{\partial\tilde\varphi}{\partial t}+
\divr\tilde{\bold A}\right).
$$
It is known that d'Alambert equation is solvable under rather weak
restrictions for its right hand side (see book \cite{1}). Hence
practically always we can fulfill the condition \thetag{4.4}. This
condition is called {\it Lorentzian gauge}.\par
     If Lorentzian gauge condition \thetag{4.4} is fulfilled, then
Maxwell equations \thetag{3.8} simplify substantially:
$$
\xalignat 2
&\square\varphi=4\,\pi\,\rho,
&&\square\bold A=\frac{4\,\pi\,\bold j}{c}.
\tag4.5
\endxalignat
$$
They look like pair of independent d'Alambert equations.
However, one shouldn't think that variables $\bold A$ and $\varphi$
are completely separated. Lorentzian gauge condition \thetag{4.4}
itself is an additional equation requiring concordant choice of
solutions for d'Alambert equations \thetag{4.5}.\par
     D'Alambert operator \thetag{4.3} is a scalar operator, in
\thetag{4.5} it acts upon each component of vector $\bold A$ separately.
Therefore operator $\square$ commutates with rotor operator and with
time derivative as well. Therefore on the base of \thetag{3.4} we
derive
$$
\xalignat 2
&\square\bold E=-4\pi\,\grad\rho-\frac{4\pi}{c^2}\,
\frac{\partial j}{\partial t},
&&\square\bold H=\frac{4\pi}{c}\rot\bold j.
\hskip -2em
\tag4.6
\endxalignat
$$
These equations \thetag{4.6} have no entries of potentials $\bold A$
and $\varphi$. They are written in terms of real physical fields 
$\bold E$ and $\bold H$, and are consequences of Maxwell equations
\thetag{1.1} and \thetag{1.2}. However, backward Maxwell equations
do not follow from \thetag{4.6}.
\head
\S\,5. Electromagnetic waves.
\endhead
\parshape 15 0cm 10.1cm 0cm 10.1cm
5cm 5.1cm 5cm 5.1cm 5cm 5.1cm 5cm 5.1cm
5cm 5.1cm 5cm 5.1cm 5cm 5.1cm 5cm 5.1cm
5cm 5.1cm 5cm 5.1cm 5cm 5.1cm 5cm 5.1cm
0cm 10.1cm
     In previous Chapter we considered static electromagnetic fields.
Such fields are uniquely determined by static configuration of charges
and currents (see formulas \thetag{3.5} and \thetag{5.6} in
Chapter~\uppercase\expandafter{\romannumeral 1}\,). They cannot exist
in the absence of charges and currents.
\vadjust{\vtop to 0pt{\hbox to 0pt{\kern -25pt
\includegraphics{fig6.eps}\hss}
\vskip  73pt\hbox{\kern 40pt {\it Fig\.~5.1 }\hss}
\vskip -30pt\hbox{\kern 20pt $z$\hss}
\vskip -38pt\hbox{\kern 5pt $\bold H_0$\hss}
\vskip -33pt\hbox{\kern 115pt $x$\hss}
\vskip -16pt\hbox{\kern 90pt $\bold k$\hss}
\vskip -16pt\hbox{\kern 23pt $\bold A_0$\hss}
\vskip -29pt\hbox{\kern 23pt $\bold E_0$\hss}
\vskip -29pt\hbox{\kern 59pt $y$\hss}
\vskip -100pt\vss}}
However, as we shall see just
below, Maxwell equations have nonzero solutions even in the case of
identically zero currents and charges in the space. Let's study one of
such solutions. We choose some right-oriented rectangular Cartesian
system of coordinates and take some constant vector $\bold k$ directed
along $x$-axis (see Fig\.~5.1). Then we choose another constant vector
$\bold A_0$ directed along $y$-axis and consider the following two
functions:
$$
\xalignat 2
&\bold A=\bold A_0\sin(k\,x-\omega\,t),
&&\varphi=0.
\tag5.1
\endxalignat
$$
Here $k=|\bold k|$. Suppose $\rho=0$ and $\bold j=0$. Then, substituting
\thetag{5.1} into \thetag{4.4} and into Maxwell equations \thetag{4.5},
we get
$$
k^2=|\bold k|^2=\frac{\omega}{c}.
\tag5.2
$$
It is not difficult to satisfy this condition \thetag{5.2}. If it is
fulfilled, then corresponding potentials \thetag{5.1} describe plane
electromagnetic wave, $\omega$ being its {\it frequency} and $\bold k$
being its {\it wave-vector}, which determines the direction of
propagation of that plane wave. Rewriting \thetag{5.1} in a little bit
different form
$$
\bold A=\bold A_0\sin(k(x-c\,t)),
\tag5.3
$$
we see that the velocity of propagating of plane electromagnetic
wave is equal to constant $c$ (see \thetag{1.5} in
Chapter~\uppercase\expandafter{\romannumeral 1}).\par
     Now let's substitute \thetag{5.1} into \thetag{3.4} and calculate
electric and magnetic fields in electromagnetic wave:
$$
\xalignat 2
&\bold E=\bold E_0\cos(k\,x-\omega\,t),
&&\bold E_0=|\bold k|\,\bold A_0,\\
\vspace{1ex}
&\bold H=\bold H_0\cos(k\,x-\omega\,t),
&&\bold H_0=[\bold k,\bold A_0],
\tag5.4\\
\vspace{1ex}
&|\bold E_0|=|\bold H_0|=|\bold k|\,|\bold A_0|.
\hskip -20em &&
\endxalignat
$$
Vectors $\bold k$, $\bold E_0$, and $\bold H_0$ are perpendicular
to each other, they form right triple. Wave \thetag{5.4} with such
vectors is called {\it plane linear polarized electromagnetic wave}.
Vector $\bold E_0$ is taken for {\it polarization vector} of this
wave. Wave
$$
\align
\bold E&=\bold E_0\cos(k\,x-\omega\,t)+
\bold H_0\sin(k\,x-\omega\,t),\\
\vspace{1ex}
\bold H&=\bold H_0\cos(k\,x-\omega\,t)-
\bold E_0\sin(k\,x-\omega\,t)
\endalign
$$
is called {\it circular polarized} wave. It is superposition
of two linear polarized waves. Natural light is also electromagnetic
wave. It has no fixed polarization, however it is not circular
polarized as well. Natural light is a superposition of numerous
plane linear polarized waves with chaotically distributed
polarization vectors.\par
\head
\S\,6. Emission of electromagnetic waves.
\endhead
     Plane wave \thetag{5.4} is an endless wave filling the whole
space. It is certainly kind of idealization. Real electromagnetic
waves fill only some restricted part of the space. Moreover, they
are not eternal in time: there are sources (radiators) and
absorbers of electromagnetic fields. Formula \thetag{5.4} is an
approximate description of real electromagnetic field in that
part of space which is far apart from radiators and 
absorbers.\par
     In this section we consider process of generation and
radiation of electromagnetic waves. Usually radiator is
a system of charges and currents, which is not static. We
describe it by means of functions $\rho(\bold r,t)$ and
$\bold j(\bold r,t)$. Let's consider Maxwell equations
transformed to the form \thetag{4.5}. These are non-homogeneous
differential equations. Their solutions are not unique: to
each solution already found one can add arbitrary solution of
corresponding homogeneous equations. However, if we assume
$\rho(\bold r,t)$ and $\bold j(\bold r,t)$ to be fast decreasing
as $\bold r\to\infty$ and apply similar condition to
$\varphi(\bold r,t)$ and $\bold A(\bold r,t)$, then we restrict
substantially the freedom in choosing solutions of the equations
\thetag{4.5}. In order to find one of such solutions we need
fundamental solution of d'Alambert operator. This is distribution
of the form:
$$
u(\bold r,t)=\frac{c}{2\pi}\,\theta(t)\,\delta(c^2t^2-
|\bold r|^2),
\tag6.1
$$
where $\theta$ and $\delta$ are Heaviside theta-function and
Dirac delta-function respectively. Function \thetag{6.1} satisfies
d'Alambert equation with distribution in right hand side:
$$
\square u=\delta(t)\delta(\bold r).
$$
In physics such objects are called {\it Green functions}. Knowing
fundamental solution \thetag{6.1} of d'Alambert operator, now we
can write solution for the equations \thetag{4.5} in form of 
contractions:
$$
\xalignat 2
&\varphi=4\pi\,u*\rho,
&&\bold A=\frac{4\pi}{c}\,u*\bold j.
\tag6.2
\endxalignat
$$
Here $*$ denotes contraction of two distributions, see \cite{1}. Due
to the properties of this operation from charge conservation law
(see formula \thetag{5.4} in
Chapter~\uppercase\expandafter{\romannumeral 1}) we derive
Lorentzian gauge condition \thetag{4.4} for scalar and vectorial
potentials \thetag{6.2}. For smooth and sufficiently fast decreasing
functions $\rho(\bold r,t)$ and $\bold j(\bold r,t)$ potentials
\thetag{6.2} are reduced to the following two integrals:
$$
\aligned
&\varphi(\bold r,t)=\int\frac{\rho(\tilde{\bold r},t-\tau)}
{|\bold r-\tilde{\bold r}|}\,d^3\tilde{\bold r},\\
\vspace{2ex}
&\bold A(\bold r,t)=\int\frac{\bold j(\tilde{\bold r},t-\tau)}
{c\,|\bold r-\tilde{\bold r}|}\,d^3\tilde{\bold r}.
\endaligned
\tag6.3
$$
Here the quantity $\tau=\tau(\bold r,\tilde{\bold r})$ is called
{\it time delay}. It is determined by the ratio $\tau=|\bold r-
\tilde{\bold r}|/c$. Potentials \thetag{6.3} are called
{\it retarded potentials}.\par
    Retarded potentials have transparent physical interpretation.
Scalar potential $\varphi$ at the point $\bold r$ at time instant
$t$ is a superposition of contributions from charges at various
points of the space, the contribution from the point $\tilde{\bold r}$
being determined not by charge density at present time instant $t$,
but at previous time instant $t-\tau$. Time delay $\tau$ is exactly
equal to the time required for the signal spreading with light velocity
$c$ from the source point $\tilde{\bold r}$ to get to the observation
point $\bold r$. Similar time delay is present in formula for vector
potential $\bold A$.\par
     Note that fundamental solution of d'Alambert equation is not unique.
For example there is a solution obtained from \thetag{6.1} by changing
$\tau$ for $-\tau$. Such solution corresponds to {\it advanced potentials}.
However, in physics advanced potentials have no meaning, since they would
break causality principle.\par
     Let's consider system of charges located in some small domain
$\Omega$ surrounding the origin. Let $R$ be maximal linear size of
this domain $\Omega$. Using formulas \thetag{6.3}, we calculate we
calculate electromagnetic field of the system of charges at the point
$\bold r$ which is far distant from the domain $\Omega$, i\.\,e\.
$|\tilde{\bold r}|\leq R\ll |\bold r|$. Due to these inequalities the
ratio $\tilde{\bold r}/|\bold r|$ is small vectorial quantity.
Therefore we have the following asymptotic expansions for
$|\bold r-\tilde{\bold r}|$ and $t-\tau$:
$$
\aligned
&|\bold r-\tilde{\bold r}|=|\bold r|-\frac{\bigl<\bold r,\,
\tilde{\bold r}\bigr>}{|\bold r|}+\ldots,\\
\vspace{1ex}
&t-\tau=t-\frac{|\bold r|}{c}+\frac{\bigl<\bold r,\,
\tilde{\bold r}\bigr>}{|\bold r|\,c}+\ldots.
\endaligned
\tag6.4
$$
The ratio $|\bold r|/c$ in \thetag{6.4} determines the time
required for electromagnetic signal to get from the domain
$\Omega$ to the observation point $|\bold r|$. Posterior
terms in the series for $t-\tau$ are estimated by small quantity
$R/c$. This is the time of propagation of electromagnetic signal
within domain $\Omega$.\par
     Denote $t'=t-|\bold r|/c$ and let $t-\tau=t'+\theta$. For the
quantity $\theta$ we have the estimate $|\theta|\leq R/c$. Then
let's consider the following Taylor expansions for
$\rho$ and $\bold j$:
$$
\aligned
&\rho(\tilde{\bold r},t-\tau)=
\rho(\tilde{\bold r},t')+
\frac{\partial\rho(\tilde{\bold r},t')}
{\partial t}\,\theta+\ldots\,,\\
\vspace{1ex}
&\bold j(\tilde{\bold r},t-\tau)=
\bold j(\tilde{\bold r},t')+
\frac{\partial\bold j(\tilde{\bold r},t')}
{\partial t}\,\theta+\ldots\,.
\endaligned
\tag6.5
$$
The condition $R\ll |\bold r|$ is not sufficient for the
expansions \thetag{6.5} to be consistent. Use of expansions
\thetag{6.5} for approximating $\rho(\tilde{\bold r},t-\tau)$
and $\bold j(\tilde{\bold r},t-\tau)$ is possible only under
some additional assumptions concerning these functions. Denote
by $T$ some specific time for which functions $\rho$ and $\bold j$
within domain $\Omega$ change substantially. In case when one can
specify such time $T$, the following quantities are of the same
order, i\.\,e\. equally large or equally small:
$$
\aligned
&\rho\approx T\,\frac{\partial\rho}{\partial t}\approx
\ldots\approx T^n\,\frac{\partial^n\rho}{\partial t^n},\\
\vspace{1ex}
&\bold j\approx T\,\frac{\partial\bold j}{\partial t}\approx
\ldots\approx T^n\,\frac{\partial^n\bold j}{\partial t^n}.
\endaligned
\tag6.6
$$
Now \thetag{6.5} can be rewritten as follows:
$$
\aligned
&\rho(\tilde{\bold r},t-\tau)=
\rho(\tilde{\bold r},t')+
T\,\frac{\partial\rho(\tilde{\bold r},t')}
{\partial t}\,\frac{\theta}{T}+\ldots,\\
\vspace{1ex}
&\bold j(\tilde{\bold r},t-\tau)=
\bold j(\tilde{\bold r},t')+
T\,\frac{\partial\bold j(\tilde{\bold r},t')}
{\partial t}\,\frac{\theta}{T}+\ldots.
\endaligned
\tag6.7
$$
Correctness of use of expansions \thetag{6.7} and \thetag{6.5}
is provided by additional condition $R/c\ll T$. This yields
$\theta/T\ll 1$.\par
     The condition $R/c\ll T$ has simple meaning: the quantity
$\omega=2\pi/T$ is a frequency of radiated electromagnetic waves,
while $\lambda=2\pi c/\omega=c\,T$ is a wavelength. Hence condition
$R/c\ll T$ means that wavelength is mach greater than the size of
radiator.\par
     Suppose that both conditions $R\ll c\,T$ and $R\ll |\bold r|$
are fulfilled. Let's calculate retarded vector potential $\bold A$
in \thetag{6.3} keeping only first term in the expansion \thetag{6.5}:
$$
\bold A=\int\limits_\Omega\frac{\bold j(\tilde{\bold r},t')}
{|\bold r|\,c}\,d^3\tilde{\bold r}+\ldots.
\tag6.8
$$
In order to transform integral in \thetag{6.8} let's choose some
arbitrary constant vector $\bold e$ and consider scalar product 
$\bigl<\bold A,\,\bold e\bigr>$. Having defined vector $\bold a$
and function $f(\tilde{\bold r})$ by the relationships 
$$
\xalignat 2
&\bold a=\frac{\bold e}{c|\bold r|}=\grad f,
&&f(\tilde{\bold r})=\bigl<\bold a,\,\tilde{\bold r}\bigr>,
\endxalignat
$$
we make calculations analogous to that of \thetag{7.5}
in Chapter~\uppercase\expandafter{\romannumeral 1}:
$$
\aligned
&\int\limits_\Omega\bigl<\,\bold j,\grad f\bigr>\,
d^3\tilde{\bold r}=
\int\limits_\Omega\divr(f\,\bold j)\,d^3\tilde{\bold r}-
\int\limits_\Omega f\,\divr\bold j\,d^3\tilde{\bold r}=\\
&=\int\limits_{\partial\Omega}f\bigl<\,\bold j,\,\bold n
\bigr>\,dS+\int\limits_\Omega f\,\frac{\partial\rho}
{\partial t}\,d^3\tilde{\bold r}=
\int\limits_\Omega \frac{\partial\rho(\tilde{\bold r},t')}
{\partial t}\frac{\bigl<\bold e,\,\tilde{\bold r}\bigr>}
{|\bold r|\,c}\,d^3\tilde{\bold r}.\hskip -2em
\endaligned
\tag6.9
$$
Since $\bold e$ is arbitrary vector, for vector potential $\bold A$
from \thetag{6.9} we derive the following formula:
$$
\bold A=
\int\limits_\Omega \frac{\partial\rho(\tilde{\bold r},t')}
{\partial t}\frac{\tilde{\bold r}}
{|\bold r|\,c}\,d^3\tilde{\bold r}+\ldots=
\frac{\dot{\bold D}}{|\bold r|\,c}+\ldots\,.
\tag6.10
$$
Here $\dot{\bold D}=\dot{\bold D}(t')$ is time derivative of dipole
moment $\bold D$ of the system of charges at time instant $t'$.\par
     In a similar way, keeping only initial terms in the expansions
\thetag{6.4} and \thetag{6.5}, for scalar potential $\varphi$ in
\thetag{6.3} we find 
$$
\varphi=\int\limits_\Omega \frac{\rho(\tilde{\bold r},t')}
{|\bold r|}\,d^3\tilde{\bold r}+\ldots=
\frac{Q}{|\bold r|}+\ldots,
\tag6.11
$$
where $Q$ is total charge enclosed within domain $\Omega$. This charge
does not depend on time since domain $\Omega$ is isolated and we have
no electric current in outer space.\par
     Let's compare the expressions under integration in \thetag{6.10}
and \thetag{6.11} taking into account \thetag{6.6}. This comparison
yields 
$$
|\bold A|\approx\frac{R}{c\,T}\,\varphi.
$$
The estimate $R/(c\,T)\ll 1$ following from $R\ll c\,T$ means that
vectorial potential is calculated with higher accuracy than scalar
potential. Hence in calculating $\varphi$ one should take into account
higher order terms in expansions \thetag{6.4} and \thetag{6.5}.
Then
$$
\aligned
\varphi&=\frac{Q}{|\bold r|}+
\int\limits_\Omega
\frac{\partial\rho(\tilde{\bold r},t')}{\partial t}\,
\frac{\bigl<\bold r,\,\tilde{\bold r}\bigr>}{|\bold r|^2\,c}
\,d^3\tilde{\bold r}+\\
&+\int\limits_\Omega\frac{\rho(\tilde{\bold r},t')}{|\bold r|}\,
\frac{\bigl<\bold r,\,\tilde{\bold r}\bigr>}{|\bold r|^2}\,
d^3\tilde{\bold r}+\ldots\,.
\endaligned
\tag6.12
$$
Calculating integrals in formula \thetag{6.12}, we transform it to
$$
\varphi=\frac{Q}{|\bold r|}+
\frac{\bigl<\dot{\bold D},\,\bold r\bigr>}{|\bold r|^2\,c}+
\frac{\bigl<\bold D,\,\bold r\bigr>}{|\bold r|^3}+\ldots\,.
\tag6.13
$$\par
     Potentials \thetag{6.10} and \thetag{6.13} are retarded potentials
of the system of charges in {\it dipole approximation}. Dependence of
$\rho$ and $\bold j$ on time variable $t$ lead to the dependence of
$\bold D$ on $t'$ in them. Let's consider asymptotics of of these
potentials as $\bold r\to\infty$. Thereby we can omit last term in
\thetag{6.13}. Then
$$
\xalignat 2
\varphi=\frac{Q}{|\bold r|}+\frac{\bigl<\dot{\bold D},\,
\bold r\bigr>}{|\bold r|^2\,c}+\ldots\,,
&&\bold A=\frac{\dot{\bold D}}{|\bold r|\,c}+\ldots\,.
\hskip-2em
\tag6.14
\endxalignat
$$
Now on the base of formulas \thetag{3.4} and \thetag{6.14} we find

asymptotics of electric and magnetic fields at far distance from
the system of charges. In calculating $\rot\bold A$ and $\grad\varphi$
we take into account that $t'=t-|\bold r|/c$ in argument of
$\dot{\bold D}(t')$ is a quantity depending on $\bold r$. This dependence
determines leading terms in asymptotics of $\bold E$ and $\bold H$: 
$$
\xalignat 2
&\bold E=\frac{[\bold r,\,[\bold r,\ddot\bold D]]}
{|\bold r|^3\,c^2}+\ldots\,,
&&\bold H=-\frac{[\bold r,\ddot{\bold D}]}
{|\bold r|^2\,c^2}+\ldots\,.
\hskip-2em
\tag6.15
\endxalignat
$$\par
Vectors $\bold E$ and $\bold H$ (more precisely, leading terms in their
asymptotics) are perpendicular to each other and both are perpendicular
to vector $\bold r$. This situation is similar to that of plane wave.
However, in present case we deal with spherical wave being radiated from
the origin. The magnitude of fields $|\bold E|\simeq|\bold H|$ decreases
as $1/|\bold r|$, which is slower than for static Coulomb field. Using
formula \thetag{2.5}, one can find the density of energy flow for
waves \thetag{6.15}:
$$
\bold S=\frac{|[\bold r,\,\ddot{\bold D}]|^2}
{4\pi\,|\bold r|^5\,c^3}\,\bold r+\ldots.
\tag6.16
$$
For modulus of vector $\bold S$ we have $|\bold S|\sim 1/|\bold r|^2$.
This means that total flow of energy through the sphere of arbitrarily
large sphere is nonzero. So we have real radiation of electromagnetic
energy. The amount of radiated energy is determined by second time
derivative of dipole moment. Therefore this case is called dipole
approximation in theory of radiation.\par
\proclaim{\bf Exercise 6.1} Applying formula \thetag{6.16}, find
angular distribution of the intensity for dipole radiation. Also
find total intensity of dipole radiation.
\endproclaim
\proclaim{\bf Exercise 6.2} Particle with charge $q$ is moving
along circular path of radius $R$ with constant velocity $v=\omega R$
for infinitely long time ($\omega$ is angular velocity). Calculate
retarding potentials and find angular distribution for intensity of
electromagnetic radiation of this particle. Also find total intensity
of such {\it cyclotronic} radiation.
\endproclaim
\proclaim{\bf Exercise 6.3} Assume that charge density $\rho$ is zero,
while current density $\bold j$ is given by the following distribution:
$$
\bold j(\bold r,t)=-c\,[\bold M(t),\,\grad\delta(\bold r)]
\tag6.17
$$
(compare with \thetag{7.16} in
Chapter~\uppercase\expandafter{\romannumeral 1}). Find retarding
potentials \thetag{6.2} for \thetag{6.17}. Also find angular distribution
and total intensity for {\it magnetic-dipole} radiation induced by
current \thetag{6.17}.
\endproclaim
\newpage
\setfirstpage
\topmatter
\title\chapter{3}
Special theory of relativity
\endtitle
\endtopmatter
\leftheadtext{CHAPTER \uppercase\expandafter{\romannumeral 3}.
SPECIAL RELATIVITY.}
\document
\head
\S\,1. Galileo transformations.
\endhead
     Classical electrodynamics based on Maxwell equations historically
was first field theory. It explained all electromagnetic phenomena and
predicted the existence of electromagnetic waves. Later on electromagnetic
waves were detected experimentally and nowadays they have broad scope of
applications in our everyday life. However, along with successful
development of this theory, some difficulties there appeared. It was
found that classical electrodynamics contradicts to {\it relativity 
principle}.
This principle in its classical form suggested by Galileo and Newton
states that two Cartesian inertial coordinate systems moving with constant
velocity with respect to each other are equivalent. All physical phenomena
in these two systems happen identically and are described by the same laws.
\par
     Let's consider two such Cartesian inertial coordinate systems
$(\bold r,t)$ and $(\tilde{\bold r},\tilde t)$. Suppose that second
system moves with velocity $\bold u$ relative to first one so that
coordinate axes in motion remain parallel to their initial positions.
The relation of radius-vectors of points then can be written in form
of the following transformations known as Galileo transformations:
$$
\xalignat 2
&t=\tilde t,
&&\bold r=\tilde\bold r+\bold u\tilde t.
\tag1.1
\endxalignat
$$
First relationship \thetag{1.1} means that watches in two systems
are synchronized and tick synchronously. Let $\tilde\bold r(\tilde t)$
be trajectory of some material point in coordinate system $(\tilde
\bold r,\tilde t)$. In first coordinate system this trajectory is
given by vector $\bold r(t)=\tilde\bold r(\tilde t)+\bold u\tilde t$.
Differentiating this relationship, due to $\tilde t=t$ in \thetag{1.1}
we get
$$
\xalignat 2
&\frac{\partial\bold r}{\partial t}=
\frac{\partial\tilde\bold r}{\partial\tilde t}+\bold u,
&&\bold v=\tilde\bold v+\bold u.
\tag1.2
\endxalignat
$$
Last relationship in \thetag{1.2} is known as {\it classical law of
velocity addition}. Differentiating \thetag{1.2} once more, we find
the relation for accelerations of material point in these two
coordinate systems:
$$
\xalignat 2
&\frac{\partial^2\bold r}{\partial t^2}=
\frac{\partial^2\tilde\bold r}{\partial\tilde t^2},
&&\bold a=\tilde\bold a.
\tag1.3
\endxalignat
$$
According to Newton's second law, acceleration of material point
is determined by force $\bold F$ acting on it and by its mass
$m\,\bold a=\bold F$. From \thetag{1.3} due to relativity principle
we conclude that force $\bold F$ is invariant quantity. It doesn't
depend on the choice of inertial coordinate system. This fact is
represented by the relationship
$$
\bold F(\tilde\bold r+\bold u\tilde t,\tilde\bold v+\bold u)=
\tilde\bold F(\tilde\bold r,\tilde\bold v).
\tag1.4
$$\par
     Now let's consider charged particle with charge $q$ being
at rest in coordinate system $(\tilde\bold r,\tilde t)$. In this
coordinate system it produces Coulomb electrostatic field. In
coordinate system $(\bold r,t)$ this particle is moving. Hence
it should produce electric field and magnetic field as well.
This indicate that vectors $\bold E$ and $\bold H$ are not
invariant under Galileo transformations \thetag{1.1}. Even if
in one coordinate system we have pure electric field, in second
system we should expect the presence of both electric and magnetic
fields. Therefore transformation rules for $\bold E$ and $\bold H$
analogous to \thetag{1.4} for $\bold F$ should be written in
the following form:
$$
\aligned
\bold E(\tilde\bold r+\bold u\tilde t,\tilde t)&=
\alpha(\tilde\bold E(\tilde\bold r,\tilde t),
\tilde\bold H(\tilde\bold r,\tilde t),\bold u),\\
\bold H(\tilde\bold r+\bold u\tilde t,\tilde t)&=
\beta(\tilde\bold E(\tilde\bold r,\tilde t),
\tilde\bold H(\tilde\bold r,\tilde t),\bold u).
\endaligned
\tag1.5
$$
Due to superposition principle, which is fulfilled in both
coordinate systems, functions $\alpha$ and $\beta$ are linear
and homogeneous with respect to $\tilde\bold E$ and $\tilde\bold H$.
Therefore \thetag{1.5} is rewritten as
$$
\aligned
\bold E(\bold r,t)&=
\alpha_1\,\tilde\bold E(\tilde\bold r,\tilde t)+
\alpha_2\,\tilde\bold H(\tilde\bold r,\tilde t),\\
\bold H(\bold r,t)&=
\beta_1\,\tilde\bold E(\tilde\bold r,\tilde t)+
\beta_2\,\tilde\bold H(\tilde\bold r,\tilde t),
\endaligned
\tag1.6
$$
where $\alpha_1$, $\alpha_2$, $\beta_1$, $\beta_2$ are some
linear operators which depend on $\bold u$ only. Vectors
$\bold E$ and $\bold H$ determine the action of electromagnetic
field upon charges in form of Lorentz force (see formula
\thetag{4.4} in Chapter~\uppercase\expandafter{\romannumeral 1}).
Substituting \thetag{1.6} into that formula and taking into account
\thetag{1.2} and \thetag{1.4}, we get
$$
\aligned
&q\alpha_1\,\tilde\bold E+q\alpha_2\,\tilde\bold H
+\frac{q}{c}\,[\tilde\bold v+\bold u,\,\beta_1\,\tilde\bold E]+\\
&+\frac{q}{c}\,[\tilde\bold v+\bold u,\,\beta_2\,\tilde\bold H]=
q\tilde\bold E+\frac{q}{c}\,[\tilde\bold v,\,\tilde\bold H].
\endaligned
\tag1.7
$$
The relationship \thetag{1.7} is an identity with three arbitrary
parameters: $\tilde\bold v$, $\tilde\bold E$, $\tilde\bold H$.
Therefore we can equate separately terms bilinear with respect
to $\tilde\bold v$ and $\tilde\bold E$. This yields $[\tilde\bold v,
\,\beta_1\,\tilde\bold E]=0$, hence $\beta_1=0$. Now let's equate
terms bilinear with respect to $\tilde\bold v$ and $\tilde\bold H$.
This yields $[\tilde\bold v,\,\beta_2\,\tilde\bold H]=[\tilde\bold v,
\,\tilde\bold H]$. Hence $\beta_2=1$. And finally we should equate
terms linear with respect to $\tilde\bold H$ and $\tilde\bold E$.
This yields the following formulas for operators $\alpha_1$ and
$\alpha_2$:
$$
\xalignat 2
&\alpha_2\,\tilde\bold H=-\frac{1}{c}\,[\bold u,\,\tilde\bold H],
&&\alpha_1=1.
\endxalignat
$$
Now, if we substitute the above expressions for operators 
$\alpha_1$, $\alpha_2$, $\beta_1$, $\beta_2$ into formula
\thetag{1.6}, we get the relationships
$$
\xalignat 2
&\bold E=\tilde\bold E-\frac{1}{c}\,[\bold u,\,\tilde\bold H],
&&\bold H=\tilde\bold H.
\tag1.8
\endxalignat
$$\par
     The relationships \thetag{1.8} should complete Galileo transformations
\thetag{1.1} in electrodynamics. However, as we shall see just below, they
cannot do this mission in non-contradictory form. For this purpose, let's
transform Maxwell equations written as \thetag{1.1} and \thetag{1.2} in
Chapter~\uppercase\expandafter{\romannumeral 2} to coordinate system
$(\tilde{\bold r},\tilde t)$. For partial derivatives due to transformations
\thetag{1.1} we have
$$
\xalignat 2
&\frac{\partial}{\partial r^i}=\frac{\partial}{\partial\tilde r^i},
&&\frac{\partial}{\partial t}=\frac{\partial}{\partial\tilde t}-
\sum^3_{k=1} u^k\frac{\partial}{\partial\tilde r^k}.
\tag1.9
\endxalignat
$$
Now, combining \thetag{1.8} and \thetag{1.9}, we derive
$$
\align
&\divr\bold H=\divr\tilde{\bold H},\\
&\divr\bold E=\divr\tilde{\bold E}+\frac{1}{c}\bigl<
 \bold u,\,\rot\tilde{\bold H}\bigr>,\\
&\rot\bold H=\rot\tilde{\bold H}\\
&\rot\bold E=\rot\tilde{\bold E}+\frac{1}{c}\{\bold u,\,
 \tilde{\bold H}\}-\frac{1}{c}\,\bold u\,\divr\tilde{\bold H},\\
&\frac{\partial\bold H}{\partial t}=
 \frac{\partial\tilde{\bold H}}{\partial\tilde t}-
 \{\bold u,\,\tilde{\bold H}\},\\
&\frac{\partial\bold E}{\partial t}=
 \frac{\partial\tilde{\bold E}}{\partial\tilde t}-
 \{\bold u,\,\tilde{\bold E}\}+\frac{1}{c}\,
 [\bold u,\,\{\bold u,\,\tilde{\bold H}\}]-\frac{1}{c}\,
 [\bold u,\partial\tilde{\bold H}/\partial\tilde t\,].
\endalign
$$
Here by curly brackets we denote commutator of vector fields
(see \cite{2}). Thereby vector $\bold u$ is treated as constant
vector field.\par
     When substituting the above expressions into Maxwell equations
we consider the case of zero charges and currents: $\rho=0$, $\bold j=0$.
This yields the following equations:
$$
\align
&\divr\tilde{\bold H}=0,\\
\vspace{2ex}
&\divr\tilde{\bold E}=-\frac{1}{c}\bigl<\bold u,\,
 \rot\tilde{\bold H}\bigr>,\\
\vspace{2ex}
&\aligned
 \rot\tilde{\bold H}&=\frac{1}{c}\,\frac{\partial\tilde{\bold E}}
 {\partial\tilde t}-\frac{1}{c}\{\bold u,\,\tilde{\bold E}\}+\\
 &+\frac{1}{c^2}\,[\bold u,\,\{\bold u,\,\tilde{\bold H}\}]-
 \frac{1}{c^2}\,[\bold u,\,\partial\tilde{\bold H}/
 \partial\tilde t\,],
 \endaligned\\
\vspace{2ex}
&\rot\tilde{\bold E}=-\frac{1}{c}\,\frac{\partial\tilde{\bold H}}
 {\partial\tilde t}.
\endalign
$$
Only two of the above four equations coincide with original Maxwell
equations. Other two equations contain the entries of vector $\bold u$
that cannot be eliminated.\par
     This circumstance that we have found is very important. In the
end of \uppercase\expandafter{\romannumeral 19}-th century it made a
dilemma for physicists. The way how this dilemma was resolved had
determined in most further development of physics in
\uppercase\expandafter{\romannumeral 20}-th century. Indeed, one had
to make the following crucial choice:
\roster
\item to admit that Maxwell equations are not invariant with respect to
      Galileo transformations, hence they require the existence of some
      marked inertial coordinate system where they have standard form
      given in the very beginning of
      Chapter~\uppercase\expandafter{\romannumeral 2};
\item or to assume that formulas \thetag{1.1} are not correct, hence
      relativity principle claiming equivalence of all inertial
      coordinate systems is realized in some different way.
\endroster\par
      First Choice had lead to {\it ether theory}. According to this
theory, marked inertial coordinate system is bound to some hypothetical
matter, which was called {\it ether}. This matter has no mass, no
color, and no smell. It fills the whole space and does not reveal itself
otherwise, but as a carrier of electromagnetic interaction. Specified
properties of ether look quite unusual, this makes ether theory too
artificial (not natural). As a compromise this theory was admitted for
a while, but later was refuted by experiments of Michaelson and Morley,
who tried to measure the Earth velocity relative to ether (ether wind).
\par
     Second choice is more crucial. Indeed, refusing formulas \thetag{1.1},
we refuse classical mechanics of Newton in whole. Nevertheless the
development of science went through this second choice.
\head
\S\,2. Lorentz transformations.
\endhead
     Having refused formulas \thetag{1.1}, one should replace them by
something else. This was done by Lorentz. Following Lorentz, now we
replace Galileo transformations \thetag{1.1} by general linear
transformations relating $(\bold r,t)$ and $(\tilde\bold r,\tilde t)$:
$$
\xalignat 2
&c\,t=S^0_0\,c\,\tilde t+\sum^3_{k=1} S^0_k\,\tilde r^k,
&&r^i=S^i_0\,c\,\tilde t+\sum^3_{k=1} S^i_k\,\tilde r^k.
\hskip -2em
\tag2.1
\endxalignat
$$
In \thetag{2.1} we introduced $c$ as a factor for time variables
$t$ and $\tilde t$ in order to equalize measure units. Upon
introducing this factor all components of matrix $S$ appear to
be purely numeric quantities that do not require measure units
at all. It is convenient to denote $c\,t$ by $r^0$ and treat this
quantity as additional (fourth) component of radius-vector:
$$
r^0=ct.
\tag2.2
$$
Then two relationships \thetag{2.1} can be united into one
relationship:
$$
r^i=\sum^3_{k=0} S^i_k\,\tilde r^k.
\tag2.3
$$
In order to have invertible transformation \thetag{2.3} one should
assume that $\det S\neq 0$. Let $T=S^{-1}$. Then inverse transformation
for \thetag{2.3} is written as follows:
$$
\tilde r^i=\sum^3_{k=0} T^i_k\,r^k.
\tag2.4
$$
By their structure transformation \thetag{2.3} and \thetag{2.4}
coincide with transformations of coordinates of four-dimensional
vector under the change of base. Soon we shall see that such
interpretation appears to be very fruitful.\par
     Now the problem of deriving Lorentz transformations can be
formulated as problem of finding components of matrix $S$ in
\thetag{2.3}. The only condition we should satisfy thereby is
the invariance of Maxwell equations with respect to transformations
\thetag{2.3} upon completing them with transformations for
$\rho$, $\bold j$, $\bold E$ and $\bold H$.\par
     For the beginning let's consider the case with no currents and
charges, i\.\,e\. the case $\rho=0$, $\bold j=0$. Instead of Maxwell
equations let's study their differential consequences written in form
of the equations \thetag{4.6} in
Chapter~\uppercase\expandafter{\romannumeral 2}:
$$
\xalignat 2
&\square\bold E=0, &&\square\bold H=0.
\tag2.5
\endxalignat
$$
Invariance of \thetag{2.5} under the transformations \thetag{2.3} and
\thetag{2.4} is necessary (but possibly not sufficient) condition for
invariance of Maxwell equations from which the equations \thetag{2.5}
were derived.\par
     Further we need the following formula for d'Alambert operator used
in the above equations \thetag{2.5}:
$$
\square=\sum^3_{i=0}\sum^3_{j=0} g^{ij}\,\frac{\partial}{\partial r^i}
\frac{\partial}{\partial r^j}.
\tag2.6
$$
Here by $g^{ij}$ we denote components of matrix
$$
g^{ij}=g_{ij}=\left(\matrix
1 &  0 &  0 &  0\\
0 & -1 &  0 &  0\\
0 &  0 & -1 &  0\\
0 &  0 &  0 & -1\endmatrix\right).
\tag2.7
$$
It is easy to see that inverse matrix $g_{ij}$ for \thetag{2.7} has
the same components, i\.\,e\. $g_{ij}=g^{ij}$.\par
     From \thetag{2.3} and \thetag{2.4} we derive the following
transformation rules for first order differential operators:
$$
\xalignat 2
&\frac{\partial}{\partial r^i}=\sum^3_{k=0}
T^k_i\,\frac{\partial}{\partial\tilde r^k},
&&\frac{\partial}{\partial\tilde r^i}=\sum^3_{k=0}
S^k_i\,\frac{\partial}{\partial r^k}.
\hskip -2em
\tag2.8
\endxalignat
$$
Substituting \thetag{2.8} into formula \thetag{2.6}, we get
$$
\square=\sum^3_{p=0}\sum^3_{q=0}\tilde g^{pq}\,
\frac{\partial}{\partial\tilde r^p}
\frac{\partial}{\partial\tilde r^q},
$$
where matrices $g^{ij}$ and $\tilde g^{pq}$ are related by
formula
$$
\tilde g^{pq}=\sum^3_{i=0}\sum^3_{j=0} T^p_i\,T^q_j\,g^{ij}.
\pagebreak
\tag2.9
$$
In terms of inverse matrices $g_{pq}$ and $\tilde g_{pq}$ this
relationship \thetag{2.9} can be rewritten as follows:
$$
g_{ij}=\sum^3_{p=0}\sum^3_{q=0} T^p_i\,T^q_j\,\tilde g_{pq}.
\tag2.10
$$
\proclaim{\bf Theorem 2.1} For any choice of operator coefficients
$\alpha_1$, $\alpha_2$, $\beta_1$, and $\beta_2$ in formulas
\thetag{1.6} the invariance of the form of equations \thetag{2.5}
under the transformations \thetag{2.3} and \thetag{2.4} is equivalent
to proportionality of matrices $g$ and $\tilde g$, i\.\,e\.
$$
\tilde g^{ij}=\lambda\, g^{ij}.
\tag2.11
$$
\endproclaim
     Numeric factor $\lambda$ in formula \thetag{2.11} is usually
chosen to be equal to unity: $\lambda=1$. In this case from
\thetag{2.10} and \thetag{2.11} we derive
$$
g_{ij}=\sum^3_{p=0}\sum^3_{q=0} T^p_i\,T^q_j\,g_{pq}.
\tag2.12
$$
In matrix form this relationship \thetag{2.12} looks like
$$
T^t\,g\,T=g.
\tag2.13
$$
Here $g$ is a matrix of the form \thetag{2.7}, while by $T^t$
in \thetag{2.13} we denote transposed matrix $T$.
\definition{\bf Definition 2.1} Matrix $T$ satisfying the relationship
\thetag{2.13} is called {\it Lorentzian matrix}.
\enddefinition
    It is easy to check up that the set of Lorentzian matrices form a
group. This group is usually denoted by $\MatGrO(1,3)$. It is called
{\it matrix Lorentz group}.\par
    From the relationship \thetag{2.13} for Lorentzian matrix we derive
the equality $(\det T)^2=1$. Hence $\det T=\pm 1$. Lorentzian matrices
with unit determinant form the group $\MatGrSO(1,3)$, it is called
{\it special matrix Lorentz group}.\par
     If $i=j=0$, from \thetag{2.12} we obtain the following formula
relating components of Lorentzian matrix $T$:
$$
(T^0_0)^2-(T^1_0)^2-(T^2_0)^2-(T^3_0)^2=1.
\tag2.14
$$
Inequality $|T^0_0|\geqslant 1$ is immediate consequence of the
relationship \thetag{2.14}. Hence $T^0_0\geqslant 1$ or $T^0_0
\leqslant -1$. Lorentzian matrix with $T^0_0\geqslant 1$ is called
{\it orthochronous}. The set of orthochronous Lorentzian matrices
form {\it orthochronous matrix Lorentz group} $\MatGrO^+(1,3)$.
Intersection $\MatGrSO^+(1,3)=\MatGrSO(1,3)\cap\MatGrO^+(1,3)$
is called {\it special orthochronous matrix Lorentz group}.\par
\proclaim{\bf Exercise 2.1} Prove theorem~2.1 under the assumption
that transformation \thetag{1.6} given by operator coefficients
$\alpha_1$, $\alpha_2$, $\beta_1$, and $\beta_2$ is invertible.
\endproclaim
\head
\S\,3. Minkowsky space.
\endhead
     In previous section we have found that each Lorentzian matrix
from group $\MatGrO(1,3)$ determines some transformation \thetag{2.1}
preserving the form the equations \thetag{2.5}. In deriving this fact
we introduced notations \thetag{2.2} and united space and time into
one four-dimensional ``space-time''. Let's denote it by $M$.
Four-dimensional space $M$ is basic object in special theory of
relativity. Its points are called {\it events}. The space of events
is equipped with quadratic form $g$ with signature $(1,3)$. This
quadratic form is called {\it Minkowsky metric}. Thereby inertial
coordinate systems are interpreted as Cartesian coordinates for which
Minkowsky metric has canonical form \thetag{2.7}.
\definition{\bf Equivalence principle} All physical laws in any two
inertial coordinate systems are written in the same form.
\enddefinition
    Let's choose some inertial coordinate system. This choice
determines separation of event space $M$ into geometric space $V$
(space of points) and time axis $T$:
$$
M=T\oplus V.
\tag3.1
$$
Matrix of Minkowsky metric in chosen coordinate system has canonic
form \thetag{2.7}. Therefore geometric space $V$ is orthogonal to
time axis $T$ with respect to Minkowsky metric $g$. Restriction of
this metric to $V$ is negative quadratic form. Changing its sign,
we get positive quadratic form. This is standard Euclidean scalar
product in $V$.\par
     Now let's consider another inertial coordinate system. Like
\thetag{3.1}, it determines second expansion of $M$ into space and
time:
$$
M=\tilde T\oplus \tilde V.
\tag3.2
$$
In general time axes $T$ and $\tilde T$ in expansions \thetag{3.1}
and \thetag{3.2} do not coincide. Indeed, bases of these two coordinate
systems are related to each other by formula
$$
\tilde\bold e_i=\sum^3_{j=0}S^j_i\,\bold e_j,
\tag3.3
$$
where $S$ is Lorentzian matrix from \thetag{2.3}. For base vector
$\tilde\bold e_0$ directed along time axis $\tilde T$ from \thetag{3.3}
we derive
$$
\tilde\bold e_0=S^0_0\,\bold e_0+S^1_0\,\bold e_1+
S^2_0\,\bold e_2+S^3_0\,\bold e_3.
\tag3.4
$$
In general components $S^1_0$, $S^2_0$, and $S^3_0$ in Lorentz matrix
$S$ are nonzero. Therefore vectors $\tilde\bold e_0$ and $\bold e_0$
are non-collinear. Hence $T\neq\tilde T$.\par
    Non-coincidence of time axes $T\neq\tilde T$ for two inertial
coordinate systems leads to non-coincidence of geometric spaces:
$V\neq \tilde V$. This fact lead to quite radical conclusion when
we interpret it physically: observers in two such inertial systems
observe {\it two different three-dimensional geometric spaces} and
have {\it two different time ticks}. However, in our everyday life
this difference is very small and never reveals.\par
     Let's calculate how big is the difference in the rate of time
ticks for two inertial coordinate systems. From \thetag{2.4} we get
$$
\tilde t=T^0_0\,t+\sum^3_{k=1} \frac{T^0_k}{c}\,r^k.
\tag3.5
$$
Let $t\to+\infty$. If Lorentzian matrix $T$ is orthochronous,
then $T^0_0>0$ and $\tilde t\to+\infty$. If matrix $T$ is not
orthochronous, then $t\to+\infty$ we get $\tilde t\to-\infty$.
Transformations \thetag{2.4} with non-orthochronous matrices
$T$ invert the direction of time exchanging the future and
the past. It would be very intriguing to have such a feature
in theory. However, presently in constructing theory of
relativity one uses more realistic approach. So we shall assume
that two physically real inertial coordinate systems can be related
only by orthochronous Lorentz matrices from $O^+(1,3)$.\par
     Restriction of the set of admissible Lorentz matrices from
$O(1,3)$ to $O^+(1,3)$ is due to the presence of additional structure
in the space of events. It is called {\it polarization}. Let's choose
some physical inertial coordinate system. Minkowsky metric in such
system is given by matrix of canonical form \thetag{2.7}. Let's
calculate scalar square of four-dimensional vector $\bold x$ in
Minkowsky metric:
$$
g(\bold x,\,\bold x)=(x^0)^2-(x^1)^2-(x^2)^2-(x^3)^2.
\tag3.6
$$
By value of their scalar square $g(\bold x,\,\bold x)$
\pagebreak in Minkowsky metric $g$ vectors of Minkowsky
space $M$ are subdivided into three parts:
\roster
\item {\it tome-like vectors}, for which $g(\bold x,
      \,\bold x)$ is positive;
\item {\it light vectors}, for which $g(\bold x,\,\bold x)=0$;
\item {\it space-like vectors}, for which $g(\bold x,\,\bold x)$
      is negative.
\endroster
Coordinates of light vectors satisfy the following equation:
$$
(x^0)^2-(x^1)^2-(x^2)^2-(x^3)^2=0.
\tag3.7
$$
It is easy to see that \thetag{3.7} is the equation of cone in
four-dimensional space (see classification of quadrics in \cite{4}).
This cone \thetag{3.7} is called {\it light cone}.\par
\parshape 14 4.6cm 5.5cm 4.6cm 5.5cm 4.6cm 5.5cm 4.6cm 5.5cm
4.6cm 5.5cm 4.6cm 5.5cm 4.6cm 5.5cm 4.6cm 5.5cm
4.6cm 5.5cm 4.6cm 5.5cm 4.6cm 5.5cm 4.6cm 5.5cm
4.6cm 5.5cm
0cm 10.1cm
     Time-like vectors fill interior of light cone, while space-like
vectors fill outer space outside this cone.
\vadjust{\vtop to 0pt{\hbox to 0pt{\kern -30pt
\includegraphics{fig7.eps}\hss}
\vskip 80pt\hbox{\kern 40pt {\it Fig\.~3.1 }\hss}
\vskip -45pt\hbox{\kern 53pt {\it past}\hss}
\vskip -90pt\hbox{\kern 53pt{\it future}\hss}
\vskip -80pt\vss}}Interior of light cone is a union of two parts:
time-like vectors with $x^0>0$ are directed to the future, others
with $x_0<0$ are directed to the past. Vector directed to the
future can be continuously transformed to any other vector directed
to the future. However, it cannot be continuously transformed to
a vector directed to the past without making it space-like vector
or zero vector at least once during transformation. This means that
the set of time-like vectors is disjoint union of two connected
components. 
\definition{\bf Definition 3.1} Geometric structure in Minkowsky
space $M$ that marks one of two connected components in the set
of time-like vectors is called {\it polarization}. It is used to
say that marked component {\it points to the future}.
\enddefinition
     Let $\bold e_0$, $\bold e_1$, $\bold e_2$, $\bold e_3$ be
{\it orthonormal base} in Minkowsky metric \footnote"*"{\
\ i\.\,e\. base for which Minkowsky metric has the form
\thetag{2.7}.}. In the space $M$ with polarization one can consider
only those such bases for which unit vector of time axis $\bold e_0$
is directed to the future. Then transition from one of such bases
to another would be given by orthochronous matrix from group
$\MatGrO^+(1,3)$.\par
\definition{\bf Definition 3.2} Four-dimensional affine space $M$
equipped with metric $g$ of signature $(1,3)$ and equipped with
orientation\footnote"**"{\ remember that orientation is geometric
structure distinguishing left and right bases (see~\cite{4}).}
and polarization is called {\it Minkowsky space}.
\enddefinition
    According to special theory of relativity Minkowsky space, which
is  equipped with orientation and polarization, is proper mathematical
model for the space of real physical events. Now we can give strict
mathematical definition of inertial coordinate system.
\definition{\bf Definition 3.3} Orthonormal right inertial coordinate
system is orthonormal right coordinate system in Minkowsky space
with time base vector directed to the future.
\enddefinition
     It is easy to verify that any two inertial coordinate systems
as defined above are related to each other by Lorentz transformation
with matrix $S$ from orthochronous Lorentz group $SO^+(1,3)$. Let's
choose one of such coordinate systems and consider related expansion
\thetag{3.1}. It is clear that $\bold e_0\in T$, while linear span of
spatial vectors $\bold e_1$,~$\bold e_2$,~$\bold e_3$ defines subspace
$V$. Taking orthonormal base $\bold e_1$,~$\bold e_2$,~$\bold e_3$
for the standard of right bases in $V$, we equip this three-dimensional
space with orientation. This is concordance with the fact that geometric
space that we observe in our everyday life possesses orientation
distinguishing left and right.
\proclaim{\bf Exercise 3.1} By analogy with definition~3.3 formulate the
definition of skew-angular inertial coordinate system.
\endproclaim
\head
\S\,4. Kinematics of relative motion.
\endhead
     Galileo transformations are used in mechanics for describing
physical processes as they are seen by two observers representing
two inertial coordinate systems. Lorentz transformations, which we
have derived from the condition of invariance of electrodynamical
equations \thetag{2.5}, are designed for the same purpose. However,
this is not immediately clear when looking at formulas \thetag{2.3}
and \thetag{2.4}. Therefore we shall bring these formulas to the
form more convenient for studying their physical nature.\par 
     Let's fix two inertial coordinate systems related by Lorentz
transformation \thetag{2.1}. First one is related with orthonormal
base $\bold e_0$,~$\bold e_1$,~$\bold e_2$,~$\bold e_3$ in Minkowsky
space and with the expansion \thetag{3.1}. Second is related with
the base $\tilde\bold e_0$,~$\tilde\bold e_1$,~$\tilde
\bold e_2$,~$\tilde\bold e_3$ and with the expansion \thetag{3.2}.
If time axes are parallel $\bold e_0=\tilde\bold e_0$, then
Lorentz matrix $S$ in \thetag{2.3} is reduced to orthogonal matrix
$O\in SO(3)$ relating two right orthonormal bases
$\bold e_1$,~$\bold e_2$,~$\bold e_3$ and
$\tilde\bold e_1$,~$\tilde\bold e_2$,~$\tilde\bold e_3$.
It has the following blockwise-diagonal shape:
$$
S=\left(\matrix
1 &  0     &  0     &  0\\
\vspace{0.5ex}
0 &  O^1_1 &  O^1_2 &  O^1_3\\
\vspace{0.5ex}
0 &  O^2_1 &  O^2_2 &  O^2_3\\
\vspace{0.5ex}
0 &  O^3_1 &  O^3_2 &  O^3_3\endmatrix\right).
\tag4.1
$$
Thus, in case if $T\parallel\tilde T$ two inertial coordinate
systems differ only in directions of spatial axes. They do not
move with respect to each other.\par
     Now let's consider the case $T\nparallel\tilde T$. Hence
$\bold e_0\neq\tilde\bold e_0$. Let $H$ be linear span of vectors
$\bold e_0$ and $\tilde\bold e_0$. Denote by $W$ intersection of
subspaces $V$ and $\tilde V$ from \thetag{3.1} and \thetag{3.2}:
$$
\xalignat 2
&H=\Span(\bold e_0,\tilde\bold e_0),
&W=V\cap\tilde V.
\tag4.2
\endxalignat
$$
\proclaim{\bf Lemma 4.1} Two-dimensional subspaces $H$ and $W$ in
\thetag{4.2} are perpendicular to each other in Minkowsky metric
$g$. Their intersection is zero: $H\cap W=\{0\}$, while direct
sum of these subspaces coincides with the whole Minkowsky space:
$H\oplus W=M$.
\endproclaim
\demo{Proof} Subspace $H$ is two-dimensional since it is linear
span of two non-collinear vectors. Subspaces $V$ and $\tilde V$
are three-dimensional and $V\neq\tilde V$. Hence their sum
$V+\tilde V$ coincides with $M$, i\.\,e\. $\dim(V+\tilde V)=4$.
Applying theorem on the dimension of sum and intersection of two
subspaces (see \cite{4}), we get
$$
\dim(W)=\dim V+\dim\tilde V-\dim(V+\tilde V)=3+3-4=2.
$$\par
     In order to prove orthogonality of $H$ and $W$ we use
orthogonality of $T$ and $V$ in the expansion \thetag{3.1} and
orthogonality of $\tilde T$ and $\tilde V$ in \thetag{3.2}.
Let $\bold y$ be an arbitrary vector in subspace $W$. Then
$\bold y\in V$. From $V\perp T$ we get $\bold y\perp\bold e_0$.
Analogously from $\bold y\in\tilde V$ we get $\bold y\perp\tilde e_0$.
Now from orthogonality of $\bold y$ to both vectors $\bold e_0$ and
$\tilde\bold e_0$ we derive orthogonality of $\bold y$ to their
linear span: $\bold y\perp H$. Since $\bold y$ is arbitrary vector
in $W$, we have $W\perp H$.\par
     Now let's prove that $H\cap W=\{0\}$. Let's consider an arbitrary
vector $\bold x\in H\cap W$. From $\bold x\in H$ and $\bold x\in W$
due to orthogonality of $H$ and $W$, which is already proved, we
get $g(\bold x,\bold x)=0$. But $\bold x\in W\subset V$, while
restriction of Minkowsky metric to subspace $V$ is negative quadratic
form of signature $(0,3)$. Therefore from $g(\bold x,\bold x)=0$ we
derive $\bold x=0$. Proposition $H\cap W=\{0\}$ is proved.\par
     From $H\cap W=\{0\}$ we conclude that sum of subspaces $H$ and
$W$ is direct sum and $\dim(H+W)=2+2=4$. Hence $H\oplus W=M$.
Lemma is proved.
\qed\enddemo
     Now let's return back to considering pair of inertial coordinate
systems with bases $\bold e_0$,~$\bold e_1$,~$\bold e_2$,~$\bold e_3$
and $\tilde\bold e_0$,~$\tilde\bold e_1$,~$\tilde\bold e_2$,~$\tilde
\bold e_3$. There is the expansion \thetag{3.4} for vector $\tilde
\bold e_0$. Let's write it as follows:
$$
\tilde\bold e_0=S^0_0\,\bold e_0+\bold v.
\tag4.3
$$
Here $\bold v=S^1_0\,\bold e_1+S^2_0\,\bold e_2+S^3_0\,\bold e_3
\in V$. Since matrix $S$ is orthochronous and since $\tilde\bold
e_0\neq\bold e_0$, we have
$$
\xalignat 2
&S^0_0>1,
&&\bold v\neq 0.
\tag4.4
\endxalignat
$$
For any real number $a>1$ there exists a number $\alpha>0$ such
that $a=\ch(\alpha)$. Let's apply this observation to $S^0_0$
in \thetag{4.3}:
$$
S^0_0=\ch(\alpha).
\tag4.5
$$
From \thetag{4.3}, from \thetag{4.5}, and from orthogonality of
vectors $\bold e_0$ and $\bold v$ in Minkowsky metric we obtain
$$
1=g(\tilde\bold e_0,\tilde\bold e_0)=(S^0_0)^2\,
g(\bold e_0,\bold e_0)+g(\bold v,\bold v)=\ch^2(\alpha)-|\bold v|^2.
$$
Using this equality we can find Euclidean length of vector $\bold v$
in three-dimensional subspace $V$:
$$
|\bold v|=\sh(\alpha),\text{ \ where \ }\alpha>0.
\tag4.6
$$
Let's replace vector $\bold v$ by vector of unit length $\bold h_1=
\bold v/|\bold v|$ and rewrite the relationship \thetag{4.3} as
follows:
$$
\tilde\bold e_0=\ch(\alpha)\,\bold e_0+
\sh(\alpha)\,\bold h_1.
\tag4.7
$$
Due to \thetag{4.7} vector $\bold h_1$ is linear combination of
vectors $\bold e_0$ and $\tilde\bold e_0$, hence $h_1\in H$.
But $h_1\in V$ as well. Therefore $h_1\in V\cap H$. Vectors
$\bold e_0$ and $\bold h_1$ are perpendicular to each other,
they form orthonormal base in two-dimensional subspace $H$:
$$
\xalignat 2
&g(\bold e_0,\bold e_0)=1,
&&g(\bold h_1,\bold h_1)=-1.
\tag4.8
\endxalignat
$$
From \thetag{4.8} we conclude that restriction of Minkowsky metric
to subspace $H$ is metric with signature $(1,1)$.\par
     Now we need another vector from subspace $H$. Let's determine
it by the following relationship:
$$
\tilde\bold h_1=\sh(\alpha)\,\bold e_0+
\ch(\alpha)\,\bold h_1.
\tag4.9
$$
It is easy to check that vectors $\tilde\bold e_0$ and $\tilde\bold h_1$
form another orthonormal base in subspace $H$. Transition matrix relating
these two bases has the following form:
$$
S_{\text{L}}=
\left(
\matrix
\ch(\alpha) &\sh(\alpha)\\
\vspace{2ex}
\sh(\alpha) &\ch(\alpha)
\endmatrix
\right).
\tag4.10
$$
Matrix \thetag{4.10} is called the matrix of {\it Lorentzian rotation}.
\par
     There is four-dimensional version of matrix \thetag{4.10}. Indeed,
vector $\bold h_1\in V$ is perpendicular to subspace $W\subset V$.
Therefore we have the expansion of subspace $V$ as a direct sum:
$$
V=\Span(\bold h_1)\oplus W.
$$
Let's choose two vectors $\bold h_2$ and $\bold h_3$ forming orthonormal
base in subspace $W$ and complementing $\bold h_1$ up to an orthonormal
right base in $V$. Then four vectors $\bold e_0$,~$\bold h_1$,~$\bold
h_2$~$\bold h_3$ constitute orthonormal right base in $M$ with time
vector $\bold e_0$ directed to the future. Transition matrix relating
this base with the base $\tilde\bold e_0$,~$\tilde \bold h_1$,~$\bold
h_2$~$\bold h_3$ has the following blockwise-diagonal form:
$$
S_{\text{L}}=
\left(
\matrix
\ch(\alpha) &\sh(\alpha) &0 &0\\
\vspace{2ex}
\sh(\alpha) &\ch(\alpha) &0 &0\\
\vspace{2ex}
0 & 0 &\ 1\  &0\\
\vspace{2ex}
0 & 0 &0 &\ 1\
\endmatrix
\right).
\tag4.11
$$
Transition from base $\bold e_0$,~$\bold e_1$,~$\bold e_2$~$\bold e_3$
to base $\bold e_0$,~$\bold h_1$,~$\bold h_2$~$\bold h_3$ is given by
a matrix of the form \thetag{4.1}. This is because their time vectors
do coincide. In a similar way transition from base $\tilde\bold
e_0$,~$\tilde\bold h_1$,~$\bold h_2$~$\bold h_3$ to base $\tilde\bold
e_0$,~$\tilde\bold e_1$,~$\tilde\bold e_2$~$\tilde\bold e_3$ is given
by a matrix of the same form \thetag{4.1}. Ultimate change of base
$\bold e_0$,~$\bold e_1$,~$\bold e_2$~$\bold e_3$ for another base
$\tilde\bold e_0$,~$\tilde\bold e_1$,~$\tilde\bold e_2$~$\tilde\bold e_3$
then can be done in three steps.
\proclaim{\bf Theorem 4.1} Each Lorentzian matrix $S\in SO^+(1,3)$
is a product of three matrices $S=S_1\,S_{\text{L}}\,S_2$, one of
which $S_{\text{L}}$ is a matrix of Lorentzian rotation \thetag{4.11},
while two others $S_1$ and $S_2$ are matrices of the form
\thetag{4.1}.
\endproclaim
     In order to clarify physical meaning of Lorentz transformations
let's first consider transformations with matrix $S$ of the form
\thetag{4.11}. Let
$ct=r^0$,~$r^1$,~$r^2$,~$r^3$ be coordinates of some vector $\bold r
\in M$ in the base $\bold e_0$,~$\bold h_1$,~$\bold h_2$,~$\bold h_3$.
By $c\tilde t=\tilde r^0$,~$\tilde r^1$,~$\tilde r^2$,~$\tilde r^3$
we denote coordinates of the same vector in the base $\tilde\bold
e_0$,~$\tilde\bold h_1$,~$\bold h_2$~$\bold h_3$. For matrix $S$ of
the form \thetag{4.11} formula \thetag{2.3} leads to relationships
$$
\aligned
&t=\ch(\alpha)\,\tilde t+\frac{\sh(\alpha)}{c}\,\tilde r^1,\\
\vspace{1ex}
&r^1=\sh(\alpha)\,c\,\tilde t+\ch(\alpha)\,\tilde r^1,\\
&r^2=\tilde r^2,\\
&r^3=\tilde r^3.
\endaligned
\tag4.12
$$
Let $\tilde r^1$,~$\tilde r^2$,~$\tilde r^3$ be coordinates of
radius-vector of some point $A$ which is at rest in inertial
coordinate system with base $\tilde\bold e_1$,~$\tilde\bold
e_2$,~$\tilde\bold e_3$. Then $\tilde r^1$,~$\tilde r^2$,~$\tilde
r^3$ are constants, they do not depend on time $\tilde t$ in this
coordinate system. Upon calculating coordinates of this point $A$
in other inertial coordinate system by means of formulas
\thetag{4.12} its first coordinate $r^1$ appears to be a function
of parameter $\tilde t$. We use first relationship \thetag{4.12}
in order to express parameter $\tilde t$ through time variable $t$
in second coordinate system:
$$
\tilde t=\frac{t}{\ch(\alpha)}-\frac{\th(\alpha)}{c}\,\tilde r^1.
\tag4.13
$$
Substituting \thetag{4.13} into other three formulas \thetag{4.12},
we get
$$
\aligned
&r^1=r^1(t)=c\,\th(\alpha)\,t+\const,\\
&r^2=r^2(t)=\const,\\
&r^3=r^3(t)=\const.
\endaligned
\tag4.14
$$
From \thetag{4.14} we see that in second coordinate system our point
$A$ is moving with constant velocity $u=c\,\th(\alpha)$ in the direction
of first coordinate axis.
\par
     In contrast to parameter $\alpha$ in matrix \thetag{4.11}, parameter
$u$ has transparent physical interpretation as magnitude of relative
velocity of one coordinate system with respect to another. Let's express
components of matrix \thetag{4.11} through $u$:
$$
\xalignat 2
&\ch(\alpha)=\frac{1}{\sqrt{1-\dsize\frac{u^2}{c^2}}},
&&\sh(\alpha)=\frac{u}{c}\,\frac{1}{\sqrt{1-\dsize\frac{u^2}{c^2}}}.
\endxalignat
$$
Let's substitute these formulas into \thetag{4.12}. As a result we get
$$
\xalignat 2
&t=\frac{\tilde t+\dsize\frac{u}{\vphantom{y}c^2}\,\tilde r^1}
{\sqrt{1-\dsize\frac{u^2}{c^2}}},
&&r^1=\frac{u\tilde t+\tilde r^1}
{\sqrt{1-\dsize\frac{u^2}{c^2}}},\\
&&&\tag4.15\\
&r^2=\tilde r^2,
&&r^3=\tilde r^3.
\endxalignat
$$\par
     Denote for a while by $\bold r$ and $\tilde\bold r$ the following
two three-dimensional vectors in subspaces $V$ and $\tilde V$:
$$
\aligned
&\bold r=r^1\,\bold h_1+r^2\,\bold h_2+r^3\,\bold h_3,\\
&\tilde\bold r=\tilde r^1\,\tilde\bold h_1+\tilde r^2\,
\bold h_2+\tilde r^3\,\bold h_3.
\endaligned
\tag4.16
$$
Then we define linear map $\theta: V\to\tilde V$ determined by its
action upon base vectors $\bold h_1$, $\bold h_2$, and $\bold h_3$:
$$
\xalignat 3
&\theta(\bold h_1)=\tilde\bold h_1,
&&\theta(\bold h_2)=\bold h_2,
&&\theta(\bold h_3)=\bold h_3.
\endxalignat
$$
This map $\theta$ is orientation preserving isometry, since it maps
right orthonormal base of subspace $V$ to right orthonormal base in
subspace $\tilde V$. Using above notations \thetag{4.16} and the map
$\theta$, we can write formulas \thetag{4.15} in vectorial form:
$$
\aligned
&t=\frac{\tilde t+\dsize\frac{\bigl<\theta\bold u,\,\tilde\bold r\bigr>}
{\vphantom{y}c^2}}
{\sqrt{\dsize{1-\frac{|\bold u|^2}{c^2}}}},\\
\vspace{2ex}
&\theta\bold r=
\frac{\theta\bold u\,\tilde t+
\dsize\frac{\bigl<\theta\bold u,\,\tilde\bold r\bigr>}
{|\bold u|^2}\,\theta\bold u}{\sqrt{\dsize{1-\frac{|\bold u|^2}{c^2}}}}
+\tilde\bold r-\frac{\bigl<\theta\bold u,\,
\tilde\bold r\bigr>}{|\bold u|^2}\,\theta\bold u.
\endaligned
\tag4.17
$$
Here $\bold u=u\,\bold h_1$ is vector of relative velocity of
second coordinate system with respect to first one. Formulas
\thetag{4.17} are irrespective to the choice of bases in
subspaces $V$ and $\tilde V$. Therefore they are applicable to
Lorentz transformations with special matrix of the form \thetag{4.11}
and to arbitrary Lorentz transformations with matrix $S=S_1\,S_{\text{L}}
\,S_2$ (see theorem~4.1).\par
      Very often the sign of map $\theta$ realizing isomorphism of
subspaces $V$ and $\tilde V$ in formulas \thetag{4.17} is omitted:
$$
\aligned
&t=\frac{\tilde t+\dsize\frac{\bigl<\bold u,\,\tilde\bold r\bigr>}
{\vphantom{y}c^2}}
{\sqrt{\dsize{1-\frac{|\bold u|^2}{c^2}}}},\\
\vspace{2ex}
&\bold r=
\frac{\bold u\,\tilde t+
\dsize\frac{\bigl<\bold u,\,\tilde\bold r\bigr>}
{|\bold u|^2}\,\bold u}{\sqrt{\dsize{1-\frac{|\bold u|^2}{c^2}}}}
+\tilde\bold r-\frac{\bigl<\bold u,\,
\tilde\bold r\bigr>}{|\bold u|^2}\,\bold u.
\endaligned
\tag4.18
$$
Formulas \thetag{4.18} represent ``conditionally three-dimensional''
understanding of Lorentz transformations when vectors $\bold r$ and
$\tilde\bold r$ treated as vectors of the same three-dimensional
Euclidean space, while $t$ and $\tilde t$ are treated as scalar
parameters. However, according to modern paradigm four-dimensional
Minkowsky space is real physical entity, not purely mathematical
abstraction convenient for shortening formulas (compare \thetag{2.3}
and \thetag{4.17}).\par
     When writing formulas \thetag{4.17} and \thetag{4.18} in
components we should expand vectors $\bold r$ and $\bold u$ in
the base of one coordinate system, while vector $\tilde\bold r$
is expanded in the base of another coordinate system. Thereby the
difference in the shape of these two formulas completely disappears.
\proclaim{\bf Exercise 4.1} Using expansions \thetag{4.16} for
vectors $\bold r$ and $\tilde\bold r$, derive the following
formulas:
$$
\xalignat 2
&\tilde r^1=\frac{\bigl<\theta\bold u,\,\tilde\bold r\bigr>}
{|\bold u|},
&&\tilde r^2\,\bold h_2+\tilde r^3\,\bold h_3=
\tilde\bold r-\frac{\bigl<\theta\bold u,\,
\tilde\bold r\bigr>}{|\bold u|^2}\,\theta\bold u.
\endxalignat
$$
Combining these formulas with \thetag{4.15}, derive formulas
\thetag{4.17}.
\endproclaim
\head
\S\,5. Relativistic law of velocity addition.
\endhead
     Classical law of velocity addition was first consequence that we
obtained from Galileo transformations:
$$
\bold v=\tilde\bold v+\bold u,
\tag5.1
$$
see formulas \thetag{1.2}. Replacing Galileo transformations by
Lorentz transformations, now we should derive new {\it relativistic}
law of velocity addition.\par
     Suppose that vector-function $\tilde\bold r(\tilde t)$ describes
the motion of a point $A$ in inertial coordinate system 
$(\tilde\bold r,\tilde t)$ and suppose that this coordinate system
moves with velocity $\bold u$ with respect to other inertial coordinate
system $(\bold r,t)$. For passing to coordinate system $(\bold r,t)$
we use Lorentz transformation given by formulas \thetag{4.18}.
As a result we get two functions
$$
\aligned
&t(\tilde t)=\frac{\tilde t+\dsize\frac{\bigl<\bold u,
\,\tilde\bold r(\tilde t)\bigr>}
{\vphantom{y}c^2}}
{\sqrt{\dsize{1-\frac{|\bold u|^2}{c^2}}}},\\
\vspace{2ex}
&\bold r(\tilde t)=
\frac{\bold u\,\tilde t+
\dsize\frac{\bigl<\bold u,\,\tilde\bold r(\tilde t)\bigr>}
{|\bold u|^2}\,\bold u}{\sqrt{\dsize{1-\frac{|\bold u|^2}{c^2}}}}
+\tilde\bold r(\tilde t)-\frac{\bigl<\bold u,\,
\tilde\bold r(\tilde t)\bigr>}{|\bold u|^2}\,\bold u.
\endaligned
\tag5.2
$$
Let's calculate first derivatives of functions \thetag{5.2}:
$$
\frac{dt}{d\tilde t}=\frac{1+\dsize\frac{\bigl<\bold u,
\,\tilde\bold v\bigr>}
{\vphantom{y}c^2}}
{\sqrt{\dsize{1-\frac{|\bold u|^2}{c^2}}}},
\tag5.3a
$$
$$
\frac{d\bold r}{d\tilde t}=
\frac{\bold u+
\dsize\frac{\bigl<\bold u,\,\tilde\bold v\bigr>}
{|\bold u|^2}\,\bold u}{\sqrt{\dsize{1-\frac{|\bold u|^2}{c^2}}}}
+\tilde\bold v-\frac{\bigl<\bold u,\,
\tilde\bold v\bigr>}{|\bold u|^2}\,\bold u.
\tag5.3b
$$
By $\tilde\bold v$ we denote the velocity of
the point $A$ in coordinates $(\tilde\bold r,\tilde t)$:
$$
\tilde\bold v=\dot{\tilde\bold r}(\tilde t)=\frac{d\tilde\bold r}
{d\tilde t}.
$$
In a similar way by $\bold v$ we denote the velocity of this
point in other coordinates $(\bold r,t)$. To calculate $\bold v$
we divide derivatives:
$$
\bold v=\frac{d\bold r}{dt}=\dot\bold r(t)=
\left(\dfrac{d\bold r}{d\tilde t}\right)\bigg/
\left(\dfrac{dt}{d\tilde t}\right).
\tag5.4
$$
Substituting \thetag{5.3a} and \thetag{5.3b} into \thetag{5.4},
we get formula
$$
\bold v=\frac{\bold u+\dsize\frac{\bigl<\bold u,\,
\tilde\bold v\bigr>}{|\bold u|^2}\,\bold u}
{1+\dsize\frac{\bigl<\bold u,\,\tilde\bold v\bigr>}{c^2}}+
\frac{\tilde\bold v-\dsize\frac{\bigl<\bold u,\,
\tilde\bold v\bigr>}{|\bold u|^2}\,\bold u}
{1+\dsize\frac{\bigl<\bold u,\,\tilde\bold v\bigr>}{c^2}}
\sqrt{\dsize{1-\frac{|\bold u|^2}{c^2}}}.
\hskip-2em
\tag5.5
$$
Formula \thetag{5.5} is relativistic law of velocity addition.
It is mach more complicated than classical law given by formula
\thetag{5.1}. However, in the limit of small velocities
$|\bold u|\ll c$ formula \thetag{5.5} reduces to formula
\thetag{5.1}.
\proclaim{\bf Exercise 5.1} Derive relativistic law of velocity addition
from formula \thetag{4.17}. Explain why resulting formula differs
from \thetag{5.5}.
\endproclaim
\head
\S\,6. World lines and private time.
\endhead
     Motion of point-size material object in arbitrary inertial
coordinate system $(\bold r,t)$ is described by vector-function
$\bold r(t)$, where $t$ is time variable and $\bold r$ is
radius-vector of point. Four-dimensional radius-vector of this
material point has the following components:
$$
r^0(t)=ct,\ \ r^1(t),\ \ r^2(t),\ \ r^3(t).
\tag6.1
$$
Vector-function with components \thetag{6.1} determines parametric
line in Minkowsky space $M$, this line is called {\it world line}
of material point. Once world line is given, motion of material point
is described completely. Let's differentiate four-dimensional
radius-vector \thetag{6.1} with respect to parameter $t$. As a result
we get four-dimensional vector tangent to world line:
$$
\bold K=(c,\dot r^1,\dot r^2,\dot r^3)=
(c,v^1,v^2,v^3).
\tag6.2
$$
Last three
components of this vector form velocity vector of material point.
Velocity of most material objects is not greater than light
velocity: $|\bold v|<c$. When applied to vector $\bold K$ in
\thetag{6.2} this means that tangent-vector of world line is
time-like vector:
$$
g(\bold K,\bold K)=c^2-|\bold v|^2>0.
\tag6.3
$$
\definition{\bf Definition 6.1} Smooth curve in Minkowsky space
is called time-like curve if tangent-vector of this curve is
time-like vector at each its point.
\enddefinition
World lines for most material objects are time-like curves.
Exception are world lines of photons (light particles) and
world lines of other elementary particles with zero mass.
For them $|\bold v|=c$, hence we get $g(\bold K,\bold K)=0$.
\par
     World line have no singular points. Indeed, even if
$g(\bold K,\bold K)=0$, tangent vector $\bold K$ in \thetag{6.2}
is nonzero since $K^0=c\neq 0$.\par
     Let's consider world \pagebreak line of material point of nonzero mass.
For this line we have the condition \thetag{6.3} fulfilled, hence
we can introduce natural parameter on this line:
$$
s(t)=\int\limits^t_{t_0}\sqrt{g(\bold K,\bold K)\,}\,dt.
\tag6.4
$$
Integral \thetag{6.4} yields invariant parameter for world lines.
For any two points $A$ and $B$ on a given world line the quantity
$s(B)-s(A)$ does not depend on inertial coordinate system used
for calculating integral \thetag{6.4}. This quantity is called
interval length of the arc $AB$ on world line.
\proclaim{\bf Theorem 6.1} Straight line segment connecting end
points of an arc on smooth time-like curve is a segment of
time-like straight line. Its interval length is greater than interval
length of corresponding arc.
\endproclaim
     Let $A$ and $B$ be two successive events in the ``life'' of
material point of nonzero mass. The answer to the question what
time interval separates these two events depend on the choice of
inertial coordinate system from which we observe the ``life'' of
this material point. So this answer is relative (not invariant).
However, there is invariant quantity characterizing time distance
between two events on world line:
$$
\tau=\frac{s(B)-s(A)}{c}.
\tag6.5
$$
This quantity $\tau$ in formula \thetag{6.5} is called interval of
{\it private time} on world line.\par
     Concept of private time determine {\it microlocal} concept
of time in theory of relativity. According to this concept each
material point lives according to its own watch, and watches
of different material points are synchronized only in very rough
way: they count time from the past to the future. This rough
synchronization is determined by polarization in Minkowsky space.
Exact synchronization of watches is possible only when material
points come to immediate touch with each other, i\.\,e\. when
their world lines intersect. However, even after such exact
synchronization in the point of next meeting watches of tho
material points will show different times. This difference is
due to different "life paths" between two meetings.\par
     Concept of private time is illustrated by so-called
twins problem, well-known from science fiction. Suppose that
one of twins goes to far-away travel in interstellar spacecraft,
while his brother stays on the Earth. Which of them will be older
when they meet each other on the Earth in the end of space
voyage.\par
     The answer is: that one who stayed on the Earth will be
older. World lines of twins intersect twice. Both intersections occur
on the Earth, one before travel and other after travel. Its known
that Coordinate system associated with the Earth can be taken for
inertial coordinate system with high degree of accuracy (indeed,
acceleration due to rotation of the Earth around its axis and due
to orbital rotation around the Sun is not sensible in our everyday
life). Therefore world line of twin stayed on the Earth is straight
line. World line of twin in spacecraft is curved. In the beginning
of travel he accelerates reaching substantial velocity comparable
with light velocity in the middle of the path. Then he experiences
backward acceleration in order to brake before reaching target
point of his travel. Then he accelerates and brakes again in
his back way to the Earth. According to theorem~6.1 interval length
of curved world line connecting two events is shorter than
interval length of straight world line connecting the same two
events. Hence twin stayed on the Earth will be older.
\proclaim{\bf Exercise 6.1} Remember proof of the fact that
the length of curved line connecting two points $A$ and $B$
in Euclidean space is greater than the length of straight line
segment $AB$. By analogy to this proof find the proof for
theorem~6.1.
\endproclaim
\head
\S\,7. Dynamics of material point.
\endhead
     Motion of material point in theory of relativity is
described by its world line in Minkowsky space. Let's choose
natural parameter on world line and consider four-dimensional
tangent vector 
$$
\bold u(s)=\frac{d\bold r(s)}{ds},
\tag7.1
$$
where $\bold r(s)$ is four-dimensional radius vector of events
on world line. Vector $\bold u$ is \thetag{7.1} is called vector
of {\it $4$-velocity}. It is time-like vector and it is unit
vector in Minkowsky metric: $g(\bold u,\bold u)=1$. Upon choosing
some inertial coordinate system we can write components of $4$-velocity
vector explicitly:
$$
\bold u=\frac{1}{\sqrt{c^2-|\bold v|^2\,}}
\Vmatrix c\\ v^1\\ v^2\\ v^3\endVmatrix.
\tag7.2
$$
Here $v^1$,~$v^2$, $v^3$ are components of three-dimensional velocity
vector $\bold v$. Note that components $u^0$,~$u^1$,~$u^2$, $u^3$ of
$4$-velocity vector are absolute numbers (without measure unit). It
is easy to see from \thetag{7.2}. Upon multiplying $\bold u$ by scalar
$mc$ with the measure unit of momentum we get vector of {\it
$4$-momentum}
$$
\bold p=\frac{m}{\sqrt{1-\dfrac{|\bold v|^2}{c^2}\,}}
\Vmatrix c\\ v^1\\ v^2\\ v^3\endVmatrix
\tag7.3
$$
for material point with mass $m$. Vector $\bold p$ plays important
role in physics since there is fundamental law of nature: {\it
the law of conservation of $4$-momentum}.
\proclaim{\bf Momentum conservation law} Vector of $4$-momentum
of material point which do not experience external action remains
unchanged.
\endproclaim
    According to the law just stated, for particle that do not
experience external action we have $\bold p=\const$. Hence
$\bold u=\const$. Integrating the equation \thetag{7.1}, for
$\bold r(s)$ we derive
$$
\bold r(s)=\bold r_0+\bold u\,s.
$$
Conclusion: in the absence of external action material point
moves uniformly along straight line.\par
    External actions causing change of $4$-momentum of material
point are subdivided into two categories:
\roster
\item continuous;
\item discrete.
\endroster
Continuous actions are applied to material particle by external
fields. They cause world line to bend making it curved line. In
this case $\bold p\neq\const$. Derivative of $4$-momentum with
respect to natural parameter $s$ is called {\it vector of $4$-force}:
$$
\frac{d\bold p}{ds}=\bold F(s).
\tag7.4
$$
Vector of $4$-force in \thetag{7.4} is quantitative characteristic
of the action of external fields upon material particle. It is
determined by parameters of particle itself and by parameters
of external fields at current position of particle as well. We know
that vector of $4$-velocity $\bold u$ is unit vector, therefore
$g(\bold p,\bold p)=m^2\,c^2$. Differentiating this relationship
with respect to $s$ and taking into account that components of
matrix \thetag{2.7} are constant, we find
$$
g(\bold u,\bold F)=0.
\tag7.5
$$
The relationship \thetag{7.5} means that vector of $4$-force
is perpendicular to vector of $4$-velocity in Minkowsky metric,
i\.\,e\. force vector is perpendicular to world line of particle.
\par
     Choosing some inertial coordinate system, we can replace natural
parameter $s$ in \thetag{7.5} by time variable $t$ of this coordinate
system. Then, taking into account \thetag{7.3}, from \thetag{7.4} we
derive
$$
\frac{dp^i}{dt}
=\sqrt{c^2-|\bold v|^2\,}\,F^i,
\text{ \ where \ }i=1,2,3.
\tag7.6
$$
Now, if we denote by $\bold f$ three-dimensional vector with components
$f^i=\sqrt{c^2-|\bold v|^2\,}\,F^i$, then for three-dimensional
vector of momentum from \thetag{7.6} we obtain differential equation
$$
\frac{d\bold p}{dt}=\bold f.
\tag7.7
$$
The equation \thetag{7.7} is treated as relativistic analog of
Newton's second law. Instead of classical formula $\bold p=m\bold v$
relating momentum and velocity vectors here we have the following
relationship:
$$
\bold p=\frac{m\bold v}{\sqrt{1-\dfrac{|\bold v|^2}{c^2}\,}}.
\tag7.8
$$
In order to write \thetag{7.8} in classical form we introduce
the quantity
$$
m_v=\frac{m}{\sqrt{1-\dfrac{|\bold v|^2}{c^2}\,}}.
\tag7.9
$$
Constant $m$ is called {\it mass at rest}, while $m_v$ in
\thetag{7.8} is called {\it dynamic mass} of moving particle.
Now $\bold p=m_v\,\bold v$, and Newton's second law is written
as follows:
$$
(m_v\,\bold v)'_t=\bold f.
\tag7.10
$$
Formulas \thetag{7.9} and \thetag{7.10} are the very ones which
are in mind when one says that mass in theory of relativity
depends on velocity. It seems to me that such terminology is not
so good. In what follows we shall mostly use four-dimensional
invariant equation \thetag{7.4} and, saying mass, we shall imply
mass at rest.\par
\parshape 13 4.6cm 5.5cm 4.6cm 5.5cm 4.6cm 5.5cm 4.6cm 5.5cm
4.6cm 5.5cm 4.6cm 5.5cm 4.6cm 5.5cm 4.6cm 5.5cm
4.6cm 5.5cm 4.6cm 5.5cm 4.6cm 5.5cm 4.6cm 5.5cm
0cm 10.1cm
    Discrete external actions appear in those situations when
$4$-momen\-tum of material particle changes abruptly in jump-like
manner. Such situation arise in particle collisions, particle
confluence, and particle decay.
\vadjust{\vtop to 0pt{\hbox to 0pt{\kern -30pt
\includegraphics{fig8.eps}\hss}
\vskip 35pt\hbox{\kern 40pt {\it Fig\.~7.1 }\hss}
\vskip -43pt\hbox{\kern 35pt $p_2$\hss}
\vskip -16pt\hbox{\kern 93pt $p_k$\hss}
\vskip -30pt\hbox{\kern 11pt $p_1$\hss}
\vskip -38pt\hbox{\kern 105pt $\tilde p_n$\hss}
\vskip -33pt\hbox{\kern 17pt $\tilde p_1$\hss}
\vskip -23pt\hbox{\kern 62pt $\tilde p_2$\hss}
\vskip -100pt\vss}}Collision of particles correspond to that point
in Minkowsky space where world lines of two or several particles
come together. After collision particles can simply fly out from
that point. But if these are molecules of ingredients in
chemical reaction, then after collision we would have new molecules
of reaction products. In a similar way in collisions of atomic
nuclei nuclear reactions occur.\par
     Let's consider simultaneous collision of $k$ particles.
Denote by $\bold p_1,\ldots,\bold p_k$ their $4$-momenta just before
the collision. Suppose that as a result of collision he have
$n$ new particles created from initial ones. Denote by $\tilde\bold p_1,
\ldots,\tilde\bold p_n$ $4$-momenta of outgoing particles just after
the collision. If $k=1$ this is particle decay process, while if
$n=1$ we have particle confluence into one composite particle.
\par
\proclaim{\bf Momentum conservation law} Total $4$-momentum of
ingoing particles before collision is equal to total $4$-momentum
of outgoing particles after collision:
$$
\sum^k_{i=1}\bold p_i=\sum^n_{i=1}\tilde\bold p_i.
\tag7.11
$$
\endproclaim
    As an example we consider process of frontal collision of two
identical particles of mass $m$ leading to creation of one particle
of mass $M$. Suppose that velocities of initial particles are equal
by magnitude but opposite to each other:
$$
\xalignat 2
&\bold p_1=\frac{m}{\sqrt{1-\dfrac{|\bold v|^2}{c^2}\,}}
\Vmatrix c\\ v^1\\ v^2\\ v^3\endVmatrix,
&&\bold p_2=\frac{m}{\sqrt{1-\dfrac{|\bold v|^2}{c^2}\,}}
\Vmatrix c\\ -v^1\\ -v^2\\ -v^3\endVmatrix.
\endxalignat
$$
For $4$-momentum of resulting particle we have
$$
\tilde\bold p_1=\frac{M}{\sqrt{1-\dfrac{|\bold w|^2}{c^2}\,}}
\Vmatrix c\\ w^1\\ w^2\\ w^3\endVmatrix.
$$
Applying momentum conservation law \thetag{7.11} to this
situation, we get $\bold w=0$ and additionally we obtain
$$
M=\frac{2\,m}{\sqrt{1-\dfrac{|\bold v|^2}{c^2}\,}}.
\tag7.12
$$
From \thetag{7.12} we see that mass at rest of resulting composite
particle is greater than sum of rest masses of its components:
$M>m+m$. Conclusion: the low of mass conservation is 
fulfilled approximately only in the limit of small velocities
$|\bold v|\ll c$.\par
     Let's multiply zeroth component of $4$-momentum of material
particle by $c$. Resulting quantity has the measure unit of energy.
Let's denote this quantity by $E$:
$$
E=\frac{mc^2}{\sqrt{1-\dfrac{|\bold v|^2}{c^2}\,}}.
\tag7.13
$$
The quantity \thetag{7.13} is called {\it kinetic energy} of
moving particle. Writing relationship \thetag{7.11} for zeroth
components of colliding particles, we get {\it energy
conservation law}:
$$
\sum^k_{i=1}E_i=\sum^n_{i=1}\tilde E_i.
\tag7.14
$$
Thus, $4$-momentum conservation law for collision includes both
energy conservation law \thetag{7.14} and the law of conservation
for three-dimensional momentum.\par
     Note that for zero velocity $\bold v=0$ the above quantity
\thetag{7.13} does not vanish, but takes nonzero value 
$$
E=mc^2.
\tag7.15
$$
This quantity is known as {\it rest energy} of material particle.
Formula \thetag{7.15} is well-known. It reflects very important fact
absent in classical physics: the ability of energy to mass and mass
to energy conversion. In practice conversion of energy to mass is
realized in particle confluence (see $M>m+m$ in formula \thetag{7.12}).
Converse phenomenon of particle decay yields mass defect
(mass decrease). Lost mass is realized in additional amount of kinetic
energy of outgoing particles. Total conversion of mass to energy is
also possible. This happens in process of {\it annihilation}, when
elementary particle meets corresponding antiparticle. Large amount
of energy released in annihilation is scattered in form of short-wave
electromagnetic radiation. 
\head
\S\,8. Four-dimensional form of Maxwell equations.
\endhead
\rightheadtext{\S\,8. Four-dimensional form \dots}
     Starting from electromagnetic equations $\square\bold E=0$
and $\square\bold H=0$ in previous sections we have constructed
and described Lorentz transformations preserving form of these
equations. We also have given geometric and physical interpretation
of Lorentz transformations and even have described dynamics of
material points on the base of new relativistic notion of space
and time. Now time has come to remember that equations $\square\bold
E=0$ and $\square\bold H=0$ are not primary equations of electrodynamics,
they were derived from Maxwell equations. To have complete picture
we should we should write Maxwell equations in four-dimensional
form. Let's begin with second pair of these equations containing
charges and currents (see equations \thetag{1.2} in
Chapter~\uppercase\expandafter{\romannumeral 2}. Let's modify them:
$$
\xalignat 2
&\frac{1}{c}\,\frac{\partial\bold E}{\partial t}-
\rot\bold H=-\frac{4\pi}{c}\,\bold j,
&&-\divr\bold E=-4\pi\rho.
\endxalignat
$$
Then rewrite these equations in components using Levi-Civita
symbol for to express rotor (see \cite{3}):
$$
\aligned
\frac{\partial E^p}{\partial r^0}
&-\sum^3_{q=1}\sum^3_{k=1}\varepsilon_{pqk}\frac{\partial H^k}
{\partial r^q}=-\frac{4\pi}{c}j^p,\\
\vspace{1.5ex}
&-\sum^3_{q=1}\frac{\partial E^q}{\partial r^q}=-4\pi\,\rho.
\endaligned
\tag8.1
$$
Here we used notation $r^0=ct$ associating time variable with
zeroth component of radius-vector in Minkowsky space.\par
     Using Levi-Civita symbol and components of vector $\bold H$,
we can construct skew-symmetric $3\times 3$ matrix with 
elements
$$
F^{pq}=-\sum^3_{k=1}\varepsilon_{pqk}\,H^k.
\pagebreak
\tag8.2
$$
Due to \thetag{8.2} we can easily write explicit form of matrix
$F$:
$$
F^{pq}=\left(\matrix
  0   &  -H^3 & \,H^2\\
\vspace{0.5ex}
\,H^3 &   0   &  -H^1\\
\vspace{0.5ex}
 -H^2 & \,H^1 &   0\endmatrix\right).
\tag8.3
$$
Let's complement the above matrix \thetag{8.3} with one additional line
and one additional column:
$$
F^{pq}=\left(\matrix
0   &  -E^1 &  -E^2 &  -E^3\\
\vspace{1ex}
E^1 &   0   &  -H^3 & \,H^2\\
\vspace{1ex}
E^2 & \,H^3 &   0   &  -H^1\\
\vspace{1ex}
E^3 &  -H^2 & \,H^1 &   0\endmatrix\right).
\tag8.4
$$
Additional line and additional column in \thetag{8.4} are indexed
by zero, i\.\,e\. indices $p$ and $q$ run over integer numbers from
$0$ to $3$. In addition, we complement three-dimensional vector of
current density with one more component
$$
j^0=\rho c.
\tag8.5
$$
By means of \thetag{8.4} and \thetag{8.5} we can rewrite
Maxwell equations \thetag{8.1} in very concise four-dimensional form:
$$
\sum^3_{q=0}\frac{\partial F^{pq}}{\partial r^q}=-
\frac{4\pi}{c}\,j^p.
\tag8.6
$$\par
    Now let's consider first pair of Maxwell equations (see equations
\thetag{1.1} in Chapter~\uppercase\expandafter{\romannumeral 2}). In
coordinates they are written as
$$
\xalignat 2
\quad\frac{\partial H^p}{\partial r^0}+
&\sum^3_{q=1}\sum^3_{k=1}\varepsilon_{pqk}\frac{\partial E^k}
{\partial r^q}=0,
&&\sum^3_{q=1}\frac{\partial H^q}{\partial r^q}=0.
\hskip -2em
\tag8.7
\endxalignat
$$
The structure of the equations \thetag{8.7} is quite similar to
that of \thetag{8.1}. However, their right hand sides are zero
and we see slight difference in signs. Main difference is that
components of vectors $\bold E$ and $\bold H$ have exchanged
their places. To exchange components of vectors $\bold E$ and
$\bold H$ in matrix \thetag{8.4} we need four-dimensional analog
of Levi-Civita symbol:
$$
\varepsilon_{pqks}=\varepsilon^{pqks}=
\cases
0,&\vtop{\hsize=4.1cm\noindent\baselineskip 0pt if among 
          $p$, $q$, $k$, $s$ there are at least two equal
          numbers;}\\
1,&\vtop{\hsize=4.1cm\noindent\baselineskip 0pt if $(p\,q\,k\,s)$
          is even permutation of numbers $(0\,1\,2\,3)$;}\\
-1,&\vtop{\hsize=4.1cm\noindent\baselineskip 0pt if $(p\,q\,k\,s)$
          is odd permutation of numbers $(0\,1\,2\,3)$.}
\endcases
$$
Let's define matrix $G$ by the following formula for its
components:
$$
G^{pq}=-\frac{1}{2}\sum^3_{k=0}\sum^3_{s=0}\sum^3_{m=0}\sum^3_{n=0}
\varepsilon^{pqks}\,g_{km}\,g_{sn}\,F^{mn}.
\tag8.8
$$
Here $g$ is matrix \thetag{2.7} determining Minkowsky metric.
Matrix $G$ with components \thetag{8.8} can be expressed in
explicit form:
$$
G^{pq}=\left(\matrix
0   &  -H^1 &  -H^2 &  -H^3\\
\vspace{1ex}
H^1 &   0   & \,E^3 &  -E^2\\
\vspace{1ex}
H^2 &  -E^3 &   0   &  \,E^1\\
\vspace{1ex}
H^3 & \,E^2 &  -E^1 &   0\endmatrix\right).
\tag8.9
$$
The structure of matrix \thetag{8.9} enable us \pagebreak
to write remaining
two Maxwell equations \thetag{8.7} in concise four-dimensional form:
$$
\sum^3_{q=0}\frac{\partial G^{pq}}{\partial r^q}=0.
\tag8.10
$$
Usage of both matrices $F$ and $G$ in theory is assumed to be
too excessive. For this reason equations \thetag{8.10} are
written as
$$
\sum^3_{q=0}\sum^3_{k=0}\sum^3_{s=0}
\varepsilon^{pqks}\frac{\partial F_{ks}}{\partial r^q}=0.
\tag8.11
$$
Matrix $F_{ks}$ is obtained from $F^{mn}$ by means of standard
index lowering procedure using matrix \thetag{2.7}:
$$
F_{ks}=\sum^3_{m=0}\sum^3_{n=0}
g_{km}\,g_{sn}\,F^{mn}.
\tag8.12
$$\par
     Four-dimensional form of Maxwell equations \thetag{8.6} and
\thetag{8.11} gives a hint for proper geometric interpretation
of these equations. Matrix \thetag{8.4} defines tensor of the
type $(2,0)$ in Minkowsky space. This tensor is called {\it
tensor of electromagnetic field}. Tensorial interpretation of
matrix \thetag{8.4} immediately yields transformation rules, which
were lacking so far:
$$
F^{pq}=\sum^3_{m=0}\sum^3_{n=0} S^p_m\,S^q_n\,\tilde F^{mn}.
\tag8.13
$$
These relationships \thetag{8.13} determine transformation rules for
components of vectors $\bold E$ and $\bold H$. Before now we express
these rules in undetermined form by the relationships \thetag{1.6}.
For special Lorentz matrices \thetag{4.11} vectors of electric and
magnetic fields $\bold E$ and $\bold H$ in two inertial coordinate
systems are related as follows:
$$
\xalignat 3
&E^1=\tilde E^1,
&&E^2=\frac{\tilde E^2+\dsize\frac{u}{\vphantom{y}c}\,\tilde H^3}
{\sqrt{1-\dsize\frac{u^2}{c^2}}},
&&E^3=\frac{\tilde E^3-\dsize\frac{u}{\vphantom{y}c}\,\tilde H^2}
{\sqrt{1-\dsize\frac{u^2}{c^2}}},\\
\vspace{2ex}
&H^1=\tilde H^1,
&&H^2=\frac{\tilde H^2-\dsize\frac{u}{\vphantom{y}c}\,\tilde E^3}
{\sqrt{1-\dsize\frac{u^2}{c^2}}},
&&H^3=\frac{\tilde H^3+\dsize\frac{u}{\vphantom{y}c}\,\tilde E^2}
{\sqrt{1-\dsize\frac{u^2}{c^2}}}.
\endxalignat
$$
According to theorem~4.1, general Lorentz matrix is a product of
special Lorentz matrix of the form \thetag{4.11} and two matrices
of spatial rotation in tree-dimensional space. The latter ones
can be excluded if one writes Lorentz transformation in
``conditionally three-dimensional'' vectorial form:
$$
\aligned
&\bold E=
\dfrac{\bigl<\bold u,\,\tilde\bold E\bigr>}{|\bold u|^2}\,\bold u
+\frac{\tilde\bold E-\dfrac{\bigl<\bold u,\,\tilde\bold E\bigr>}
{|\bold u|^2}\,\bold u-\dfrac{1}{c}\,[\bold u,\,\tilde\bold H]}
{\sqrt{\dsize{1-\frac{|\bold u|^2}{c^2}}}},\\
\vspace{2ex}
&\bold H=
\dfrac{\bigl<\bold u,\,\tilde\bold H\bigr>}{|\bold u|^2}\,\bold u
+\frac{\tilde\bold H-\dfrac{\bigl<\bold u,\,\tilde\bold H\bigr>}
{|\bold u|^2}\,\bold u+\dfrac{1}{c}\,[\bold u,\,\tilde\bold E]}
{\sqrt{\dsize{1-\frac{|\bold u|^2}{c^2}}}}.
\endaligned
\tag8.14
$$\par
     From \thetag{8.13} we derive the following rule for transforming
covariant components of the tensor of electromagnetic field:
$$
F_{pq}=\sum^3_{m=0}\sum^3_{n=0} T^m_p\,T^n_q\,\tilde F_{mn}.
\tag8.15
$$
This relationship \thetag{8.15} provides invariance of the form of
Maxwell equations \thetag{8.11} under Lorentz transformation
\thetag{2.3}. In order to verify this fact it is sufficient to apply
relationships \thetag{2.8} for transforming derivatives and then
remember well-known property of four-dimensional Levi-Civita symbol
$\varepsilon^{pqks}$:
$$
\sum^3_{a=0}\sum^3_{b=0}\sum^3_{c=0}\sum^3_{d=0}
T^p_a\,T^q_b\,T^k_c\,T^s_d\,\varepsilon^{abcd}=
\det T\,\varepsilon^{pqks}.
\tag8.16
$$
The condition of invariance of Maxwell equations \thetag{8.6}
with respect to Lorentz transformations leads to the following
transformation rule for components of four-dimensional current
density:
$$
j^p=\sum^3_{m=0}S^p_m\,\tilde j^m.
\tag8.17
$$
In \thetag{8.17} it is easy to recognize the transformation rule
for components of four-dimensional vector. In case of special
Lorentz matrix of the form \thetag{4.11}, taking into account
\thetag{8.5}, one can write the above relationship \thetag{8.17}
as follows:
$$
\xalignat 2
&\rho=\frac{\tilde \rho+\dsize\frac{u}{\vphantom{y}c^2}\,\tilde j^1}
{\sqrt{1-\dsize\frac{u^2}{c^2}}},
&&j^1=\frac{u\tilde\rho+\tilde j^1}
{\sqrt{1-\dsize\frac{u^2}{c^2}}},\\
&&&\tag8.18\\
&j^2=\tilde j^2,
&&j^3=\tilde j^3.
\endxalignat
$$
Remember that here $u=c\th(\alpha)$ is a magnitude of relative
velocity of one inertial coordinate system with respect to another.
In vectorial form relationships \thetag{8.18} are written as
$$
\aligned
&\rho=\frac{\tilde\rho+\dsize\frac{\bigl<\bold u,\,\tilde\bold j\bigr>}
{\vphantom{y}c^2}}
{\sqrt{\dsize{1-\frac{|\bold u|^2}{c^2}}}},\\
\vspace{2ex}
&\bold j=
\frac{\bold u\,\tilde\rho+
\dsize\frac{\bigl<\bold u,\,\tilde\bold j\bigr>}
{|\bold u|^2}\,\bold u}{\sqrt{\dsize{1-\frac{|\bold u|^2}{c^2}}}}
+\tilde\bold j-\frac{\bigl<\bold u,\,
\tilde\bold j\bigr>}{|\bold u|^2}\,\bold u.
\endaligned
\tag8.19
$$
In such form they give transformation rule for charge density
$\rho$ and three-dimensional current density $\bold j$ under
Lorentz transformations with arbitrary Lorentz matrix.\par
\proclaim{\bf Exercise 8.1} Prove the relationship \thetag{8.16},
assuming $T$ to be an arbitrary $4\times 4$ matrix.
\endproclaim
\proclaim{\bf Exercise 8.2} Using \thetag{2.12}, derive the
relationship \thetag{8.15} from \thetag{8.12} and \thetag{8.13}.
\endproclaim
\proclaim{\bf Exercise 8.3} Using \thetag{8.15}, \thetag{8.16} 
and \thetag{2.8}, transform Max\-well equations \thetag{8.11} 
from one inertial coordinate system to another. Verify that the
form of these equations is invariant.
\endproclaim
\proclaim{\bf Exercise 8.4} Using \thetag{8.13}, \thetag{8.17} 
and \thetag{2.8}, transform Max\-well equations \thetag{8.6} 
from one inertial coordinate system to another. Verify that the
form of these equations is invariant.
\endproclaim
\head
\S\,9. Four-dimensional vector-potential.
\endhead
     Due to special structure of Maxwell equations one can introduce
vector-potential $\bold A$ and scalar potential $\varphi$. This was
done in \S\,3 of Chapter~\uppercase\expandafter{\romannumeral 2}. Here
are formulas for components of $\bold E$ and $\bold H$:
$$
\aligned
&E^p=-\frac{\partial\varphi}{\partial r^p}-\frac{1}{c}
\frac{\partial A^p}{\partial t},\\
\vspace{1ex}
&H^p=\sum^3_{q=1}\sum^3_{k=1}\varepsilon^{pqk}\frac{\partial A^k}
{\partial r^q},
\endaligned
\tag9.1
$$
(see formulas \thetag{3.4} in
Chapter~\uppercase\expandafter{\romannumeral 2}). Denote
$A^0=\varphi$ and consider four-dimensional vector $\bold A$
with components $A^0$, $A^1$, $A^2$, $A^3$. This is
{\it four-dimensional vector-potential} of electromagnetic field.
By lowering index procedure we get covector $\bold A$:
$$
A_p=\sum^3_{q=0} g_{pq}\,A^q.
\tag9.2
$$
Taking into account relationships \thetag{2.7} for components
of matrix $g_{pq}$, from formula \thetag{9.2} we derive
$$
\xalignat 2
&A_0=A^0,&&A_1=-A^1,\\
\vspace{-1.4ex}
&&&\tag9.3\\
\vspace{-1.4ex}
&A_2=-A^2,&&A_3=-A^3.
\endxalignat
$$
Moreover, let's write explicitly covariant components for the tensor
of electromagnetic field:
$$
F_{pq}=\left(\matrix
 0   & \,E^1 & \,E^2 & \,E^3\\
\vspace{1ex}
-E^1 &   0   &  -H^3 & \,H^2\\
\vspace{1ex}
-E^2 & \,H^3 &   0   &  -H^1\\
\vspace{1ex}
-E^3 &  -H^2 & \,H^1 &   0\endmatrix\right).
\pagebreak
\tag9.4
$$
Due to \thetag{9.3} and \thetag{9.4} first relationship
\thetag{9.1} can be written as
$$
F_{0q}=\frac{\partial A_q}{\partial r^0}-
\frac{\partial A_0}{\partial r^q}.
\tag9.5
$$
In order to calculate other components of tensor $F_{pq}$ let's
apply \thetag{8.2} and second relationship \thetag{9.1}. Thereby
let's take into account that $F_{pq}=F^{pq}$ and $A_p=-A^p$ for
$p,q=1,2,3$:
$$
F_{pq}=-\sum^3_{k=1}\varepsilon_{pqk}\,H^k=
\sum^3_{k=1}\sum^3_{m=1}\sum^3_{n=1}\varepsilon_{pqk}\,
\varepsilon^{kmn}\frac{\partial A_n}{\partial r^m}.
\tag9.6
$$
Further transformation of \thetag{9.6} is based on one of the
well-known contraction identities for Levi-Civita symbol:
$$
\sum^3_{k=1}\varepsilon_{pqk}\,\varepsilon^{kmn}=
\delta^m_p\delta^n_q-\delta^m_q\delta^n_p.
\tag9.7
$$
Applying \thetag{9.7} to \thetag{9.6}, we get
$$
F_{pq}=\sum^3_{m=1}\sum^3_{n=1}
(\delta^m_p\delta^n_q-\delta^m_q\delta^n_p)\,
\frac{\partial A_n}{\partial r^m}=
\frac{\partial A_q}{\partial r^p}-
\frac{\partial A_p}{\partial r^q}.
\tag9.8
$$
Combining \thetag{9.8} and \thetag{9.5}, we obtain the following
formula for all covariant components of the tensor of electromagnetic
field:
$$
F_{pq}=\frac{\partial A_q}{\partial r^p}-
\frac{\partial A_p}{\partial r^q}.
\tag9.9
$$
In essential, formula \thetag{9.9} is four-dimensional form of the
relationships \thetag{9.1}. It unites these two relationships into
one.\par
     Remember that vectorial and scalar potentials of electromagnetic
field are not unique. They are determined up to a gauge transformation
(see formula \thetag{4.1} in
Chapter~\uppercase\expandafter{\romannumeral 2}). This uncertainty
could be included into transformation rule for component of
four-dimensional potential $\bold A$. However, if we assert that
$A^0$, $A^1$, $A^2$, $A^3$ are transformed as components of
four-dimensional vector
$$
A^p=\sum^3_{q=0}S^p_q\,\tilde A^q,
\tag9.10
$$
and $A_0$, $A_1$, $A_2$, $A_3$ are obtained from them by index
lowering procedure \thetag{9.2}, then we find that quantities
$F_{pq}$ defined by formula \thetag{9.9} are transformed exactly
by formula \thetag{8.15}, as they actually should.\par
    From \thetag{9.10} one can easily derive explicit transformation
formulas for scalar potential $\varphi$ and for components of
three-dimensional vector-potential $\bold A$. For special Lorentz
transformations with matrix \thetag{4.11} they are written as
follows:
$$
\xalignat 2
&\varphi=\frac{\tilde\varphi+\dsize\frac{u}{\vphantom{y}c}\,\tilde A^1}
{\sqrt{1-\dsize\frac{u^2}{c^2}}},
&&A^1=\frac{
\dsize\frac{u}{\vphantom{y}c}\,\tilde\varphi+\tilde A^1}
{\sqrt{1-\dsize\frac{u^2}{c^2}}},\\
&&&\tag9.11\\
&A^2=\tilde A^2,
&&A^3=\tilde A^3.
\endxalignat
$$
Note that one can rederive transformation rules for components
of electric and magnetic fields (see \S\,8 above). However, we
shall not do it now.\par
    In case of Lorentz transformations with arbitrary Lorentz
matrix the relationships \thetag{9.11} should be written in
vectorial form:
$$
\varphi=\frac{\tilde\varphi+\dsize\frac{\bigl<\bold u,\,\tilde\bold j\bigr>}
{\vphantom{y}c}}
{\sqrt{\dsize{1-\frac{|\bold u|^2}{c^2}}}},
\tag9.12a
$$
$$
\bold A=
\frac{\dfrac{\bold u}{c}\,\tilde\varphi+
\dsize\frac{\bigl<\bold u,\,\tilde\bold A\bigr>}
{|\bold u|^2}\,\bold u}{\sqrt{\dsize{1-\frac{|\bold u|^2}{c^2}}}}
+\tilde\bold A-\frac{\bigl<\bold u,\,
\tilde\bold A\bigr>}{|\bold u|^2}\,\bold u.
\tag9.12b
$$
\proclaim{\bf Theorem 9.1} Each skew-symmetric tensor field
$\bold F$ of type $(0,2)$ in four-dimensional space satisfying
differential equations \thetag{8.11} is determined by some
covector field $\bold A$ according to the above formula \thetag{9.9}.
\endproclaim
\demo{Proof} Each skew-symmetric tensor field $\bold F$ of type
$(0,2)$ in four-dimensional space can be identified with pair of
three-dimensional vector fields $\bold E$ and $\bold H$ depending
on additional parameter $r^0=ct$. In order to do this one should
use \thetag{9.4}. Then equations \thetag{8.11} are written as
Maxwell equations for $\bold E$ and $\bold H$:
$$
\xalignat 2
&\divr\bold H=0,&&\rot\bold E=-\frac{1}{c}\,\frac{\partial
\bold H}{\partial t}.
\endxalignat
$$
Further construction of covector field $\bold A$ is based on
considerations from \S\,3 of
Chapter~\uppercase\expandafter{\romannumeral 2}, where three-dimensional
vector-potential and scalar potential were introduced. Then we denote
$A^0=\varphi$ and thus convert three-dimensional vector-potential into
four-dimensional vector. And the last step is index lowering procedure
given by formula \thetag{9.2}.
\qed\enddemo
     Choice of vector field $\bold A$ in formula \thetag{9.9}, as
we noted above, has gauge uncertainty. In four-dimensional formalism
this fact is represented by gauge transformations
$$
A_k\to A_k+\frac{\partial\psi}{\partial r^k},
\tag9.13
$$
where $\psi$ --- is some arbitrary scalar field. Formula \thetag{9.13}
is four-dimensional version of gauge transformations \thetag{4.1}
\pagebreak
considered in Chapter~\uppercase\expandafter{\romannumeral 2}. It is easy
to verify that gauge transformations \thetag{9.13} do not break
transformation rules \thetag{9.10} for contravariant components of vector
$\bold A$.
\proclaim{\bf Exercise 9.1} Prove theorem~9.1 immediately in
four-dimen\-sional form without passing back to three-dimensional
statements and constructions.
\endproclaim
\head
\S\,10. The law of charge conservation.
\endhead
    Earlier we have noted that charge conservation law can be
derived from Maxwell equations (see \S\,1 in
Chapter~\uppercase\expandafter{\romannumeral 2}). To prove this
fact in four-dimensional formalism is even easier. Let's differentiate
the relationship \thetag{8.6} with respect to $r^p$ and add one more
summation with respect to index $p$:
$$
\sum^3_{p=0}\sum^3_{q=0}\frac{\partial^2 F^{pq}}
{\partial r^p\partial r^q}=-\frac{4\pi}{c}
\sum^3_{p=0}\frac{\partial j^p}{\partial r^p}.
\tag10.1
$$
Double differentiation in \thetag{10.1} is symmetric operation, while
tensor of electromagnetic field $F^{pq}$, to which it is applied, is
skew-symmetric. Therefore the expression under summation in left hand
side of \thetag{10.1} is skew-symmetric with respect to indices $p$
and $q$. This leads to vanishing of double sum in left hand side of
formula \thetag{10.1}. Hence we obtain
$$
\sum^3_{p=0}\frac{\partial j^p}{\partial r^p}=0.
\tag10.2
$$
The equality \thetag{10.2} is four-dimensional form of charge
conservation law. If we remember that $j^0=c\rho$ and $r^0=ct$,
we see that this equality coincides with \thetag{5.4} in
Chapter~\uppercase\expandafter{\romannumeral 1}.\par
    Conservation laws for scalar quantities (those like electric
char\-ge) in theory of relativity are expressed by equations analogous
to \thetag{10.2} in form of vanishing of four-dimensional divergencies
for corresponding four-dimensional currents. For vectorial quantities
corresponding current densities are tensors. Thus the law of con\-servation
of $4$-momentum for fields is represented by the equation
$$
\sum^3_{p=0}\frac{\partial T^{qp}}{\partial r^p}=0.
\tag10.3
$$
Tensor $T^{qp}$ in \thetag{10.3} playing the role of current density
of $4$-mo\-mentum is called {\it energy-momentum tensor}.
\proclaim{\bf Theorem 10.1} For any vector field $\bold j$ in
$n$-dimensional space ($n\geqslant 2$) if its divergency is zero
$$
\sum^n_{p=1}\frac{\partial j^p}{\partial r^p}=0,
\tag10.4
$$
then there is skew-symmetric tensor field $\boldsymbol\psi$ of type
$(2,0)$ such that 
$$
j^p=\sum^n_{q=1}\frac{\partial\psi^{pq}}{\partial r^q}.
\tag10.5
$$
\endproclaim
\demo{Proof} Choosing some Cartesian coordinate system, we shall
construct matrix $\psi^{pq}$ of the following special form:
$$
\psi^{pq}=
\left(\matrix
0          & \hdots & 0              & \psi^{1n}\\
\vspace{1ex}
\vdots     & \ddots & \vdots         & \vdots\\
\vspace{1ex}
0          & \hdots & 0              & \psi^{n-1\,n}\\
\vspace{1ex}
-\psi^{1n} & \hdots & -\psi^{n-1\,n} & 0
\endmatrix\right).
\hskip -2em
\tag10.6
$$
Matrix \thetag{10.6} is skew-symmetric, it has $(n-1)$
independent components. From \thetag{10.5} for these
components we derive 
$$
\aligned
&\frac{\partial\psi^{kn}}{\partial r^n}=j^k,\text{ \ where \ }
k=1,\ldots,n-1,\\
\vspace{1ex}
&\sum^{n-1}_{k=1}\frac{\partial\psi^{kn}}{\partial r^k}=-j^n.
\endaligned
\tag10.7
$$
Let's define functions $\psi^{kn}$ in \thetag{10.7} by the
following integrals:
$$
\aligned
\psi^{kn}&=\int\limits^{\ r^n}_0 j^k(r^1,\ldots,r^{n-1},y)\,dy+\\
&+\frac{1}{n-1}\int\limits^{r^k}_0 j^n(r^1,\ldots,y,\ldots,r^{n-1},0)
\,dy.\hskip -2em
\endaligned
\tag10.8
$$
It is easy to verify that functions \thetag{10.8} satisfy first series
of differential equations \thetag{10.7}. Under the condition \thetag{10.4}
they satisfy last equation \thetag{10.7} as well. Thus, theorem is
proved.\qed\enddemo
     Theorem~10.1 can be easily generalized for arbitrary tensorial
currents. Its prove thereby remains the same in most.
\proclaim{\bf Theorem 10.2} For any tensorial field $\bold T$ of
type $(m,s)$ in the space of dimension $n\geqslant 2$ if its 
divergency is zero
$$
\pagebreak
\sum^n_{p_m=1}\frac{\partial T^{p_1\,\ldots\, p_m}_{q_1\,\ldots\,q_s}}
{\partial r^{p_m}}=0,
$$
then there is tensorial field $\boldsymbol\psi$ of type $(m+1,s)$
skew-symmetric in last pair of upper indices and such that
$$
T^{p_1\,\ldots\, p_m}_{q_1\,\ldots\,q_s}
=\sum^n_{p_{m+1}=1}\frac{\partial
\psi^{p_1\,\ldots\, p_m\,p_{m+1}}_{q_1\,\ldots\,q_s}}
{\partial r^{p_{m+1}}}.
$$
\endproclaim
\proclaim{\bf Exercise 10.1} Verify that the equation \thetag{10.4}
provides last equation \thetag{10.7} to be fulfilled for the
functions \thetag{10.8}.
\endproclaim
\proclaim{\bf Exercise 10.2} Clarify the relation of theorem~10.1
and theorem on vortex field in case of dimension $n=3$.
\endproclaim
\head
\S\,11. Note on skew-angular and\\ curvilinear coordinates.
\endhead
\rightheadtext{\S\,11. Note on coordinates.}
     In previous three sections we have managed to write in
four-dimensional form all Maxwell equations, charge conservation law,
and the relation of $\bold E$, $\bold H$ and their potentials. The
relationships \thetag{8.6}, \thetag{8.11}, \thetag{9.9}, \thetag{9.13},
\thetag{10.2}, which were obtained there, preserve their shape when we
transfer from one rectangular Cartesian coordinate system to another.
Such transitions are interpreted as Lorentz transformations, they are
given by Lorentz matrices. However, all these relationships \thetag{8.6},
\thetag{8.11}, \thetag{9.9}, \thetag{9.13}, \thetag{10.2} possess
transparent tensorial interpretation. Therefore they can be transformed
to any skew-angular Cartesian coordinate system as well. Thereby we
would have minor differences: the shape of matrix $g$ would be
different and instead of $\varepsilon^{pqks}$ in \thetag{8.11} we
would require volume tensor with components 
$$
\pagebreak
\omega^{pqks}=\pm\sqrt{-\det\hat g}\ \varepsilon^{pqks}.
\tag11.1
$$\par
     Matrix $g_{pq}$ in skew-angular coordinate system is not given by
formula \thetag{2.7}, here it is arbitrary symmetric matrix determining
quadratic form of signature $(1,3)$. Therefore differential equations
$\square\bold E=0$ and $\square\bold H=0$, which we are started from,
have not their initial form. They are written as $\square F^{pq}=0$,
where d'Alambert operator is given by formula \thetag{2.6} with
non-diagonal matrix $g^{ij}$.\par
     In arbitrary skew-angular coordinate system none of axes
should have time-like direction. Therefore none of them can be
interpreted as time axis. Three-dimensional form of electrodynamics
equations, even if we could write them, would not have proper
physical interpretation in such coordinate system. In particular,
interpretation of components of tensor $F^{pq}$ as components of
electric and magnetic fields in formula \thetag{8.4} would not be
physically meaningful.\par
     Tensorial form of four-dimensional electrodynamics equations
enables us to make one more step toward increasing arbitrariness
in the choice of coordinate system: we can use not only skew-angular,
but curvilinear coordinates as well. To make this step we need to
replace partial derivatives by covariant derivatives:
$$
\frac{\partial}{\partial r^p}\to\nabla_p
\tag11.2
$$
(see \cite{3} for more details). Connection components required for
passing to covariant derivatives \thetag{11.2} are determined by
components of metric tensor. The latter ones in curvilinear coordinate
system do actually depend on $r^0$, $r^1$, $r^2$, $r^3$:
$$
\Gamma^k_{ij}=\frac{1}{2}\sum^3_{s=0} g^{ks}
\left(\frac{\partial g_{sj}}{\partial r^i}+
\frac{\partial g_{is}}{\partial r^j}-
\frac{\partial g_{ij}}{\partial r^s}\right).
\tag11.3
$$
No we give list of all basic equations,
which we derived above, in covariant form. Maxwell equations
are written as follows:
$$
\aligned
&\sum^3_{q=0}\nabla_qF^{pq}=-\frac{4\pi}{c}\,j^p,\\
\vspace{1ex}
&\sum^3_{q=0}\sum^3_{k=0}\sum^3_{s=0}
\omega^{pqks}\,\nabla_qF_{ks}=0.
\endaligned
\tag11.4
$$
Here components of volume tensor $\omega^{pqks}$ are given by formula
\thetag{11.1}. Tensor of electromagnetic field is expressed through
four-dimensional vector-potential by formula 
$$
F_{pq}=\nabla_pA_q-\nabla_qA_p,
\tag11.5
$$
while gauge uncertainty in the choice of vector-potential itself is
described by the relationship
$$
A_k\to A_k+\nabla_k\psi,
\tag11.6
$$
where $\psi$ is arbitrary scalar field. Charge conservation law
in curvilinear coordinates is written as
$$
\sum^3_{p=0}\nabla_pj^p=0.
\tag11.7
$$
Instead of formula \thetag{2.6} for D'Alambert operator here we have
$$
\square=\sum^3_{i=0}\sum^3_{j=0} g^{ij}\,\nabla_i\nabla_j.
\tag11.8
$$
Dynamics of material point of \pagebreak nonzero mass $m\neq 0$ is
described by ordinary differential equations of Newtonian type:
$$
\xalignat 2
&\dot\bold r=\bold u,
&&\nabla_{\!s}\bold u=\frac{\bold F}{mc}.
\tag11.9
\endxalignat
$$
Here dot means standard differentiation with respect to natural parameter
$s$ on world line, while $\nabla_{\!s}$ is covariant derivative with respect
to the same parameter.
\proclaim{\bf Exercise 11.1} Using symmetry of Christoffel symbols
\thetag{11.3} with respect to lower pair of indices $i$ and $j$, show
that the relationship \thetag{11.5} can be brought to the form
\thetag{9.9} in curvilinear coordinate system as well.
\endproclaim
\newpage
\setfirstpage
\topmatter
\title\chapter{4}
Lagrangian formalism in theory of relativity
\endtitle
\endtopmatter
\leftheadtext{CHAPTER \uppercase\expandafter{\romannumeral 4}.
LAGRANGIAN FORMALISM \dots}
\document
\head
\S\,1. Principle of minimal action\\
for particles and fields.
\endhead
\rightheadtext{\S\,1. Principle of minimal action \dots}
\parshape 14 0cm 10.1cm 0cm 10.1cm
3.8cm 6.3cm 3.8cm 6.3cm 3.8cm 6.3cm 3.8cm 6.3cm
3.8cm 6.3cm 3.8cm 6.3cm 3.8cm 6.3cm 3.8cm 6.3cm
3.8cm 6.3cm 3.8cm 6.3cm 3.8cm 6.3cm 
0cm 10.1cm 
     Dynamics of material points in theory of relativity
is described by their world lines. These are time-like lines
in Minkowsky space.
\vadjust{\vtop to 0pt{\hbox to 0pt{\kern -30pt
\includegraphics{fig9.eps}\hss}
\vskip 100pt\hbox{\kern 30pt {\it Fig\.~1.1 }\hss}
\vskip -55pt\hbox{\kern 10pt $A$\hss}
\vskip -65pt\hbox{\kern 53pt $B$\hss}
\vskip -70pt\vss}}
Let's consider some world line corresponding to real motion
of some particle under the action of external fields. Let's
fix two points $A$ and $B$ on this world line not too far
from each other. Then consider small deformation of world line
in the range between these two points $A$ and $B$. Suppose that we
have some coordinate system in Minkowsky space (either Cartesian,
or curvilinear, no matter). Then our world line is given
in parametric form by four functions
$$
r^0(s),\quad r^1(s),\quad r^2(s),\quad r^3(s),
\tag1.1
$$
where $s$ is natural parameter. \pagebreak Then deformed curve can be given
by the following four functions:
$$
\hat r^i(s)=r^i(s)+h^i(\varepsilon,s),
\ i=0,\,\ldots,\,3.
\tag1.2
$$
Here $s$ is original natural parameter on initial non-deformed
world line \thetag{1.1}, while $h^i(\varepsilon,s)$ are smooth
functions which are nonzero only within the range between points
$A$ and $B$. Note that functions $h^i(\varepsilon,s)$ in
\thetag{1.2} depend on additional parameter $\varepsilon$ which
is assumed to be small. Moreover, we shall assume that
$$
h^i(\varepsilon,s)\to 0\text{ \ as \ }\varepsilon\to 0.
\tag1.3
$$
Thus, in \thetag{1.2} we have whole family of deformed lines. This
family of lines is called {\it variation} of world line \thetag{1.1}.
Due to \thetag{1.3} we have the following Taylor expansion
for $h^i(\varepsilon,s)$:
$$
h^i(\varepsilon,s)=\varepsilon\,h^i(s)+\ldots\,.
\tag1.4
$$
Under the change of one curvilinear coordinate system for another
quantities $h^i(s)$ are transformed as components of four-dimensional
vector. This vector is called {\it vector of variation} of world line,
while quantities 
$$
\delta r^i(s)=\varepsilon h^i(s)
\tag1.5
$$
are called {\it variations of point coordinates}. It is clear that
they also are transformed as components of four-dimensional vector.
Due to formulas \thetag{1.4} and \thetag{1.5} parametric equations
of deformed curves \thetag{1.2} are written as follows:
$$
\hat r^i(s)=r^i(s)+\delta r^i(s)+\ldots\,.
\tag1.6
$$
By this formula we emphasize that terms other than linear with
respect to small parameter $\varepsilon$ are of no importance.
\par
    By varying functions $h^i(\varepsilon,s)$ in \thetag{1.2} and
by varying parameter $\varepsilon$ in them we can surround segment
of initial world line by a swarm of its variations. Generally
speaking, these variations do not describe real dynamics of points.
However, they are used in statement of {\it minimal action principle}.
Within framework of Lagrangian formalism {\it functional of action}
$S$ is usually introduced, this is a map that to each line connecting
two points $A$ and $B$ put into correspondence some real number $S$.
\proclaim{\bf Principle of minimal action for particles} World line
connecting two points $A$ and $B$ describes real dynamics of material
point if and only if action functional $S$ reaches local minimum on
it among other lines being its small variations.
\endproclaim
     Action functional $S$ producing number by each line should
depend only on that line (as geometric set of points in $M$),
but it should not depend on coordinate system $(r^0,r^1,r^2,r^3)$
in $M$. By tradition this condition is called {\it Lorentz invariance},
though changes of one curvilinear coordinate system by another form
much broader class of transformations than Lorentz transformations
relating two rectangular Cartesian coordinate systems in Minkowsky
space.\par
     Action functional in most cases is integral. For single point
of mass $m$ in electromagnetic field with potential $\bold A$ it is
written as
$$
S=-mc\int\limits^{\vphantom{y}\ s_2}_{s_1} ds-
\frac{q}{c}\int\limits^{\vphantom{y}\ s_2}_{s_1}
g(\bold A,\bold u)\,ds.
\tag1.7
$$
Here $q$ is electric charge of particle, while $\bold u=\bold u(s)$
is vector of its $4$-velocity (unit tangent vector of world line).
First integral in \thetag{1.7} yields action for free particle
(in the absence of external fields), second integral describes
interaction of particle with electromagnetic field.\par
     If we consider system of $N$ particles, then we should write
integral \thetag{1.7} for each of them and we should add all these
integrals. And finally, in order to get the action functional for
total system of field and particles we should add integral of
action for electromagnetic field itself:
$$
\aligned
S=\sum^N_{i=1}&\left(-m_i\,c\int\limits^{\vphantom{y}\ s_2(i)}_{s_1(i)} ds-
\frac{q_i}{c}\int\limits^{\vphantom{y}\ s_2(i)}_{s_1(i)}
g(\bold A,\bold u)\,ds\right)-\\
\vspace{1ex}
&-\frac{1}{16\,\pi\,c}\int\limits^{\ V_2}_{V_1}\sum^3_{p=0}
\sum^3_{q=0}F_{pq}\,F^{pq}\sqrt{-\det g\,}\,d^4r.
\endaligned
\tag1.8
$$\par
\parshape 14 0cm 10.1cm 0cm 10.1cm
4.5cm 5.6cm 4.5cm 5.6cm
4.5cm 5.6cm 4.5cm 5.6cm 4.5cm 5.6cm 4.5cm 5.6cm
4.5cm 5.6cm 4.5cm 5.6cm 4.5cm 5.6cm 4.5cm 5.6cm
0cm 10.1cm 0cm 10.1cm
Last integral in \thetag{1.8} deserves special consideration.
This is four-dimensional volume integral over the domain enclosed
between two three-dimensional hypersurfaces $V_1$ and $V_2$.
\vadjust{\vtop to 0pt{\hbox to 0pt{\kern -30pt
\includegraphics{fig10.eps}\hss}
\vskip 70pt\hbox{\kern 35pt {\it Fig\.~1.2 }\hss}
\vskip -40pt\hbox{\kern 30pt{\it past} \hss}
\vskip -28pt\hbox{\kern 23pt $V_1$\hss}
\vskip -42pt\hbox{\kern 23pt $V_2$\hss}
\vskip -30pt\hbox{\kern 40pt{\it future}\hss}
\vskip -75pt\vss}}Hypersurfaces $V_1$ and $V_2$ are space-like,
i\.\,e\. their normal vectors are time-like vectors. These
hypersurfaces determine the fissure between the future and the
past, and over this fissure we integrate in \thetag{1.8}. Thereby
change of field functions (here these are components of
vector-potential $\bold A$) when passing from $V_1$ to $V_2$
reflects evolution of electromagnetic field from the past to the
future.\par
     Electromagnetic field is described by field functions. Therefore
variation of field is defined in other way than that of particles.
Suppose that $\Omega$ is some restricted four-dimensional domain enclosed
between hypersurfaces $V_1$ and $V_2$. Let's consider four smooth
functions $h^i(\varepsilon,\bold r)=h^i(\varepsilon,r^0,r^1,r^2,r^3)$
being identically zero outside the domain $\Omega$ and vanishing
for $\varepsilon=0$. Let's define 
$$
\hat A^i(\bold r)=A^i(\bold r)+h^i(\varepsilon,\bold r)
\tag1.9
$$
and consider Taylor expansion of $h^i$ at the point $\varepsilon=0$:
$$
h^i(\varepsilon,\bold r)=\varepsilon\,h^i(\bold r)+\ldots\,.
\tag1.10
$$
The following functions determined by linear terms in the above
Taylor expansions \thetag{1.10}
$$
\delta A^i(\bold r)=\varepsilon\,h^i(\bold r)
\tag1.11
$$
are called {\it variations of field functions} for electromagnetic field.
Deformation of vector-potential \thetag{1.9} now can be written as
$$
\hat A^i(\bold r)=A^i(\bold r)+\delta A^i(\bold r)+
\ldots\,.
\tag1.12
$$
\proclaim{\bf Principle of minimal action for fields} Field functions
determine actual configuration of physical fields if and only if they
realize local minimum of action functional $S$ in class of all variations
with restricted support\footnote"*"{ \ Variations with restricted support
are those which are identically zero outside some restricted domain
$\Omega$.}. 
\endproclaim
      The condition of minimum of action for actual field configuration
and for actual world lines of particles, as a rule, is not used. In order
to derive dynamical equations for fields and particles it is sufficient
to have extremum condition (no matter minimum, maximum, or saddle point).
For this reason minimal action principle often is stated as {\it principle
of extremal action}.
\proclaim{\bf Exercise 1.1} Verify that $h^i(s)$ in \thetag{1.4} are
transformed as components of vector under the change of coordinates.
\endproclaim
\proclaim{\bf Exercise 1.2} Prove that under gauge transformations
\thetag{11.6} from Chapter~\uppercase\expandafter{\romannumeral 3}
action functional \thetag{1.8} is transformed as follows:
$$
S\to S-\sum^N_{i=1}\left(\frac{q_i}{c}\,\psi(\bold r(s_2(i)))-
\frac{q_i}{c}\,\psi(\bold r(s_1(i)))\right).
\tag1.13
$$
Explain why terms added to action functional in \thetag{1.13}
are not sensitive to variation of word lines \thetag{1.2}.
\endproclaim
\head
\S\,2. Motion of particle in electromagnetic field.
\endhead
\rightheadtext{\S\,2. Motion of particle \dots}
     In order to find world line of relativistic particle in
external electromagnetic field we shall apply particle version of
extremal action principle to functional \thetag{1.8}. Let's choose
one of $N$ particles in \thetag{1.8} and consider deformation
\thetag{1.6} of its world line. When we substitute deformed world
line into \thetag{1.8} in place of initial non-deformed one the
value of last integral remains unchanged. Thereby in first term
containing sum of integrals only one summand changes its value,
that one which represent the particle we have chosen among others.
Therefore writing extremity condition for \thetag{1.8} we can
use action functional in form of \thetag{1.7}. The value of
\thetag{1.7} for deformed world line is calculated as follows:
$$
S_{\text{def}}=-mc\int\limits^{\vphantom{y}\ s_2}_{s_1}
\sqrt{g(\bold K,\bold K)\,}\,ds-
\frac{q}{c}\int\limits^{\vphantom{y}\ s_2}_{s_1}
g(\bold A,\bold K)\,ds.
\tag2.1
$$
Formula \thetag{2.1} visually differs from formula \thetag{1.7}
because $s$ is natural parameter on initial world line, but it
is not natural parameter on deformed line. Here tangent vector  
$$
\bold K(s)=\frac{d\hat\bold r(s)}{ds}=
\bold u(s)+\varepsilon\,\frac{d\hat\bold h(s)}{ds}+\ldots
\tag2.2
$$
is not unit vector. Therefore first integral \thetag{1.7}
is rewritten as length integral (see \thetag{6.4} in
Chapter~\uppercase\expandafter{\romannumeral 3}). In second
integral \thetag{1.7} unit tangent vector is replaced by
vector $\bold K$.\par 
     Let's write in coordinate form both expressions which are
under integration in \thetag{2.1} taking into account that we
deal with general curvilinear coordinate system in Minkowsky
space:
$$
\aligned
&\sqrt{g(\bold K,\bold K)\,}=\sqrt{
\sum^3_{i=0}\sum^3_{j=0} g_{ij}(\hat r(s))\, K^i(s)\,K^j(s)},
\hskip-2em\\
&g(\bold A,\bold K)=\sum^3_{i=0}A_i(\hat\bold r(s))\,K^i(s).
\endaligned
\tag2.3
$$
Let's substitute \thetag{2.2} into \thetag{2.3} and take into
account \thetag{1.2} and the expansion \thetag{1.4}. As a result
for the expressions \thetag{2.3} we get the following power
expansions with respect to small parameter $\varepsilon$:
$$
\align
&\aligned
\sqrt{g(\bold K,\bold K)\,}&=\sqrt{g(\bold u,\bold u)\,}+
\frac{\varepsilon}{\sqrt{g(\bold u,\bold u)\,}}\left(
\shave{\sum^3_{i=0}} u_i(s)\,\frac{dh^i(s)}{ds}+\right.\\
&\left.+\frac{1}{2}\shave{\sum^3_{i=0}\sum^3_{j=0}\sum^3_{k=0}}
\frac{\partial g_{ij}}{\partial r^k}\,u^i(s)\,u^j(s)
\,h^k(s)\right)+\ldots\,,
\endaligned\\
\vspace{1ex}
&\aligned
g(\bold A,\bold K)=g(\bold A,\bold u)+\varepsilon
&\sum^3_{i=0}A_i(\bold r(s))\,\frac{dh^i(s)}{ds}+\\
&+\varepsilon\sum^3_{i=0}\sum^3_{k=0}\frac{\partial A_i}{\partial r^k}
\,u^i(s)\,h^k(s)+\ldots\,.
\endaligned
\endalign
$$
When substituting these expansions into \thetag{2.1} we should
remember that $\bold u$ is unit vector. Then for $S_{\text{def}}$
we get
$$
\aligned
&S_{\text{def}}=S-\varepsilon\int\limits^{\vphantom{y}\ s_2}_{s_1}
\sum^3_{k=0}\left(m\,c\,u_k(s)+
\frac{q}{c}\,A_k(\bold r(s))\right)\,\frac{dh^k(s)}{ds}\,ds\,-\\
\vspace{1.5ex}
&-\varepsilon\int\limits^{\vphantom{y}\ s_2}_{s_1}
\sum^3_{k=0}\left(\frac{q}{c}\shave{\sum^3_{i=0}}
\frac{\partial A_i}{\partial r^k}\,u^i+\frac{mc}{2}
\shave{\sum^3_{i=0}\sum^3_{j=0}}\frac{\partial g_{ij}}
{\partial r^k}u^iu^j\right)\!h^k(s)\,ds+\ldots\,.
\endaligned
$$
Let's apply integration by parts to first integral above. As a result
we get the expression without derivatives of functions $h^k(s)$:
$$
\aligned
&\qquad S_{\text{def}}=S-\varepsilon\sum^3_{k=0}\left(m\,c\,u_k(s)+
\frac{q}{c}\,A_k(\bold r(s))\right)\,h^k(s)\,
{\vrule height 19pt depth 14pt}^{\,s_2}_{\,s_1}+\\
\vspace{1.5ex}
&\quad\ +\varepsilon\int\limits^{\vphantom{y}\ s_2}_{s_1}
\sum^3_{k=0}\frac{d}{ds}\left(m\,c\,u_k(s)+
\frac{q}{c}\,A_k(\bold r(s))\right)\,h^k(s)\,ds\,-\\
\vspace{1.5ex}
&-\varepsilon\int\limits^{\vphantom{y}\ s_2}_{s_1}
\sum^3_{k=0}\left(\frac{q}{c}\shave{\sum^3_{i=0}}
\frac{\partial A_i}{\partial r^k}\,u^i+\frac{mc}{2}
\shave{\sum^3_{i=0}\sum^3_{j=0}}\frac{\partial g_{ij}}
{\partial r^k}u^iu^j\right)\!h^k(s)\,ds+\ldots\,.
\endaligned
$$
Remember that function $h^k(s)$ vanish at the ends of integration
path $h^k(s_1)=h^k(s_2)=0$ (see \S\,1). This provides vanishing of
non-integral terms in the above formula for $S_{\text{def}}$.\par
     Now in order to derive differential equations for world line
of particle we apply extremity condition for $S$. It means that
term linear with respect to $\varepsilon$ in power expansion for
$S_{\text{def}}$ should be identically zero irrespective to the
choice of functions $h^k(s)$:
$$
\gathered
\frac{d}{ds}\left(m\,c\,u_k(s)+\frac{q}{c}\,A_k(\bold r(s))\right)=\\
\vspace{1ex}
=\frac{q}{c}\shave{\sum^3_{i=0}}\frac{\partial A_i}{\partial r^k}\,u^i
+\frac{mc}{2}\,\shave{\sum^3_{i=0}\sum^3_{j=0}}\frac{\partial g_{ij}}
{\partial r^k}\,u^i\,u^j.
\endgathered
\tag2.4
$$
Let's calculate derivative in left hand side of \thetag{2.4}. Then
let's rearrange terms so that those with $m\,c$ factor are in left
hand side, while others with $q/c$ factor are in right hand side:
$$
mc\,\left(\frac{du_k}{ds}-\frac{1}{2}
\shave{\sum^3_{i=0}\sum^3_{j=0}}
\frac{\partial g_{ij}}{\partial r^k}\,u^i\,u^j\right)=
\frac{q}{c}\shave{\sum^3_{i=0}}
\left(\frac{\partial A_i}{\partial r^k}
-\frac{\partial A_k}{\partial r^i}\right)\,u^i.
$$
Now in right hand side of this equation we find tensor of
electromagnetic field (see formula \thetag{9.9} in
Chapter~\uppercase\expandafter{\romannumeral 3}). For transforming
left hand side of this equation we use formula \thetag{11.3}
from Chapter~\uppercase\expandafter{\romannumeral 3}. As a result
we get the following equation for world line:
$$
m\,c\left(\frac{du_k}{ds}-
\shave{\sum^3_{i=0}\sum^3_{j=0}}\Gamma^i_{kj}\,u_i\,u^j
\right)=
\frac{q}{c}\shave{\sum^3_{i=0}}F_{ki}\,u^i.
\tag2.5
$$
In left hand side of the equation \thetag{2.5} we find covariant
derivative with respect to parameter $s$ along world line:
$$
m\,c\,\nabla_{\!s}u_k=\frac{q}{c}\sum^3_{i=0} F_{ki}\,u^i.
\tag2.6
$$
Comparing \thetag{2.6} with the equations \thetag{11.9} from
\pagebreak
Chapter~\uppercase\expandafter{\romannumeral 3}, we get formula
for the vector of four-dimensional force acting on a particle
with charge $q$ in electromagnetic field:
$$
F_k=\frac{q}{c}\sum^3_{i=0} F_{ki}\,u^i.
\tag2.7
$$\par
    Suppose that we have rectangular Cartesian coordinate system
in Minkowsky space. Then we can subdivide $\bold F$ into spatial and
temporal parts and can calculate components of three-dimensional
force vector: $f^i=\sqrt{c^2-|\bold v|^2}\,F^i$ (see formula
\thetag{7.6} in Chapter~\uppercase\expandafter{\romannumeral 3}).
Upon easy calculations with the use of formulas \thetag{7.2} and
\thetag{9.4} from Chapter~\uppercase\expandafter{\romannumeral 3}
for force vector $\bold f$ we get
$$
\bold f=q\,\bold E+\frac{q}{c}\,[\bold v,\,\bold H].
\tag2.8
$$
This formula \thetag{2.8} is exactly the same as formula for
Lorentz force (see \thetag{4.4} in
Chapter~\uppercase\expandafter{\romannumeral 1}). Thus formula
\thetag{2.7} is four-dimensional generalization of formula for
Lorentz force. Orthogonality condition for $4$-force and $4$-velocity
(see \thetag{7.5} in Chapter~\uppercase\expandafter{\romannumeral 3})
for \thetag{2.7} is fulfilled due to skew symmetry of tensor of
electromagnetic field.
\proclaim{\bf Exercise 2.1} Prove that gauge transformation
of action functional \thetag{1.13} does not change dynamic
equations of material point in electromagnetic field \thetag{2.6}.
\endproclaim
\proclaim{\bf Exercise 2.2} Verify that the relationship
\thetag{7.5} from Chapter~\uppercase\expandafter{\romannumeral 3}
holds for Lorentz force.
\endproclaim
\head
\S\,3. Dynamics of dust matter.
\endhead
     Differential equation \thetag{2.6} describes motion of charged
particles in electromagnetic field. If the number of particles is
not large, then we can follow after the motion of each of them. When
describing extremely large number of particles continual limit is
used, particles are replaced by continuous medium modeling their
collective behavior. Simplest model describing large number of
non-colliding particles is a model of {\it dust cloud}.
\vadjust{\vtop to 5.2cm{\hbox to 0pt{\kern -15pt
\includegraphics{fig11a.eps}\kern 150pt
\includegraphics{fig11b.eps}\hss}
\vskip 125pt\hbox{\kern 30pt{\it Fig\.~3.1 }\kern 130pt
{\it Fig\.~3.2 }\hss}
\vskip -70pt\vss}}In this model particles of cloud
move regularly (not chaotically).
Their world lines can be modeled by regular family of lines filling
the whole space (see Fig\.~3.1).\par
\vskip 0pt plus 1pt minus 1pt
     Another model is a model of {\it ideal gas}. Here particles also
do not collide each other, i\.\,e\. their world lines do not intersect.
However, their motion is chaotic (see Fig\.~3.2). Therefore if we fill
the whole space with their world lines, they would intersect.\par
     Besides two models considered mentioned above, there are models
describing liquids and solid materials. Points of liquid and solid
media move regularly (as on Fig\.~3.1). However, in these media
interaction of particles is essential. Therefore when describing such
media one should either use detailed microscopic analysis and get
macroscopic parameters by statistical averaging, or should use some
heuristic assumptions based on experiment.\par
     In this book we consider only most simple model of dust cloud.
In this case one should assume Minkowsky space to be filled by
regular family of world lines. Some of them are world lines of
real particles, others are imaginary ones obtained by extrapolation
in continual limit. Therefore at each point of $M$ we have unit
vector $\bold u$, this is tangent vector to world line passing through
this point. This means that dynamics of dust cloud can be described
by vector field $\bold u(\bold r)$.\par
     Apart from vector field $\bold u$, below we need scalar parameter
$\nu(\bold r)$ which means the density of dust cloud. We define it as
follows. Let's choose some small fragment of three-dimensional
hypersurface in $M$ orthogonal to vector $\bold u(\bold r)$ at the point
$\bold r$. The number of dust particles whose world lines cross this
fragment is proportional to its three-dimensional volume:
$N=\nu(\bold r)\,V$, parameter $\nu(\bold r)$ is coefficient of
proportionality. Parameter $\nu(\bold r)$ has measure unit of concentration,
it can be treated as concentration of particles in small fragment of
dust cloud near the point $\bold r$ measured in that inertial coordinate
system for which particles of this small fragment are at rest. By means
of $\nu(\bold r)$ and $\bold u(\bold r)$ we compose new four-dimensional
vector
$$
\boldsymbol\eta(\bold r)=c\,\nu(\bold r)\,\bold u(\bold r).
\tag3.1
$$
Vector \thetag{3.1} is called four-dimensional {\it flow density}
for particles in cloud. If we choose some inertial coordinate
system, then $\eta^0/c$ is interpreted as concentration of particles
in dust cloud, while other three components of four-dimensional vector
$\boldsymbol\eta$ form three-dimensional vector of flow density.\par
     Suppose that dust cloud is formed by identical particles with
mass $m$ and charge $q$. Then four-dimensional current density vector
can be represented as follows:
$$
\bold j(\bold r)=q\,\boldsymbol\eta(\bold r).
\tag3.2
$$
By analogy with \thetag{3.2} one can define {\it mass flow density vector}:
$$
\boldsymbol\mu(\bold r)=m\,\boldsymbol\eta(\bold r).
\tag3.3
$$
Total number of particles in cloud is fixed. This conservation law is
written as the following equality for $\boldsymbol\eta$: 
$$
\sum^3_{p=0}\nabla_{\!p}\eta^p=0.
\tag3.4
$$
From \thetag{3.4} and \thetag{3.2} one can derive charge conservation
law in form of the relationship \thetag{11.7} from
Chapter~\uppercase\expandafter{\romannumeral 3}. Taking into account
\thetag{3.3}, we get rest mass conservation law:
$$
\sum^3_{p=0}\nabla_{\!p}\mu^p=0.
\tag3.5
$$
Rest mass conservation law here is fulfilled due to the absence
of collisions when heavy particles can be produced from light
ones (see \S\,7 in Chapter~\uppercase\expandafter{\romannumeral 3}).
\par
     Let's consider dynamics of particles composing dust cloud.
Vector field $\bold u$ is constituted by tangent vectors to world
lines of dust particles. Therefore these world lines can be determined
as integral curves of vector field $\bold u$, i\.\,e\. by solving the
following system of ordinary differential equations:
$$
\frac{dr^i}{ds}=u^i(\bold r(s)),
 \ i=0,\,\ldots,\,3.
\tag3.6
$$
Having determined world line of particle from differential equations,
we know vector of its $4$-velocity $\bold u(s)$. Now let's calculate
covariant derivative of vector $\bold u(s)$ with respect to parameter
$s$:
$$
\nabla_{\!s}u^p=\frac{du^p(s)}{ds}+
\sum^3_{k=0}\sum^3_{n=0}\Gamma^p_{nk}\,u^k(s)\,u^n(s).
\tag3.7
$$
Calculating derivative $du^p/ds$ in \thetag{3.7} we take into account
\thetag{3.6} and the equality $\bold u(s)=\bold u(\bold r(s))$. As a
result we get
$$
\frac{du^p(s)}{ds}=\sum^3_{k=0}u^k\,\frac{\partial u^p}{\partial r^k}.
\tag3.8
$$
Substituting \thetag{3.8} into \thetag{3.7}, we derive the following
formula:
$$
\nabla_{\!s}u^p=\sum^3_{k=0}u^k\,\nabla_{\!k}u^p.
\tag3.9
$$
Right hand side of \thetag{3.9} is covariant derivative of vector field
$\bold u(\bold r)$ along itself (see more details in \cite{3}).
Substituting \thetag{3.9} into the equations of the dynamics of material
point, we get:
$$
\nabla_{\!\bold u}\bold u=\frac{\bold F}{mc}.
\tag3.10
$$
Here $\bold F=\bold F(\bold r,\bold u)$ is some external force field
acting on particles of dust matter. For example in the case of charged
dust in electromagnetic field the equation \thetag{3.10} looks like
$$
\sum^3_{k=0}u^k\,\nabla_{\!k}u_p=
\frac{q}{mc^2}\sum^3_{k=0} F_{pk}\,u^k.
\tag3.11
$$\par
     In contrast to the equations \thetag{11.9} from
Chapter~\uppercase\expandafter{\romannumeral 3}, which describe
dynamics of separate particle, here \thetag{3.10} are partial
differential equations with respect to components of vector field
$\bold u(\bold r)$. They describe dynamics of dust cloud in
continual limit. The equation for scalar field $\nu(\bold r)$
is derived from conservation law \thetag{3.4} for the number of
particles. Combining these two equations, we obtain a system
of differential equations:
$$
\aligned
&\sum^3_{k=0}u^k\,\nabla_ku_p=\frac{F_p}{mc},\\
\vspace{1ex}
&\sum^3_{k=0}u^k\,\nabla_k\nu=-\nu\sum^3_{k=0}\nabla_ku^k.
\endaligned
\tag3.12
$$
System of partial differential equations \thetag{3.12} yields
complete description for the dynamics of dust cloud.\par
     Model of dust matter cam be generalized a little bit. We
can consider mixture of particles of different sorts. For
each sort of particles we define its own vector field $\bold u(i,
\bold r)$ and its own scalar field of concentration $\nu(i,\bold r)$.
Then formulas \thetag{3.2} and \thetag{3.3} for $\bold j$ and
$\boldsymbol\mu$ are generalized as follows:
$$
\xalignat 2
&\bold j(\bold r)=\sum^n_{i=1}q(i)\,
\boldsymbol\eta(i,\bold r),
&&\boldsymbol\mu(\bold r)=\sum^n_{i=1}
m(i)\,\boldsymbol\eta(i,\bold r).
\endxalignat
$$
Here $\boldsymbol\eta(i,\bold r)=c\,\nu(i,\bold r)\,
\bold u(i,\bold r)$. Each pair of fields $\bold u(i,\bold r)$ and
$\nu(i,\bold r)$ satisfies differential equations \thetag{3.12}.
We can derive mass and charge conservation laws from these equations.
\head
\S\,4. Action functional for dust matter.
\endhead
    Let's study the dynamics of dust matter in electromagnetic field
within framework of Lagrangian formalism. Fort this purpose we need
to pass to continual limit in action functional \thetag{1.8}. For the
sake of simplicity we consider dust cloud with identical particles.
Omitting details of how it was derived, now we write ultimate formula
for action functional \thetag{1.8} in continual limit:
$$
\aligned
S=-&m\int\limits^{\ V_2}_{V_1}\sqrt{g(\boldsymbol\eta,
\boldsymbol\eta)\,}\sqrt{-\det g\,}\,d^4r-\\
\vspace{1ex}
-&\frac{q}{c^2}\int\limits^{\ V_2}_{V_1}g(\boldsymbol\eta,
\bold A)\sqrt{-\det g\,}\,d^4r-\\
&-\frac{1}{16\pi\,c}\int\limits^{\ V_2}_{V_1}\sum^3_{p=0}
\sum^3_{k=0}F_{pk}\,F^{pk}\sqrt{-\det g\,}\,d^4r.
\endaligned
\tag4.1
$$
Instead of deriving formula \thetag{4.1} from \thetag{1.8} we
shall verify this formula indirectly. For this purpose we shall
derive dynamical equation \thetag{3.11} from principle of
extremal action applied to action functional \thetag{4.1}.\par
     For describing dust matter in \thetag{4.1} we have chosen
vector field $\boldsymbol\eta(\bold r)$ defined in \thetag{3.1}.
Other two fields $\bold u(\bold r)$ and $\nu(\bold r)$ can be expressed
though vector field $\boldsymbol\eta(\bold r)$:
$$
\xalignat 2
&c\,\nu=|\boldsymbol\eta|=\sqrt{g(\boldsymbol\eta,
\boldsymbol\eta)\,},
&&\bold u=\frac{\boldsymbol\eta}{c\,\nu}.
\hskip -2em
\tag4.2
\endxalignat
$$
Dealing with variation of vector field $\boldsymbol\eta(\bold r)$
we should always remember that components of this field are not
independent functions. They satisfy differential equation \thetag{3.4}.
In order to resolve tis equation \thetag{3.4} we use slightly modified
version of theorem~10.1 from
Chapter~\uppercase\expandafter{\romannumeral 3}.
\proclaim{\bf Theorem 4.1} Let $M$ be some $n$-dimensional
manifold, where $n\ge 2$, equipped with metric $g_{ij}$.
For each vector field $\boldsymbol\eta$ \pagebreak with zero
divergency with respect to metric connection
$$
\sum^n_{p=1}\nabla_{\!p}\eta^p=0
\tag4.3
$$
there is skew-symmetric tensor field $\boldsymbol\varphi$
of type  $(2,0)$ such that the following relationships are
fulfilled
$$
\eta^p=\sum^n_{q=1}\nabla_{\!q}\varphi^{pq}.
\tag4.4
$$
\endproclaim
\demo{Proof} Writing relationships \thetag{4.3}, we use well-known
formula for components of metric connection, see formula \thetag{11.3}
in Chapter~\uppercase\expandafter{\romannumeral 3}. As a result we get
$$
\align
&\sum^n_{p=1}\nabla_{\!p}\eta^p=\sum^n_{p=1}\frac{\partial\eta^p}
{\partial r^p}+\sum^n_{p=1}\sum^n_{s=1}\Gamma^p_{ps}\,
\eta^s=\sum^n_{p=1}\frac{\partial\eta^p}{\partial r^p}+\\
\vspace{1ex}
&+\frac{1}{2}\sum^n_{p=1}\sum^n_{s=1}\sum^n_{k=1}
g^{pk}\left(\frac{\partial g_{pk}}{\partial r^s}+
\frac{\partial g_{ks}}{\partial r^p}-
\frac{\partial g_{ps}}{\partial r^k}\right)\eta^s.
\endalign
$$
Note that last two derivatives of metric tensor in round brackets
are canceled when we sum over indices $p$ and $k$. This is because
$g^{pk}$ is symmetric. Hence
$$
\aligned
&\sum^n_{p=1}\nabla_{\!p}\eta^p=\sum^n_{p=1}\frac{\partial\eta^p}
{\partial r^p}+\frac{1}{2}\sum^n_{p=1}\sum^n_{s=1}\sum^n_{k=1}
g^{sk}\frac{\partial g_{ks}}{\partial r^p}\eta^p=\\
\vspace{1ex}
&=\sum^n_{p=1}\frac{\partial\eta^p}{\partial r^p}+
\frac{1}{2}\sum^n_{p=1}\tr\left(g^{-1}\,\frac{\partial g}
{\partial r^p}\right)\eta^p.
\endaligned
\tag4.5
$$
For further transforming of this expression \thetag{4.5} we use
well known formula for logarithmic derivative of determinant:
$$
\frac{\partial\ln|\det g|}{\partial r^p}=
\tr\left(g^{-1}\,\frac{\partial g}
{\partial r^p}\right).
\tag4.6
$$
Substituting \thetag{4.6} into \thetag{4.5}, we transform
\thetag{4.5} so that 
$$
\sum^n_{p=1}\nabla_{\!p}\eta^p=\frac{1}{\sqrt{|\det g|}}
\sum^n_{p=1}\frac{\partial(\eta^p\sqrt{|\det g|})}
{\partial r^p}.
\tag4.7
$$
Let's carry out analogous calculations for right hand side
of \thetag{4.4} taking into account skew symmetry of the
field $\varphi^{pq}$ and symmetry of connection components
$\Gamma^k_{pq}$. These calculations yield
$$
\sum^n_{q=1}\nabla_{\!q}\varphi^{pq}=
\frac{1}{\sqrt{|\det g|}}
\sum^n_{q=1}\frac{\partial(\varphi^{pq}\sqrt{|\det g|})}
{\partial r^q}.
\tag4.8
$$
Denote $j^p=\sqrt{|\det g|}\,\eta^p$ and $\psi^{pq}=\sqrt{|\det g|}\,
\varphi^{pq}$. Now on the base of \thetag{4.7} and \thetag{4.8}
it is easy to understand that proof of theorem~4.1 is reduced to
theorem~10.1 from Chapter~\uppercase\expandafter{\romannumeral 3}.
\qed\enddemo
     {\bf Remark.} Generally speaking, theorem~10.2 has no direct
generalization for the case of spaces with metric. It is generalized
only for metric spaces with zero curvature tensor $R^s_{kpq}=0$.\par
     Let's define deformation of the field $\boldsymbol\eta$ in a way
similar to that we used for vector-potential $\bold A$ in \S\,1:
$$
\hat\eta^p(\bold r)=\eta^p(\bold r)+\varepsilon\,\zeta^p(\bold r)
+\ldots\,.
\tag4.9
$$
Both fields $\hat{\boldsymbol\eta}$ and $\boldsymbol\eta$ satisfy
differential equation \thetag{3.4}. Hence vector field
$\boldsymbol\zeta$ defined in \thetag{4.9} also satisfy this
equation. \pagebreak Let's apply theorem~4.1 to vector field
$\boldsymbol\zeta$:
$$
\zeta^p=\sum^3_{k=0}\nabla_{\!k}\varphi^{pk}.
\tag4.10
$$
Theorem~4.1 does not specify tensor field $\varphi^{pk}$ in
\thetag{4.10}, this can be any skew-symmetric tensor field.
However, we choose it in very special form as follows:
$$
\varphi^{pk}=\eta^p\,h^k-h^p\,\eta^k.
\tag4.11
$$
This choice can be motivated by the following theorem.
\proclaim{\bf Theorem 4.2} For any two vector fields
$\boldsymbol\zeta$ and $\boldsymbol\eta$, where $\boldsymbol\eta\neq 0$,
both satisfying differential equation \thetag{3.4} there is vector field
$\bold h$ such that vector field $\boldsymbol\zeta$ is given by formula
$$
\zeta^p=\sum^3_{k=0}\nabla_{\!k}(\eta^p\,h^k-h^p\,\eta^k).
$$
\endproclaim
\noindent Our choice \thetag{4.11} leads to the following expression for
the field $\hat{\boldsymbol\eta}$:
$$
\hat\eta^p(\bold r)=\eta^p(\bold r)+\varepsilon\,
\sum^3_{k=0}\nabla_{\!k}(\eta^p\,h^k-h^p\,\eta^k)
+\ldots\,.
\tag4.12
$$
Quantities $h^i(\bold r)$ in \thetag{4.12} are chosen to be smooth
functions being nonzero only within some restricted domain $\Omega$
in Minkowsky space.\par
    When substituting \thetag{4.12} into action functional \thetag{4.1}
we use the following expansion for $\sqrt{g(\hat{\boldsymbol\eta},
\hat{\boldsymbol\eta})\,}$:
$$
\sqrt{g(\hat{\boldsymbol\eta},\hat{\boldsymbol\eta})\,}=
\sqrt{g(\boldsymbol\eta,\boldsymbol\eta)\,}+
\frac{\varepsilon}{\sqrt{g(\boldsymbol\eta,\boldsymbol\eta)\,}}
\sum^3_{p=0}\sum^3_{q=0}\eta_p\,\nabla_k\varphi^{pk}+\ldots\,.
$$
We have analogous power expansion for the expression under second
integral in formula \thetag{4.1}:
$$
g(\hat{\boldsymbol\eta},\bold A)=g(\boldsymbol\eta,\bold A)+
\varepsilon\sum^3_{p=0}\sum^3_{k=0}A_p\,\nabla_k\varphi^{pk}
+\ldots\,.
$$
Substituting these two expansions into \thetag{4.1}, we take into
account \thetag{4.2}. For the action $S_{\text{def}}$ this yields
$$
\aligned
S_{\text{def}}&=S-\varepsilon\,m\int\limits_\Omega
\sum^3_{p=0}\sum^3_{k=0}u_p\,\nabla_{\!k}\varphi^{pk}
\sqrt{-\det g\,}d^4r-\\
&-\frac{\varepsilon\,q}{c^2}\int\limits_\Omega
\sum^3_{p=0}\sum^3_{k=0}A_p\,\nabla_{\!k}\varphi^{pk}
\sqrt{-\det g\,}d^4r+
\ldots\,.
\endaligned
\tag4.13
$$\par
     Further in order to transform the above expression \thetag{4.13}
we use Ostrogradsky-Gauss formula. In the space equipped with metric
this formula is written as follows:
$$
\int\limits_\Omega\sum^3_{k=0}\nabla_{\!k}z^k\sqrt{-\det g\,}d^4r=
\int\limits_{\partial\Omega}g(\bold z,\bold n)\,dV.
\tag4.14
$$
Here $z^0$, $z^1$, $z^2$, $z^3$ are components of smooth vector field
$\bold z$, while $\bold n$ is unit normal vector for the boundary of
the domain $\Omega$. In order to transform first integral in formula
\thetag{4.13} we take $z^k=\sum^3_{p=0}u_p\,\varphi^{pk}$. Then in
right hand side of \thetag{4.14} we obtain
$$
\sum^3_{k=0}\nabla_{\!k}z^k=\sum^3_{p=0}\sum^3_{k=0}u_p\,
\nabla_{\!k}\varphi^{pk}+\sum^3_{p=0}\sum^3_{k=0}\nabla_{\!k}u_p\,
\varphi^{pk}.
$$
Right hand side of \thetag{4.14} vanishes since $\varphi^{pk}$ do
vanish on the boundary of $\Omega$. Hence we have the equality
$$
\align
\int\limits_\Omega&\sum^3_{p=0}\sum^3_{k=0}u_p\,\nabla_{\!k}\varphi^{pk}
\sqrt{-\det g\,}d^4r=\\
&-\int\limits_\Omega\sum^3_{p=0}\sum^3_{k=0}\nabla_{\!k}u_p\,\varphi^{pk}
\sqrt{-\det g\,}d^4r.
\endalign
$$
In a similar way we transform second integral in \thetag{4.13}. In whole
for the action $S_{\text{def}}$ we get the following expression
$$
\aligned
S_{\text{def}}&=S+\varepsilon\,m\int\limits_\Omega
\sum^3_{p=0}\sum^3_{k=0}\nabla_{\!k}u_p\,\varphi^{pk}
\sqrt{-\det g\,}d^4r\,+\\
&+\frac{\varepsilon\,q}{c^2}\int\limits_\Omega
\sum^3_{p=0}\sum^3_{k=0}\nabla_{\!k}A_p\,\varphi^{pk}
\sqrt{-\det g\,}d^4r+
\ldots\,.
\endaligned
\tag4.15
$$\par
Extremity of action $S$ means that linear part with respect to
$\varepsilon$ in formula \thetag{4.15} should vanish:
$$
\int\limits_\Omega\sum^3_{p=0}\sum^3_{k=0}
\left(m\nabla_{\!k}u_p+\frac{q}{c^2}\nabla_{\!k}A_p\right)
\varphi^{pk}\sqrt{-\det g\,}d^4r=0.
$$
Let's substitute formula \thetag{4.11} for $\varphi^{pk}$ into the
above equality. Then it is transformed to the following one
$$
\allowdisplaybreaks
\align
&\int\limits_\Omega\sum^3_{p=0}\sum^3_{k=0}
\left(m\nabla_{\!k}u_p+\frac{q}{c^2}\nabla_{\!k}A_p\right)
\eta^ph^k\,\sqrt{-\det g\,}d^4r=\\
&=\int\limits_\Omega\sum^3_{p=0}\sum^3_{k=0}
\left(m\nabla_{\!k}u_p+\frac{q}{c^2}\nabla_{\!k}A_p\right)
\eta^kh^p\,\sqrt{-\det g\,}d^4r.
\endalign
$$
Let's exchange indices $k$ and $p$ in second integral. Thereafter
integrals can be united into one integral:
$$
\aligned
&\int\limits_\Omega\sum^3_{k=0}\sum^3_{p=0}
\left(m\nabla_{\!k}u_p-m\nabla_{\!p}u_k+\frac{q}{c^2}\nabla_{\!k}A_p
-\right.\\
\vspace{1.5ex}
&\quad\qquad\qquad\left.-\frac{q}{c^2}\nabla_{\!p}A_k\right)
\eta^ph^k\,\sqrt{-\det g\,}d^4r=0.
\hskip-2em
\endaligned
\tag4.16
$$
Now let's take into account that in resulting equality $h^k=h^k(\bold r)$
are arbitrary smooth functions vanishing outside the domain $\Omega$.
Therefore vanishing of integral \thetag{4.16} means vanishing of each
summand in sum over index $k$ in the expression under integration:
$$
\sum^3_{p=0}\left(m\,\nabla_{\!k}u_p-m\,\nabla_{\!p}u_k+\frac{q}{c^2}\,F_{kp}
\right)\eta^p=0.
\tag4.17
$$
Here we used the relationship \thetag{11.5} from
Chapter~\uppercase\expandafter{\romannumeral 3}. It relates
tensor of electromagnetic field and four-dimensional
vector-potential.\par
     In order to bring the equation \thetag{4.17} just derived
to its ultimate form we use the relationships \thetag{4.2},
which relate vector field $\boldsymbol\eta$ and vector field
$\bold u$: $\eta^p=c\,\nu\,u^p$. Since $\bold u$ is unit vector,
we have
$$
\pagebreak
\sum^3_{p=0}u^p\nabla_{\!k}u_p=0.
\tag4.18
$$
Taking into account \thetag{4.18}, we bring \thetag{4.17} to the
following form:
$$
\sum^3_{p=0}\,u^p\,\nabla_{\!p}u_k=\frac{q}{mc^2}\sum^3_{p=0}
F_{kp}\,u^p.
\tag4.19
$$
Now it is easy to see that \thetag{4.19} exactly coincides with
the equation \thetag{3.11}, which we have derived earlier. This
result approves the use of the action \thetag{4.1} for describing
charged dust matter in electromagnetic field.
\proclaim{\bf Exercise 4.1} Prove that for any skew-symmetric tensor
field $\varphi^{pq}$ vector field $\boldsymbol\eta$ determined by
formula \thetag{4.4} has zero divergency, i\.\,e\. is satisfies
differential equation \thetag{3.4}.
\endproclaim
\proclaim{\bf Exercise 4.2} Prove theorem~4.2. For this purpose use
the following fact known as theorem on rectification of vector field.
\endproclaim
\proclaim{\bf Theorem 4.3} For any vector field $\boldsymbol\eta\neq 0$
there exists some curvilinear coordinate system $r^0$, $r^1$, $r^2$,
$r^3$ such that $\eta^0=1$, $\eta^1=0$, $\eta^2=0$, $\eta^3=0$ in
this coordinate system.
\endproclaim
\proclaim{\bf Exercise 4.3} Prove theorem~4.3 on rectification of
vector field.
\endproclaim
\proclaim{\bf Exercise 4.4} Derive Ostrogradsky-Gauss formula
\thetag{4.14} for the space equipped with metric on the base of the
following integral relationship in standard space $\Bbb R^n$:
$$
\int\limits_\Omega\frac{\partial f(\bold r)}{\partial r^i}
d^nr=\int\limits_{\partial\Omega}f(\bold r)\,dr^1\ldots\,dr^{i-1}
dr^{i+1}\ldots\,dr^n.
$$
\endproclaim
\head
\S\,5. Equations for electromagnetic field.
\endhead
    In this section we continue studying action functional
\thetag{4.1}. This functional describes dust cloud composed of
particles with mass $m$ and charge $q$ in electromagnetic field.
In previous section we have found that applying extremal action
principle to $S$ with respect to the field $\boldsymbol\eta$
one can derive dynamical equations for velocity field in dust
cloud. Now we shall apply extremal action principle to $S$ with
respect to vector-potential $\bold A$. Deformation of\linebreak
vector-potential is defined according to \thetag{1.9},
\thetag{1.10}, \thetag{1.11}, \thetag{1.12}:
$$
\hat A_i(\bold r)=A_i(\bold r)+\varepsilon h_i(\bold r)+
\ldots\,.
\tag5.1
$$
For components of tensor of electromagnetic field we derive
$$
\hat F_{ij}=F_{ij}+\varepsilon\,(\nabla_{\!i}h_j-\nabla_{\!j}h_i)+
\ldots\,.
\tag5.2
$$
When substituting \thetag{5.2} into action functional \thetag{4.1} we
carry out the following calculations:
$$
\align
&\sum^3_{p=0}\sum^3_{k=0}\hat F_{pk}\,\hat F^{pk}=
\sum^3_{i=0}\sum^3_{j=0}\sum^3_{p=0}\sum^3_{k=0}
\hat F_{pk}\,\hat F_{ij}\,g^{pi}\,g^{kj}=\\
&=\sum^3_{p=0}\sum^3_{k=0}F_{pk}\,F^{pk}+
2\,\varepsilon\sum^3_{p=0}\sum^3_{k=0}F^{pk}(\nabla_{\!p}h_k
-\nabla_{\!k}h_p)+
\ldots\,.
\endalign
$$
Taking into account skew symmetry of tensor $F^{pk}$, this expansion
can be simplified more and can be brought to the form
$$
\sum^3_{p=0}\sum^3_{k=0}\hat F_{pk}\,\hat F^{pk}=
\sum^3_{p=0}\sum^3_{k=0}F_{pk}\,F^{pk}+
4\,\varepsilon\sum^3_{p=0}\sum^3_{k=0}F^{pk}\nabla_{\!p}h_k+
\ldots\,.
$$
Analogous calculations in substituting \thetag{5.1} into \thetag{4.1}
yield 
$$
g(\boldsymbol\eta,\hat\bold A)=g(\boldsymbol\eta,\bold A)+
\varepsilon\sum^3_{k=0}\eta^k\,h_k+\ldots\,.
$$
As a result for deformation of action functional \thetag{4.1} we
get 
$$
\gather
S_{\text{def}}=S-\frac{\varepsilon q}{c^2}\int\limits_{\Omega}
\sum^3_{k=0}\eta^k\,h_k\,\sqrt{-\det g\,}\,d^4r\,-\\
-\frac{\varepsilon}{4\pi\,c}\int\limits_{\Omega}
\sum^3_{p=0}\sum^3_{k=0}F^{pk}\nabla_{\!p}h_k\,\sqrt{-\det g\,}\,d^4r
+\ldots\,.
\endgather
$$
Let's transform second integral in the above expansion for $S_{\text{def}}$
by means of Ostrogradsky-Gauss formula \thetag{4.14}. For this purpose
let's choose $z^p=\sum^3_{k=0} F^{pk}h_k$ and take into account vanishing
of $h_k$ on the boundary of the domain $\Omega$. Then for $S_{\text{def}}$
we get
$$
S_{\text{def}}=S+\varepsilon\int\limits_{\Omega}
\sum^3_{k=0}\left(-\frac{q\eta^k}{c^2}+
\shave{\sum^3_{p=0}}\frac{\nabla_{\!p}F^{pk}}{4\pi\,c}\right)
h_k\,\sqrt{-\det g\,}\,d^4r+\ldots\,.
$$
Extremal action principle means that linear in $\varepsilon$ part of
the above expansion for $S_{\text{def}}$ should vanish. Note also that
$\Omega$ is an arbitrary domain and $h_k(\bold r)$ are arbitrary functions
within $\Omega$. This yield the following equations for the tensor of
electromagnetic field:
$$
\sum^3_{p=0}\nabla_{\!p}F^{pk}=\frac{4\pi\,q}{c}\,\eta^k.
\tag5.3
$$
Remember that $\boldsymbol\eta(\bold r)$ is related to current density
by means of \thetag{3.2}. Then \thetag{5.3} can be written as
$$
\sum^3_{p=0}\nabla_{\!p}F^{kp}=-\frac{4\pi}{c}\,j^k.
\tag5.4
$$
It is easy to see that \thetag{5.4} are exactly Maxwell equations
written in four-dimensional form (see \thetag{11.4} in
Chapter~\uppercase\expandafter{\romannumeral 3}). Another pair of
Maxwell equations written in four-dimensional form
$$
\sum^3_{q=0}\sum^3_{k=0}\sum^3_{s=0}
\omega^{pqks}\,\nabla_{\!q}F_{ks}=0
$$
is a consequence of the relationship $F_{pq}=\nabla_{\!p}A_q
-\nabla_{\!q}A_p$ (see formula \thetag{11.5} in 
Chapter~\uppercase\expandafter{\romannumeral 3}).
\proclaim{\bf Exercise 5.1} Which form will have differential equations
\thetag{5.3} if we consider dust cloud composed by particles of
several sorts with masses $m(1),\ldots,\,m(N)$ and charges $q(1),\ldots,\,
q(N)$\,? Will this change differential equations \thetag{5.4}\,?
\endproclaim
\newpage
\setfirstpage
\topmatter
\title\chapter{5}
General theory of relativity
\endtitle
\endtopmatter
\startpage{145}
\leftheadtext{CHAPTER \uppercase\expandafter{\romannumeral 5}.
GENERAL RELATIVITY.}
\document
\head
\S\,1. Transition to non-flat metrics and curved Minkowsky
space.
\endhead
\rightheadtext{\S\,1. Transition to non-flat metrics \dots}
     Passing from classical electrodynamics to special theory
of re\-lativity, in previous two chapters we have successively
geometrized many basic physical concepts. Having denoted $r^0=ct$
and combining $r^0$ with components of three-dimensional
radius-vector in inertial coordinate system, we have constructed
four-dimensional space of events (Minkowsky space). This space
appears to be equipped with metric of signature $(1,3)$, which
is called Min\-kowsky metric. Thereby inertial coordinate systems
are interpreted as orthonormal bases in Minkowsky metric.\par
     In four-dimensional formalism dynamics of material point is
described by vectorial differential equations, while Maxwell
equations for electromagnetic field are written in tensorial
form. Due to this circumstance in previous two chapters we managed
to include into consideration skew-angular and even curvilinear
coordinate systems in Minkowsky space. Thereby we got explicit
entries of metric tensor components $g_{ij}$, metric connection
components $\Gamma^k_{ij}$, and covariant derivatives
$\nabla_{\!i}$ in all our equations.\par
     Next step in this direction is quite natural. One should keep
the shape of all equations and pass from flat Minkowsky metric to
metric of signature $(1,3)$ with nonzero curvature tensor:
$$
R^k_{qij}=\frac{\partial\Gamma^k_{jq}}{\partial r^i}
-\frac{\partial\Gamma^k_{iq}}{\partial r^j}
+\sum^3_{s=0}\Gamma^k_{is}\,\Gamma^s_{jq}-
\sum^3_{s=0}\Gamma^k_{js}\,\Gamma^s_{iq}.
\tag1.1
$$
This crucial step was first made by Einstein. Theory he had discovered
in this way later was called {\it Einstein's theory of gravitation} or
{\it general theory of relativity}.
\definition{\bf Definition 1.1} Four-dimensional affine space equipped
with orientation, polarization, and with metric of signature $(1,3)$
and nonzero curvature \thetag{1.1} is called {\it curved Minkowsky space}.
\enddefinition
     In non-flat Minkowsky space we loose some structures available in
flat case. In such space there are no coordinates for which Minkowsky
metric is given by matrix \thetag{2.7} from
Chapter~\uppercase\expandafter{\romannumeral 3}, i\.\,e\. here we have
no inertial coordinate systems. This is substantial loss, but it is not
catastrophic since dynamic equation for material points and Maxwell
equations rewritten in vectorial and tensorial form are not bound to
inertial coordinate systems.\par 
     Geodesic lines in curved Minkowsky space do not coincide with affine
straight lines. Therefore affine structure becomes excessive restriction
in general relativity. As appears, one can give up topologic structure
of flat space $\Bbb R^4$ as well. Even in two-dimensional case, as we
know, apart from deformed (curved) plain, there are surfaces with more
complicated topology: sphere, torus and sphere with several handles
glued to it (see \cite{5}). In multidimensional case these objects are
generalized in concept of {\it smooth manifold} (see details in
\cite{2}, \cite{5}, and \cite{6}).\par
     Smooth manifold $M$ of dimension $n$ is a topologic space each point
of which has a neighborhood ({\it a chart}) identical to some neighborhood
of a point in $\Bbb R^n$. In other words $M$ is covered by a family of
charts $U_\alpha$, each of which is diffeomorphic to some open set
$V_\alpha$ in $\Bbb R^n$. Such chart maps (chart diffeomorphisms) define
local curvilinear coordinate systems within their chart domains $U_\alpha$.
At those points of manifold $M$, where two chart domains are overlapping,
transition functions arise. They relate one curvilinear coordinate system
with another:
$$
\aligned
&\tilde r^i=\tilde r^i(r^1,\ldots,\,r^n),
\text{ \ where \ } i=1,\ldots,\,n,\\
&r^i=r^i(\tilde r^1,\ldots,\,\tilde r^n),
\text{ \ where \ } i=1,\ldots,\,n.
\endaligned
\tag1.2
$$
According to definition of smooth manifold, transition functions
\thetag{1.2} are smooth functions (of class $C^\infty$). Transition
functions determine transition matrices $S$ and $T$:
$$
\xalignat 2
&T^i_j=\frac{\partial\tilde r^i}{\partial r^j},
&&S^i_j=\frac{\partial r^i}{\partial\tilde r^j}.
\tag1.3
\endxalignat
$$
Presence of transition matrices \thetag{1.3} lead to full-scale
theory of tensors, which is almost literally the same as theory
of tensors for curvilinear coordinates in $\Bbb R^n$ (see \cite{3}).
The only difference is that here we cannot choose Cartesian
coordinates at all. This is because in general there is no smooth
diffeomorphic map from manifold $M$ to $\Bbb R^n$.
\definition{\bf Definition 1.1} Four-dimensional smooth manifold
equipped with orientation, polarization, and with metric of
signature $(1,3)$ is called {\it generalized Minkowsky space} or
{\it Minkowsky manifold}.
\enddefinition
\head
\S\,2. Action for gravitational field.\\ Einstein equation.
\endhead
\rightheadtext{\S\,2. Action for gravitational field.}
    Space of events in general relativity is some smooth Minkowsky
manifold $M$. This circumstance provides additional arbitrariness
consisting in choosing $M$ and in choosing metric on $M$. Nonzero
curvature described by tensor \thetag{1.1} is interpreted as
{\it gravitational field}. Gravitational field acts upon material
bodies and upon electromagnetic field enclosed within $M$. This
action is due to the presence of covariant derivatives in dynamic
equations. The magnitude of gravitational field itself should be
determined by presence of matter in $M$ in form of massive particles
or in form of electromagnetic radiation, i\.\,e\. we should have
backward relation between geometry of the space and its content.
\par
     In order to describe backward relation between gravitational
field and other physical fields we use Lagrangian formalism and
extremal action principle. Let's start from action functional
\thetag{4.1} in Chapter~\uppercase\expandafter{\romannumeral 4}.
It is sum of three integrals:
$$
S=S_{\text{mat}}+S_{\text{int}}+S_{\text{el}}.
\tag2.1
$$
First integral $S_{\text{mat}}$ is responsible for material particles
in form of dust cloud, second integral describes interaction of dust
cloud and electromagnetic field, third term in \thetag{2.1} describes
electromagnetic field itself. In order to describe gravitational field
one more summand in \thetag{2.1} is added:
$$
S=S_{\text{gr}}+S_{\text{mat}}+S_{\text{int}}+S_{\text{el}}.
\tag2.2
$$
This additional term is chosen in the following form:
$$
S_{\text{gr}}=-\frac{c^3}{16\pi\gamma}\int\limits^{\ V_2}_{V_1}
R\,\sqrt{-\det g\,}\,d^4r.
\tag2.3
$$
Here $\gamma$ gravitational constant same as in Newton's universal
law of gravitation (see formula \thetag{1.11} in
Chapter~\uppercase\expandafter{\romannumeral 1}). Scalar quantity $R$
in \thetag{2.3} is {\it scalar curvature} determined by curvature
tensor \thetag{1.1} according to the following formula:
$$
R=\sum^3_{q=0}\sum^3_{k=0}\sum^3_{j=0}g^{qj}R^k_{qkj}.
\tag2.4
$$
{\it Ricci tensor} is an intermediate object relating curvature
tensor \thetag{1.1} and scalar quantity \thetag{2.4}. Here are
its components:
$$
R_{qj}=\sum^3_{k=0}R^k_{qkj}.
\tag2.5
$$
Ricci tensor is symmetric (see \cite{3}). Scalar curvature $R$
is obtained by contracting Ricci tensor and metric tensor $g^{qj}$
with respect to both indices $q$ and $j$. This fact is obvious
due to \thetag{2.5} and \thetag{2.4}.\par
     Note that sometimes in the action for gravitational field
\thetag{2.3} one more constant parameter $\Lambda$ is added:
$$
S_{\text{gr}}=-\frac{c^3}{16\pi\gamma}\int\limits^{\ V_2}_{V_1}
(R+2\,\Lambda)\,\sqrt{-\det g\,}\,d^4r.
$$
This parameter is called {\it cosmological constant}. However,
according to contemporary experimental data the value of this
constant is undetectably small or maybe is exactly equal to
zero. Therefore further we shall use action $S_{\text{gr}}$
in form of \thetag{2.3}.\par
    Note also that metric tensor describing gravitational field
enters in implicit form into all summand in \thetag{2.2}. Therefore
we need not add special terms describing interaction of gravitational
field with material particles and electromagnetic field. Moreover, such
additional terms could change the form of dynamical equations for matter
and form of Maxwell equations for electromagnetic field thus contradicting
our claim that these equations are the same in general and in special
relativity.\par
     Now let's begin with deriving dynamical equations for gravitational
field. For this purpose we consider deformation of components of metric
tensor given by the following relationship:
$$
\hat g^{ij}(\bold r)=g^{ij}(\bold r)+\varepsilon\,
h^{ij}(\bold r)+\ldots\,.
\tag2.6
$$
Functions $h^{ij}(\bold r)$ in \thetag{2.6} are assumed to be smooth
functions vanishing outside some restricted domain $\Omega\subset M$.
Deformation of matrix $g^{ij}$ lead to deformation of inverse matrix
$g_{ij}$:
$$
\aligned
\hat g_{ij}&=g_{ij}-\varepsilon\,h_{ij}+\ldots=\\
\vspace{1ex}
&=g_{ij}-\varepsilon\,\sum^3_{p=0}\sum^3_{q=0}
g_{ip}\,h^{pq}\,g_{qj}+\ldots\,.
\endaligned
\tag2.7
$$
Let's differentiate the relationship \thetag{2.7} and let's
express partial derivatives through covariant derivatives in
resulting formula:
$$
\aligned
\frac{\partial\hat g_{ij}}{\partial r^k}=
\frac{\partial g_{ij}}{\partial r^k}-
&\varepsilon\,\frac{\partial h_{ij}}{\partial r^k}+\ldots=
\frac{\partial g_{ij}}{\partial r^k}-\varepsilon\,\nabla_{\!k}h_{ij}+\\
\vspace{1ex}
&+\varepsilon\,\sum^3_{p=0}\Gamma^p_{ki}\,h_{pj}+
\varepsilon\,\sum^3_{p=0}\Gamma^p_{kj}\,h_{ip}+\ldots\,.
\endaligned
\tag2.8
$$
In \thetag{2.8} we used covariant derivatives corresponding to
non-deformed metric $g_{ij}$. Now on the base of \thetag{2.8}
we calculate the following combination of derivatives:
$$
\aligned
&\frac{\partial\hat g_{kj}}{\partial r^i}+
\frac{\partial\hat g_{ik}}{\partial r^j}-
\frac{\partial\hat g_{ij}}{\partial r^k}=
\frac{\partial g_{kj}}{\partial r^i}+
\frac{\partial g_{ik}}{\partial r^j}-
\frac{\partial g_{ij}}{\partial r^k}-\\
\vspace{1ex}
&-\varepsilon\left(\nabla_{\!i}{h_{kj}}+
\nabla_{\!j}{h_{ik}}-
\nabla_{\!k}{h_{ij}}-2\shave{\sum^3_{p=0}}
\Gamma^p_{ij}\,h_{pk}\right)+\ldots\,.
\endaligned
\tag2.9
$$
Let's use the relationships \thetag{2.6} and \thetag{2.8} in
calculating deformation of connection components. For this
purpose let's apply well-known formula to $\hat\Gamma^p_{ij}$
(see formula \thetag{11.3} in
Chapter~\uppercase\expandafter{\romannumeral 3}):
$$
\hat\Gamma^p_{ij}=\Gamma^p_{ij}+
\frac{\varepsilon}{2}\sum^3_{k=0}g^{pk}
\left(\nabla_{\!i}{h_{kj}}+\nabla_{\!j}{h_{ik}}-\nabla_{\!k}{h_{ij}}
\right)+\ldots\,.
$$
This expansion for $\hat\Gamma^p_{ij}$ can be written in
symbolic concise form
$$
\hat\Gamma^p_{ij}=\Gamma^p_{ij}+\varepsilon\,Y^p_{ij}
+\ldots\,
\tag2.10
$$
by introducing the following quite natural notation:
$$
Y^p_{ij}=\frac{1}{2}\sum^3_{k=0}g^{pk}
\left(\nabla_{\!i}{h_{kj}}+\nabla_{\!j}{h_{ik}}-\nabla_{\!k}{h_{ij}}
\right).\hskip-2em
\tag2.11
$$\par
     Now let's substitute the expansion \thetag{2.10} into the
formula \thetag{1.1} for curvature tensor. This yields
$$
\hat R^k_{qij}=R^k_{qij}+
\varepsilon\,\left(\nabla_{\!i}Y^k_{jq}-\nabla_{\!j}Y^k_{iq}\right)+
\ldots\,.
\tag2.12
$$
Upon contracting \thetag{2.12} with respect to one pair of indices
we get similar expansion for deformation of Ricci tensor:
$$
\hat R_{qj}=R_{qj}+\varepsilon\sum^3_{k=0}
\left(\nabla_{\!k}Y^k_{jq}-\nabla_{\!j}Y^k_{kq}\right)+\ldots\,.
\tag2.13
$$
We multiply \thetag{2.13} by $g_{qj}$ using formula \thetag{2.6}.
Then we carry out complete contraction with respect to both indices
$q$ and $j$. This yields deformation of scalar curvature:
$$
\hat R=R+\varepsilon\,\sum^3_{j=0}\sum^3_{q=0}
\left(R_{qj}\,h^{qj}+\shave{\sum^3_{k=0}}
g^{qj}(\nabla_{\!k}Y^k_{jq}-\nabla_{\!j}Y^k_{kq})\right)
+\ldots\,.
$$
Let's introduce vector field with the following components:
$$
Z^k=\sum^3_{j=0}\sum^3_{q=0}
\left(Y^k_{jq}g^{qj}-Y^j_{jq}g^{qk}\right).
$$
Then we can rewrite deformation of scalar curvature $\hat R$ as 
$$
\hat R=R+\varepsilon\,\sum^3_{j=0}\sum^3_{q=0}
R_{qj}\,h^{qj}+\varepsilon\,\sum^3_{k=0}\nabla_{\!k}Z^k+
\ldots\,.
\hskip-2em
\tag2.14
$$
When substituting \thetag{2.14} into action integral \thetag{2.3}
we should note that second sum in \thetag{2.14} is exactly covariant
divergency of vector field $\bold Z$. Components of $\bold Z$ are
smooth functions vanishing outside the domain $\Omega$. Therefore
integral of such sum is equal to zero:
$$
\int\limits_\Omega\sum^3_{k=0}\nabla_{\!k}Z^k\,\sqrt{-\det g\,}\,d^4r=
\int\limits_{\partial\Omega}g(\bold Z,\bold n)\,dV=0.
$$
This follows from Ostrogradsky-Gauss formula (see \thetag{4.14} in
Chapter~\uppercase\expandafter{\romannumeral 4}). Hence for
deformation of $S_{\text{gr}}$ we get
$$
S_{\text{def}}=S_{\text{gr}}-\frac{\varepsilon\,c^3}{16\pi\gamma}
\int\limits_\Omega\sum^3_{j=0}\sum^3_{q=0}
\left(\!R_{qj}-\frac{R}{2}g_{qj}\!\right)h^{qj}\sqrt{-\det g\,}\,d^4r+
\ldots\,.
$$
In deriving this formula we also used the following expansion:
$$
\sqrt{-\det\hat g\,}=\sqrt{-\det g\,}\left(1-
\varepsilon\shave{\sum^3_{j=0}\sum^3_{q=0}}
\frac{g_{qj}\,h^{qj}}{2}\right)+\ldots\,.
\hskip-2em
\tag2.15
$$
It follows from \thetag{2.6}. Now we shall not calculate deformations
of other three terms in \thetag{2.2} in explicit form. This will be
done in \S\,4 and \S\,5 below. However, we introduce notation
$$
S_{\text{m.f.}}=S_{\text{mat}}+S_{\text{int}}+S_{\text{el}}.
\tag2.16
$$
Here $S_{\text{m.f.}}$ denotes overall action for all material
fields other than gravitation. The number of terms in the sum
\thetag{2.16} could be much more than three, if one consider
more complicated models for describing matter. But in any case
action of gravitational field is excluded from this sum since
gravitational field plays exceptional role in general relativity.
Now we shall write deformation of the action \thetag{2.16} in
the following conditional form:
$$
S_{\text{def}}=S_{\text{m.f.}}+\frac{\varepsilon}{2c}
\int\limits_\Omega\sum^3_{q=0}\sum^3_{j=0} T_{qj}\,h^{qj}
\sqrt{-\det g\,}\,d^4r+\ldots\,.
\hskip-1.5em
\tag2.17
$$
Then extremity condition for total action \thetag{2.2} is written
as
$$
R_{qj}-\frac{R}{2}g_{qj}=\frac{8\pi\gamma}{c^4}T_{qj}.
\tag2.18
$$
This equation \thetag{2.18} is known as {\it Einstein equation}.
It is basic equation describing dynamics of metric tensor
$g_{ij}$ in general theory of relativity.\par
\proclaim{\bf Exercise 2.1} Derive the relationships \thetag{2.7}
and \thetag{2.15} from the expansion \thetag{2.6} for deformation
of tensor $g^{ij}$.
\endproclaim
\head
\S\,3. Four-dimensional momentum conservation law for fields.
\endhead
\rightheadtext{\S\,3. Conservation law \dots}
     Tensor $\bold T$ in right hand side of Einstein equation
\thetag{2.18} is called energy-momentum tensor for material fields.
It is determined by the relationship \thetag{2.17} and comprises
contributions from all material fields and their interactions.
In the model of dust matter in electromagnetic field tensor $T$
is composed of three parts (see formula \thetag{2.16}).\par
     Energy-momentum tensor is related with $4$-momentum conservation
law for material fields. In order to derive this conservation law we
use well-known Bianchi identity:
$$
\nabla_{\!k}R^p_{sij}+\nabla_{\!i}R^p_{sjk}+\nabla_{\!j}R^p_{ski}=0.
\tag3.1
$$
More details concerning Bianchi identity \thetag{3.1} can be found in
\cite{2} and \cite{6}. Let's contract this identity with respect to
$i$ and $p$:
$$
\nabla_{\!k}R_{sj}+\sum^3_{p=0}\nabla_{\!p}R^p_{sjk}
-\nabla_{\!j}R_{sk}=0.
\tag3.2
$$
Here we used skew symmetry of curvature tensor with respect
to last pair of indices (see \cite{3}). Let's multiply
\thetag{3.2} by $g^{sj}$ and contract it with respect to double
indices $s$ and $j$. Upon slight transformation based on
skew symmetry $R^{ps}_{ij}=-R^{sp}_{ij}$ we get
$$
\sum^3_{s=0}\nabla_{\!s}R^s_k-\frac{1}{2}\nabla_{\!k}R=0.
\tag3.3
$$
Now let's raise index $j$ in the equation \thetag{2.18}, then apply
covariant differentiation $\nabla_{\!j}$ and contract with respect to
double index $j$:
$$
\sum^3_{j=0}\nabla_{\!j}R^j_q-\frac{1}{2}\nabla_{\!q}R=
\frac{8\pi\gamma}{c^4}\sum^3_{j=0}\nabla_{\!j}T^j_q.
\tag3.4
$$
Comparing \thetag{3.3} and \thetag{3.4}, we get the following
equation for energy-momentum tensor of material fields:
$$
\sum^3_{j=0}\nabla_{\!j}T^j_q=0.
\tag3.5
$$
The equation \thetag{3.5} expresses {\it $4$-momentum conservation
law} for the whole variety of material fields. It is usually written
in the following form with raised index $q$: 
$$
\sum^3_{j=0}\nabla_{\!j}T^{qj}=0.
\tag3.6
$$
Energy-momentum tensor is symmetric therefore the order of indices
$q$ and $j$ in \thetag{3.6} is unessential.
\head
\S\,4. Energy-momentum tensor\\ for electromagnetic field.
\endhead
\rightheadtext{\S\,4. Energy-momentum tensor \dots}
    Energy-momentum tensor for whole variety of material fields
is defined by the relationship \thetag{2.17}. By analogy with 
\thetag{2.17} we define energy-momentum tensor for electromagnetic
field:
$$
S_{\text{def}}=S_{\text{el}}+\frac{\varepsilon}{2c}
\int\limits_\Omega\sum^3_{q=0}\sum^3_{j=0} T_{qj}\,h^{qj}
\sqrt{-\det g\,}\,d^4r+\ldots\,.
\hskip -2em
\tag4.1
$$
Basic fields in the action $S_{\text{el}}$ are covariant components of
vector-potential $A_i(\bold r)$. Covariant components of tensor of
electromagnetic field are defined by formula
$$
F_{ij}=\nabla_iA_j-\nabla_jA_i=\frac{\partial A_j}{\partial r^i}-
\frac{\partial A_i}{\partial r^j}
\tag4.2
$$
(see also formula \thetag{11.5} in
Chapter~\uppercase\expandafter{\romannumeral 3}). Ultimate expression
in right hand side of \thetag{4.2} has no entry of connection components
$\Gamma^k_{ij}$. Therefore covariant components $F_{ij}$ are not changed
by deformation of metric \thetag{2.6}. Upon raising indices we get
$$
\hat F^{pk}=\sum^3_{i=0}\sum^3_{j=0}
\hat g^{pi}\,\hat g^{kj}\,F_{ij}
$$
and, using this formula, for contravariant components $F^{pq}$ of tensor
of electromagnetic field we derive the expansion
$$
\pagebreak
\hat F^{pk}=F^{pk}+
\varepsilon\sum^3_{i=0}\sum^3_{j=0}(h^{pi}\,g^{kj}+g^{pi}\,h^{kj})
\,F_{ij}+\ldots\,.
\tag4.3
$$
Substituting $\hat F^{pk}$ and $\hat g$ into action functional
$S_{\text{el}}$, we get 
$$
S_{\text{def}}=-\frac{1}{16\pi\,c}\int\limits^{\ V_2}_{V_1}
\sum^3_{p=0}\sum^3_{k=0} F_{pk}\,\hat F^{pk}
\sqrt{-\det\hat g\,}\,d^4r
$$
Then, taking into account \thetag{4.3} and \thetag{2.15}, we derive
formula
$$
\aligned
S_{\text{def}}=S_{\text{el}}&-\frac{\varepsilon}{16\pi\,c}
\int\limits_\Omega\sum^3_{q=0}\sum^3_{j=0}\left(\,
\shave{\sum^3_{p=0}\sum^3_{i=0}}2\,F_{pq}\,g^{pi}\,F_{ij}-\right.\\
&-\frac{1}{2}\left.\shave{\sum^3_{p=0}\sum^3_{i=0}}
F_{pi}\,F^{pi}\,g_{qj}\right)h^{qj}
\sqrt{-\det g\,}\,d^4r+\ldots\,.
\endaligned
$$
Comparing this actual expansion with expected expansion \thetag{4.1}
for $S_{\text{def}}$, we find components of energy-momentum tensor
for electromagnetic field in explicit form:
$$
T_{qj}=-\frac{1}{4\pi}\shave{\sum^3_{p=0}\sum^3_{i=0}}
\left(F_{pq}\,g^{pi}\,F_{ij}-\frac{1}{4}
F_{pi}\,F^{pi}\,g_{qj}\right).
\hskip-2em
\tag4.4
$$
Raising indices $q$ and $j$ in \thetag{4.4}, for contravariant
components of energy-momentum tensor $\bold T$ we derive
$$
T^{qj}=-\frac{1}{4\pi}\shave{\sum^3_{p=0}\sum^3_{i=0}}
\left(F^{pq}\,g_{pi}\,F^{ij}-\frac{1}{4}
F_{pi}\,F^{pi}\,g^{qj}\right).
\hskip-2em
\tag4.5
$$
By means of formula \thetag{4.5} one can calculate covariant
divergency for energy-momentum tensor of electromagnetic field:
$$
\pagebreak
\sum^3_{s=0}\nabla_{\!s}T^{ps}=-\frac{1}{c}\sum^3_{s=0}F^{ps}\,j_s.
\tag4.6
$$
Formula \thetag{4.6} shows that $4$-momentum conservation law for
separate electromagnetic field is not fulfilled. This is due to
momentum exchange between electromagnetic field and other forms
of matter, e\.\,g\. dust matter.
\proclaim{\bf Exercise 4.1} Verify the relationship \thetag{4.6}.
For this purpose use well-known formula for commutator of covariant
derivatives
$$
(\nabla_i\nabla_j-\nabla_i\nabla_j)A_k=
-\sum^3_{s=0}R^s_{kij}\,A_s
$$
and properties of curvature tensor (see details in \cite{3}).
\endproclaim
\proclaim{\bf Exercise 4.2} Calculate components of energy-momentum
\linebreak
tensor \thetag{4.5} in inertial coordinate system for flat Minkowsky
metric. Compare them with components of Maxwell tensor, with density
of energy, and with vector of energy flow for electromagnetic field
(see formulas \thetag{2.5} and \thetag{2.15} in
Chapter~\uppercase\expandafter{\romannumeral 2}).
\endproclaim
\head
\S\,5. Energy-momentum tensor\\ for dust matter.
\endhead
\rightheadtext{\S\,5. Energy-momentum tensor \dots}
     Let's consider energy-momentum tensor related with last two
terms $S_{\text{mat}}$ and $S_{\text{int}}$ in the action
\thetag{2.16}. They contain entries of vector field $\boldsymbol\eta$
whose components satisfy differential equation
$$
\sum^3_{p=0}\nabla_{\!p}\eta^p=0,
\tag5.1
$$
see \thetag{3.4} in Chapter~\uppercase\expandafter{\romannumeral 4}.
This circumstance differs them from components of vector-potential
$\bold A$. Metric tensor $g_{ij}$ enters differential equation
\thetag{5.1} through connection components $\Gamma^k_{ij}$ of metric
connection. Therefore by deformation of metric $g_{ij}\to\hat g_{ij}$
one cannot treat $\eta^p$ as metric independent quantities.\par
     In order to find truly metric independent variables for dust matter
we use formula \thetag{4.7} from
Chapter~\uppercase\expandafter{\romannumeral 4} and rewrite differential
equation \thetag{5.1} as follows:
$$
\sum^3_{p=0}\frac{\partial(\eta^p\dsize\sqrt{-\det\hat g\,})}
{\partial r^p}=0.
$$
Denote $\dsize\hat\eta^p=\eta^p\sqrt{-\det g\,}$. These quantities
$\hat\eta^p$ can be treated as metric independent ones since
differential constraint for them is written in form of the equation
that does not contain metric:
$$
\sum^3_{p=0}\frac{\partial\hat\eta^p}{\partial r^p}=0.
\tag5.2
$$\par
     Expressing $\eta^p$ through $\hat\eta^p$, for action functional
$S_{\text{int}}$ describing interaction of dust matter and electromagnetic
field we get
$$
S_{\text{int}}=-\frac{q}{c^2}\int\limits^{\ V_2}_{V_1}
\sum^3_{p=0}\hat\eta^p\,A_p\,d^4r.
\tag5.3
$$
It is easy to see that integral \thetag{5.3} does not depend on metric
tensor. Therefore action functional $S_{\text{int}}$ makes no contribution
to overall energy-momentum tensor.\par
     Now let's express $\eta^p$ through $\hat\eta^p$ in action functional
$S_{\text{mat}}$  for dust matter. As a result we get formula
$$
S_{\text{mat}}=-m\int\limits^{\ V_2}_{V_1}\sqrt{\ \sum^3_{p=0}
\sum^3_{q=0}g_{pq}\,\hat\eta^p\,\hat\eta^q\ }
\,d^4r.
\tag5.4
$$
The dependence of this functional on metric tensor is completely determined
by explicit entry of $g_{pq}$ under square root sign in right hand side of
\thetag{5.4}. Therefore power extension for $S_{\text{mat}}$ is easily
calculated on the base of the expansion \thetag{2.7}:
$$
S_{\text{def}}=S_{\text{mat}}+\frac{\varepsilon}{2}
\int\limits_\Omega
\left(\shave{\sum^3_{p=0}\sum^3_{q=0}}\frac{m\,\eta_p\,\eta_q}
{\sqrt{g(\boldsymbol\eta,\boldsymbol\eta)\,}}\right)h^{pq}\,
\sqrt{-\det g\,}\,d^4r+
\ldots\,.
$$
Let's compare this expansion with expected expansion for
$S_{\text{def}}$:
$$
S_{\text{def}}=S_{\text{mat}}+\frac{\varepsilon}{2c}
\int\limits_\Omega\sum^3_{p=0}\sum^3_{q=0} T_{pq}\,h^{pq}
\sqrt{-\det g\,}\,d^4r+\ldots\,.
\hskip -2em
$$
By this comparison we find explicit formula for components of
energy-momentum tensor for dust matter:
$$
T_{pq}=\frac{mc\ \eta_p\,\eta_q}
{\sqrt{g(\boldsymbol\eta,\boldsymbol\eta)\,}}
=mc\,\sqrt{g(\boldsymbol\eta,\boldsymbol\eta)\,}\,
\,u_p\,u_q.
\tag5.5
$$\par
    Contravariant components of energy-momentum tensor \thetag{5.5}
are obtained by raising indices $p$ and $q$:
$$
T^{pq}=\frac{mc\ \eta^p\,\eta^q}
{\sqrt{g(\boldsymbol\eta,\boldsymbol\eta)\,}}
=mc\,\sqrt{g(\boldsymbol\eta,\boldsymbol\eta)\,}\,
\,u^p\,u^q.
\tag5.6
$$
Using collinearity of vectors $\bold u$ and $\boldsymbol\eta$
(see formula \thetag{3.1} in
Chapter~\uppercase\expandafter{\romannumeral 4}) and recalling that
$\bold u$ is unit vector, we can bring formula \thetag{5.6} to the
following simple form:
$$
T^{pk}=mc\,u^p\,\eta^k.
\tag5.7
$$
Formula \thetag{5.7} is convenient for calculating covariant divergency
of energy-momentum tensor for dust matter:
$$
\sum^3_{s=0}\nabla_{\!s}T^{ps}=\frac{q}{c}\sum^3_{s=0}
F^{ps}\,\eta_s.
\tag5.8
$$
Now, applying formula \thetag{3.2} from
Chapter~\uppercase\expandafter{\romannumeral 4}, we can transform
formula \thetag{5.8} and write it as follows:
$$
\sum^3_{s=0}\nabla_{\!s}T^{ps}=\frac{1}{c}\sum^3_{s=0}
F^{ps}\,j_s.
\tag5.9
$$\par
     Let's compare \thetag{5.9} with analogous formula \thetag{4.6}
for energy-momentum tensor of electromagnetic field. Right hand sides
of these two formulas differ only in sign. This fact has transparent
interpretation. It means that in our model the overall energy-momentum
tensor for matter
$$
\bold T_{\text{m.f.}}=\bold T_{\text{mat}}+\bold T_{\text{el}}
$$
satisfies differential equation \thetag{3.6}. This fact is in complete
agreement with $4$-momentum conservation law.\par
     Another important conclusion, which follows from of \thetag{4.6}
and \thetag{5.9}, is that $4$-momentum conservation law for the whole
variety of material fields can be derived from dynamical equations for
these fields. Therefore this law is valid also in special relativity,
where Einstein equation \thetag{2.18} is not considered and where in
general case for flat Minkowsky metric it is not fulfilled.
\proclaim{\bf Exercise 5.1} Derive the relationship \thetag{5.8}
on the base of equations \thetag{3.4} and \thetag{4.19} from
Chapter~\uppercase\expandafter{\romannumeral 4}.
\endproclaim
\head
\S\,6. Concluding remarks.
\endhead
     Event space in general theory of relativity is some Minkowsky
manifold $M$ with Minkowsky metric of signature $(1,3)$. This metric
is determined by material content of the space according to Einstein
equation \thetag{2.18}. However, topology of the manifold $M$ has
great deal of arbitrariness. This manifold can have local singularities
at the points with extremely high concentration of matter. Such
objects are called {\it black holes}. Moreover, global topology of
$M$ also can be nontrivial (other than topology of $\Bbb R^4$).
In contemporary physics most popular models of $M$ include {\it big bang}
in the very beginning of times. According to these models in far past
times our Universe $M$ was extremely small, while density of matter in
it was extremely high. In further evolution our Universe was expanding
up to its present size. Will this expansion last infinitely long or it
will change for contraction\,? This problem is not yet solved. The answer
to this question depends on estimates of total amount of matter in
the Universe.\par
     In this book we cannot consider all these fascinating problems
of modern astrophysics and cosmology. However, I think the above 
theoretical material makes sufficient background for to continue
studying these problems e\.\,g\. in books \cite{2}, \cite{7}, and
\cite{8}. I would like also to recommend the book \cite{9} of
popular genre, where these problems are discussed in commonly
understandable and intriguing manner.\par
\newpage
\setfirstpage
\topmatter
\title
References
\endtitle
\endtopmatter
\document
\par\noindent
\hangindent=1.4em
\hangafter=1
1.~Vladimirov~V.~S. {\it Equations of mathematical physics},
Nauka publishers, Moscow, 1981.
\par\noindent
\hangindent=1.4em
\hangafter=1
2.~Dubrovin~B.~A., Novikov~S.~P., Fomenko~A.~T. {\it Modern geometry,
vol.~\uppercase\expandafter{\romannumeral 1}},
Nauka publishers, Moscow, 1986.
\par\noindent
\hangindent=1.4em
\hangafter=1
3. Sharipov~R.~A. {\it Course of differential geometry},
Publication of Bashkir State University, Ufa, 1996.
\par\noindent
\hangindent=1.4em
\hangafter=1
4. Sharipov~R.~A. {\it Course of linear algebra and multidimensional
geometry}, Publication of Bashkir State University, Ufa, 1996.
\par\noindent
\hangindent=1.4em
\hangafter=1
5.~Borisovich~Yu.~G., Bliznyakov~N.~M., Izrailevich~Ya.~A.,
Fomenko~T.~N. {\it Introduction to topology}, Nauka publishers, Moscow,
1995.
\par\noindent
\hangindent=1.4em
\hangafter=1
6. Kobayashi~Sh., Nomizu~K. {\it Foundations of differential geometry},
Interscience publishers, New York, London, 1963.
\par\noindent
\hangindent=1.4em
\hangafter=1
7.~Landau~L.~D., Lifshits~E.~M. {\it Course of theoretical physics,
vol.~\uppercase\expandafter{\romannumeral 2}, Field theory},
Nauka publishers, Moscow, 1988.
\par\noindent
\hangindent=1.4em
\hangafter=1
8. Bogoyavlensky~O.~I. {\it Methods of qualitative theory of dynamical
systems in astrophysics and in gas dynamics},
Nauka publishers, Moscow, 1980.
\par\noindent
\hangindent=1.4em
\hangafter=1
9.~Davis~P. {\it Superforce. The search for a grand unified theory of nature},
Symon and Schuster publishers, New York, 1984.
\par\noindent
\hangindent=1.4em
\hangafter=1
\par
\newpage
\setfirstpage
\topmatter
\title
Contacts
\endtitle
\endtopmatter
\document
\line{
\vtop{\hsize 5cm
{\bf Address:}\medskip\noindent
Ruslan A. Sharipov,\newline
Math. Department,\newline
Bashkir State University,\newline
Frunze street 32,\newline
450074, Ufa, Bashkortostan,\newline
Russia
\medskip
{\bf Phone:}\medskip
\noindent
7-(3472)-23-67-18\newline
7-(3472)-23-67-74 (FAX)
}\hss
\vtop{\hsize 4.3cm
{\bf Home address:}\medskip\noindent
Ruslan A. Sharipov,\newline
Rabochaya street 5,\newline
450003, Ufa, Bashkortostan,\newline
Russia
\vskip 1cm
{\bf E-mails:}\medskip
\noindent
R\hskip 0.5pt\_\hskip 1.5pt Sharipov\@ic.bashedu.ru\newline
r-sharipov\@mail.ru\newline
ra\hskip 0.5pt\_\hskip 1.5pt sharipov\@hotmail.com
}
}

\par
\newpage
\enddocument
\end